\documentclass[twocolumn,prd,superscriptaddress,amsmath,amssymb,nofootinbib]{revtex4-1}

\usepackage[english]{babel}
\usepackage{graphicx}
\usepackage{xspace}
\usepackage{siunitx}
\usepackage[utf8]{inputenc}
\usepackage{dcolumn}

\usepackage[colorlinks=false,urlcolor=blue, linkcolor=blue, citecolor=blue,pdfborder={0 0 0}]{hyperref}
\usepackage[caption=false,labelfont=up,justification=justified]{subfig}

\usepackage{booktabs}
\usepackage{placeins}

\addto\captionsenglish{}

\newcommand{\corsika}{\textsc{Corsika}}
\newcommand{\geant}{\textsc{Geant4}}
\newcommand{\fluka}{\textsc{Fluka}}

\begin{document}

\title{Density of GeV muons in air showers measured with IceTop}

\affiliation{III. Physikalisches Institut, RWTH Aachen University, D-52056 Aachen, Germany}
\affiliation{Department of Physics, University of Adelaide, Adelaide, 5005, Australia}
\affiliation{Dept. of Physics and Astronomy, University of Alaska Anchorage, 3211 Providence Dr., Anchorage, AK 99508, USA}
\affiliation{Dept. of Physics, University of Texas at Arlington, 502 Yates St., Science Hall Rm 108, Box 19059, Arlington, TX 76019, USA}
\affiliation{CTSPS, Clark-Atlanta University, Atlanta, GA 30314, USA}
\affiliation{School of Physics and Center for Relativistic Astrophysics, Georgia Institute of Technology, Atlanta, GA 30332, USA}
\affiliation{Dept. of Physics, Southern University, Baton Rouge, LA 70813, USA}
\affiliation{Dept. of Physics, University of California, Berkeley, CA 94720, USA}
\affiliation{Lawrence Berkeley National Laboratory, Berkeley, CA 94720, USA}
\affiliation{Institut f{\"u}r Physik, Humboldt-Universit{\"a}t zu Berlin, D-12489 Berlin, Germany}
\affiliation{Fakult{\"a}t f{\"u}r Physik {\&} Astronomie, Ruhr-Universit{\"a}t Bochum, D-44780 Bochum, Germany}
\affiliation{Universit{\'e} Libre de Bruxelles, Science Faculty CP230, B-1050 Brussels, Belgium}
\affiliation{Vrije Universiteit Brussel (VUB), Dienst ELEM, B-1050 Brussels, Belgium}
\affiliation{Department of Physics and Laboratory for Particle Physics and Cosmology, Harvard University, Cambridge, MA 02138, USA}
\affiliation{Dept. of Physics, Massachusetts Institute of Technology, Cambridge, MA 02139, USA}
\affiliation{Dept. of Physics and The International Center for Hadron Astrophysics, Chiba University, Chiba 263-8522, Japan}
\affiliation{Department of Physics, Loyola University Chicago, Chicago, IL 60660, USA}
\affiliation{Dept. of Physics and Astronomy, University of Canterbury, Private Bag 4800, Christchurch, New Zealand}
\affiliation{Dept. of Physics, University of Maryland, College Park, MD 20742, USA}
\affiliation{Dept. of Astronomy, Ohio State University, Columbus, OH 43210, USA}
\affiliation{Dept. of Physics and Center for Cosmology and Astro-Particle Physics, Ohio State University, Columbus, OH 43210, USA}
\affiliation{Niels Bohr Institute, University of Copenhagen, DK-2100 Copenhagen, Denmark}
\affiliation{Dept. of Physics, TU Dortmund University, D-44221 Dortmund, Germany}
\affiliation{Dept. of Physics and Astronomy, Michigan State University, East Lansing, MI 48824, USA}
\affiliation{Dept. of Physics, University of Alberta, Edmonton, Alberta, Canada T6G 2E1}
\affiliation{Erlangen Centre for Astroparticle Physics, Friedrich-Alexander-Universit{\"a}t Erlangen-N{\"u}rnberg, D-91058 Erlangen, Germany}
\affiliation{Physik-department, Technische Universit{\"a}t M{\"u}nchen, D-85748 Garching, Germany}
\affiliation{D{\'e}partement de physique nucl{\'e}aire et corpusculaire, Universit{\'e} de Gen{\`e}ve, CH-1211 Gen{\`e}ve, Switzerland}
\affiliation{Dept. of Physics and Astronomy, University of Gent, B-9000 Gent, Belgium}
\affiliation{Dept. of Physics and Astronomy, University of California, Irvine, CA 92697, USA}
\affiliation{Karlsruhe Institute of Technology, Institute for Astroparticle Physics, D-76021 Karlsruhe, Germany }
\affiliation{Karlsruhe Institute of Technology, Institute of Experimental Particle Physics, D-76021 Karlsruhe, Germany }
\affiliation{Dept. of Physics, Engineering Physics, and Astronomy, Queen's University, Kingston, ON K7L 3N6, Canada}
\affiliation{Dept. of Physics and Astronomy, University of Kansas, Lawrence, KS 66045, USA}
\affiliation{Department of Physics and Astronomy, UCLA, Los Angeles, CA 90095, USA}
\affiliation{Centre for Cosmology, Particle Physics and Phenomenology - CP3, Universit{\'e} catholique de Louvain, Louvain-la-Neuve, Belgium}
\affiliation{Department of Physics, Mercer University, Macon, GA 31207-0001, USA}
\affiliation{Dept. of Astronomy, University of Wisconsin{\textendash}Madison, Madison, WI 53706, USA}
\affiliation{Dept. of Physics and Wisconsin IceCube Particle Astrophysics Center, University of Wisconsin{\textendash}Madison, Madison, WI 53706, USA}
\affiliation{Institute of Physics, University of Mainz, Staudinger Weg 7, D-55099 Mainz, Germany}
\affiliation{Department of Physics, Marquette University, Milwaukee, WI, 53201, USA}
\affiliation{Institut f{\"u}r Kernphysik, Westf{\"a}lische Wilhelms-Universit{\"a}t M{\"u}nster, D-48149 M{\"u}nster, Germany}
\affiliation{Bartol Research Institute and Dept. of Physics and Astronomy, University of Delaware, Newark, DE 19716, USA}
\affiliation{Dept. of Physics, Yale University, New Haven, CT 06520, USA}
\affiliation{Dept. of Physics, University of Oxford, Parks Road, Oxford OX1 3PU, UK}
\affiliation{Dept. of Physics, Drexel University, 3141 Chestnut Street, Philadelphia, PA 19104, USA}
\affiliation{Physics Department, South Dakota School of Mines and Technology, Rapid City, SD 57701, USA}
\affiliation{Dept. of Physics, University of Wisconsin, River Falls, WI 54022, USA}
\affiliation{Dept. of Physics and Astronomy, University of Rochester, Rochester, NY 14627, USA}
\affiliation{Department of Physics and Astronomy, University of Utah, Salt Lake City, UT 84112, USA}
\affiliation{Oskar Klein Centre and Dept. of Physics, Stockholm University, SE-10691 Stockholm, Sweden}
\affiliation{Dept. of Physics and Astronomy, Stony Brook University, Stony Brook, NY 11794-3800, USA}
\affiliation{Dept. of Physics, Sungkyunkwan University, Suwon 16419, Korea}
\affiliation{Institute of Basic Science, Sungkyunkwan University, Suwon 16419, Korea}
\affiliation{Institute of Physics, Academia Sinica, Taipei, 11529, Taiwan}
\affiliation{Dept. of Physics and Astronomy, University of Alabama, Tuscaloosa, AL 35487, USA}
\affiliation{Dept. of Astronomy and Astrophysics, Pennsylvania State University, University Park, PA 16802, USA}
\affiliation{Dept. of Physics, Pennsylvania State University, University Park, PA 16802, USA}
\affiliation{Dept. of Physics and Astronomy, Uppsala University, Box 516, S-75120 Uppsala, Sweden}
\affiliation{Dept. of Physics, University of Wuppertal, D-42119 Wuppertal, Germany}
\affiliation{DESY, D-15738 Zeuthen, Germany}

\author{R. Abbasi}
\affiliation{Department of Physics, Loyola University Chicago, Chicago, IL 60660, USA}
\author{M. Ackermann}
\affiliation{DESY, D-15738 Zeuthen, Germany}
\author{J. Adams}
\affiliation{Dept. of Physics and Astronomy, University of Canterbury, Private Bag 4800, Christchurch, New Zealand}
\author{J. A. Aguilar}
\affiliation{Universit{\'e} Libre de Bruxelles, Science Faculty CP230, B-1050 Brussels, Belgium}
\author{M. Ahlers}
\affiliation{Niels Bohr Institute, University of Copenhagen, DK-2100 Copenhagen, Denmark}
\author{M. Ahrens}
\affiliation{Oskar Klein Centre and Dept. of Physics, Stockholm University, SE-10691 Stockholm, Sweden}
\author{J.M. Alameddine}
\affiliation{Dept. of Physics, TU Dortmund University, D-44221 Dortmund, Germany}
\author{A. A. Alves Jr.}
\affiliation{Karlsruhe Institute of Technology, Institute for Astroparticle Physics, D-76021 Karlsruhe, Germany }
\author{N. M. Amin}
\affiliation{Bartol Research Institute and Dept. of Physics and Astronomy, University of Delaware, Newark, DE 19716, USA}
\author{K. Andeen}
\affiliation{Department of Physics, Marquette University, Milwaukee, WI, 53201, USA}
% ...

\date{\today}

\collaboration{IceCube Collaboration}
\thanks{Contact: \href{mailto:analysis@icecube.wisc.edu}{analysis@icecube.wisc.edu}}
\noaffiliation

\begin{abstract}

We present a measurement of the density of GeV muons in near-vertical
air showers using three years of data recorded by the IceTop array at
the South Pole. Depending on the shower size, the muon
densities have been measured at lateral distances between \SI{200}{m} and \SI{1000}{m}. 
From these lateral distributions, we derive the muon densities as
functions of energy at reference distances of \SI{600}{m} and \SI{800}{m} for
primary energies between \SI{2.5}{PeV} and \SI{40}{PeV} and between \SI{9}{PeV} and \SI{120}{PeV},
respectively. The muon densities are determined using, as a
baseline, the hadronic interaction model Sibyll 2.1 together with various
composition models. The measurements are consistent with the
predicted muon densities within these baseline interaction and composition models. The measured muon
densities have also been compared to simulations using
the post-LHC models EPOS-LHC and QGSJet-II.04. The result of this comparison 
is that the post-LHC models together with any given composition model yield higher muon densities than observed. 
This is in contrast to the observations above \SI{1}{EeV} where all model simulations 
yield for any mass composition lower muon densities than the measured ones. The post-LHC 
models in general  feature higher muon densities so that the agreement with experimental data 
at the highest energies is improved but the muon densities are not correct in the 
energy range between \SI{2.5}{PeV} and about \SI{100}{PeV}.
\end{abstract}

\maketitle

\section{Introduction}

Cosmic rays with energies above \SI{1}{PeV} enter the Earth’s atmosphere where they produce extensive air showers that can be measured with large detectors at the ground. The properties of the initial cosmic ray, such as energy and mass, are inferred indirectly from the particles measured at the ground and their interpretation strongly relies on Monte Carlo simulations of the shower development and thus on theoretical models. Although the energy spectrum of cosmic rays has been measured with high precision over many orders of magnitude, the sources of cosmic rays are still unknown, their acceleration mechanisms and mass composition are uncertain, and several features observed in the energy spectrum are not well understood~\cite{Kampert:2012mx}. One of the main challenges in understanding cosmic-ray induced extended air showers
is the accurate description of hadronic interactions over several decades in
center-of-mass energy, from a few tens of GeV to more than \SI{e9}{GeV}. The
relevant interactions are in the forward fragmentation
region, with most of the energy beamed into pseudo-rapidity ranges that are
difficult to study in colliders since they lie very close to the incident beam~\cite{Engel:2011zzb,Albrecht:2021yla}.
Their cross-sections cannot be computed from quantum chromodynamics, because the
strong coupling for these interactions is too large for a perturbative solution and the 
interactions are too complex for current lattice calculations.
Instead, they are calculated using phenomenological models tuned to a variety of
data sets from collider and fixed-target experiments, and are extrapolated to the ranges of phase space
relevant for air showers.

Several hadronic interaction models are available. The most recent versions are
colloquially called \emph{post-LHC}, since they take high-energy data from the
LHC into account. These are Sibyll 2.3~\cite{Riehn:2015oba,Engel:2019dsg}, EPOS-LHC~\cite{Pierog:2013ria}, and
QGSJet-II.04~\cite{Ostapchenko:2013pia}.
Sibyll 2.1~\cite{Ahn:2009wx} is a \emph{pre-LHC} model that is still popular for
simulations of air showers.
Air shower experiments can help improve the hadronic interaction models by testing them.
Tests can be performed by making various complementary measurements of the muons in air
showers, covering different ranges in primary energy, primary arrival direction,
muon energy, muon lateral separation, and stage of the shower development.

Correctly accounting for muon production in air showers is of vital importance
to the study of cosmic rays. The muon content of an air shower, together with a
measure of its electromagnetic component, can be used to simultaneously estimate
the energy and mass of the primary cosmic ray. Since cosmic ray arrival
directions do not point back to their sources, the cosmic-ray spectrum and mass
composition are key observables for testing astrophysical models. If the energy spectrum were
accurately determined for different primary cosmic ray species, several
competing astrophysical models for the origin and propagation of cosmic rays could
be disfavored or excluded~\cite{Kampert:2012mx}. However, our incomplete knowledge of high-energy
hadronic interactions causes differences between simulated and measured air showers. As the analysis of indirect measurements relies on simulations, the uncertainty in hadronic interactions weakens otherwise powerful techniques
for inferring the mass of cosmic rays.

A cosmic-ray air shower consists of a chain of hadronic interactions
whose main characteristics can be understood using the simple Heitler-Matthews
model~\cite{Heitler,Matthews:2005sd}. In this simple picture, if the primary particle is a proton the number of muons scales sub-linearly with the energy:
\begin{equation}\label{eq:muon_scaling}
N_{\mu,p}\propto E^\beta \text{ with } \beta \simeq 0.85,
\end{equation}
An air shower initiated by a nucleus with $A$ nucleons is approximated by a superposition of
proton showers, each with energy $E/A$. In this case the number of muons is
\begin{equation}
N_{\mu,A}\propto A \left(\frac{E}{A}\right)^\beta = A^{1-\beta} E^\beta.
\end{equation}
In this model, an iron primary with 56 nucleons produces about
50\,\% more muons than a proton shower with the same energy.
The proportionality factor in Eq.~\ref{eq:muon_scaling} depends
on a complex interplay of many factors, and varies greatly depending on the
hadronic interaction model used to simulate the interactions~\cite{Ulrich:2010rg}.
Among the relevant factors are the
amount of energy channeled into the electromagnetic cascade at each stage, the
inelasticity of the interactions, their multiplicity and the energy/angular
distributions of the secondary mesons.

The earliest measurements of the muon component of air showers date to the 1960s~\cite{Greisen:1960wc,PhysRev.124.1982,RossiICRC1960}. At the time, Greisen proposed a parameterization for the lateral distribution of muons in air showers with energies in the \SI{e6}{}-\SI{e8}{GeV}
range to describe the measurements performed by the MIT
air shower program.
The number of muons in a wide range of primary energies was subsequently measured with the
Akeno extensive air-shower experiment~\cite{Akeno_1984,Akeno_muons_1995}.
Recently, the KASCADE-Grande collaboration published the energy spectra of
elemental groups of cosmic rays~\cite{Apel:2013dga} and the lateral
distributions of muons ($E_\mu>\SI{800}{MeV}$)~\cite{Apel:2014qqa} at radial distances of \SI{100}{m} to \SI{610}{m} in the primary energy range between \SI{10}{PeV} and \SI{200}{PeV} for zenith angles $\theta<\SI{18}{\degree}$. Neither of these measurements show a discrepancy between the abundance of muons in real showers compared to
simulated showers using QGSJet~II as hadronic model. However, surface detector arrays like Akeno and
KASCADE-Grande rely on simulations for inferring both the shower energy and
the muon number. An over- or under-abundance of muons in simulated air showers
affects both variables, and therefore surface arrays have not been able to
test for a discrepancy in the number of muons at a given energy.

Using air fluorescence techniques~\cite{CornellFlourescence,Baltrusaitis:1985mx},
it is possible to simultaneously measure the fluorescence profile of air showers, which
is used to determine the energy and mass of the primary, and the signals produced in an array of detectors on the ground, including the contributions from muons. Using these techniques, a discrepancy between the simulated and measured
number of muons in air showers between \SI{100}{PeV} and \SI{1}{EeV} was first reported by the HiRes-MIA
collaboration~\cite{PhysRevLett.84.4276}.
The density of muons ($E_\mu>\SI{850}{MeV}$) at \SI{600}{m} from the shower axis was found
to be larger than the density of muons in a sample of iron
primaries simulated using the QGSJet and Sibyll hadronic models~\cite{Kalmykov:1997te,Fletcher:1994bd}, while
the fluorescence measurements showed a composition much
lighter than pure iron. The Pierre Auger Observatory measured the muon
abundance ($E_\mu>\SI{300}{MeV}$) in air showers above \SI{4}{EeV} in two complementary analyses using \emph{inclined} ($\theta>\SI{65}{\degree}$)
and \emph{vertical} ($\theta < \SI{65}{\degree}$) air showers, respectively, which probe the muon flux at a different atmospheric mass overburden. Both measurements reported a deficit of muons in simulations for all hadronic interaction models~\cite{Aab:2014pza,Aab:2016hkv}. More recently, this deficit has also been confirmed in vertical air showers between \SI{200}{PeV} and \SI{2}{EeV} ($\theta<45^\circ$) by an analysis using dedicated underground muon detectors at the Pierre Auger Observatory~\cite{Aab:2020frk}. 
The measurement of the cosmic ray flux around \SI{1}{EeV} and higher, based on muon densities several meters from the shower core ($E_\mu>\SI{2}{GeV}$), by the
NEVOD-DECOR experiment can also be interpreted as evidence for a
muon deficit in simulations~\cite{Kokoulin:2009zz,SaavedraSanMartin:2017ito}.

Other results are mixed. The measurement of muons ($E_\mu>\SI{1}{GeV}$) by the Yakutsk array at primary energies of above \SI{20}{EeV} and zenith angles up to \SI{45}{\degree} appears to support the hypothesis of a deficit of muons in simulations~\cite{Glushkov:2007gd}. However, measurements of muons ($E_\mu>\SI{10}{GeV}$)  by the EAS-MSU
surface detector at radial distances up to \SI{200}{m} from air showers with energies between \SI{100}{} and \SI{500}{PeV} ($\theta<\SI{30}{\degree}$) do not appear to yield a discrepancy~\cite{Fomin:2016kul}.  Comparisons of measurements of the muon flux at energies between \SI{1}{TeV} and \SI{10}{TeV} with simulations show a deficit or excess, depending on the model~\cite{Dedenko:2015qga,Dedenko:2015hna}.  To further complicate the matter, the KASCADE-Grande collaboration has measured the attenuation of the muon number ($E_\mu>\SI{230}{MeV}$)  in air showers between \SI{10}{PeV} and \SI{1}{EeV} as a function of the zenith angle and thus in dependence of the mass overburden~\cite{Arteaga-Velazquez:2015jpa}. This study found a larger attenuation length of muons when compared to several hadronic interaction models which indicates that the simulated muon energy spectra are steeper than in experimental data.

The IceCube Neutrino Observatory~\cite{Aartsen:2016nxy} is uniquely suited
to probe the muon content of air showers. The deep portion of
the detector has been used to study the spectrum of atmospheric muons~\cite{Aartsen:2015nss}
and the lateral distribution of high-energy muons in air
showers~\cite{Abbasi:2012kza,Soldin:2018vak}, because high-energy muons are
able to penetrate the ice shield above the detector. Low-energy muons are detected with IceTop,
the surface component of IceCube~\cite{IceCube:2012nn}. IceTop has been used to measure
the energy spectrum of cosmic rays in the energy range from \SI{250}{TeV}
up to about \SI{1}{EeV}~\cite{IceCube:2020yct,Aartsen:2013wda,IceCube:2019}. The high altitude of IceTop, at \SI{2.8}{km}
above sea level and a depth of around \SI{690}{g/cm^2}, places it close to the shower maximum.
The proximity to the shower maximum allows for an energy resolution better than 0.1 in $\log_{10}(E)$.
Hybrid measurements, which used both detector components, have been used
to infer the mass composition of cosmic rays~\cite{IceCube:2012vv,IceCube:2019}.

This article reports on the density of muons measured with IceTop above about \SI{100}{MeV}, mostly in the several 
GeV region, in the following referred to as \emph{GeV muons}.
IceTop does not have dedicated muon counters, but can identify
muon hits far away from the shower axis by exploiting the difference
in detector response of water-Cherenkov detectors to muons and to
electrons, positrons and gammas. Muons form a distinctive peak in the IceTop signal distributions, which can be used to determine the  density of muons in air showers. We first estimate the muon density without
relying on Monte Carlo simulations of air showers. A correction of
less than 10\% is then applied which is estimated using Monte Carlo
simulations. This correction takes into account the finite resolution of the experiment, attenuation in the
snow, and detection and selection efficiencies. By comparing our
result with the expectations according to simulations, we constrain
the current hadronic interaction models.

The experimental setup is described in Section~\ref{section:general}, the event selection in
Section~\ref{section:datasets}, and the analysis in Section~\ref{section:analysis}.
The statistical model for signals produced by
muons is described in Section~\ref{section:muon_model}. Two kinds of backgrounds are considered:
signals without muons are described in Section~\ref{section:em_model} and accidental coincidences
in Section~\ref{section:background}. The effect of a growing snow cover on top of the IceTop detectors
is  discussed in Section~\ref{section:snow}. The correction factors obtained from simulations of the detector 
response are presented in Section~\ref{section:mc_verification}. The final result is
presented in Section~\ref{section:results}.

\section{IceTop}
\label{section:general}

IceTop is an air shower array of 81 stations deployed in a triangular grid with a typical
separation of \SI{125}{m}~\cite{IceCube:2012nn}. It was completed in December 2010. The detector covers an area of
approximately one square kilometer and is located above the IceCube detector at the
geographic South Pole. Each station consists of two Cherenkov detectors
separated by ten meters. The Cherenkov
detectors are cylindrical tanks with an inner radius of \SI{0.91}{m}. The detectors
contain two \emph{Digital Optical Modules} (DOMs) and are filled with clear ice up to
\SI{0.9}{m} depth. The \SI{40}{cm} between the ice surface and the lid are filled with
expanded perlite (amorphous volcanic glass, expanded to low density with grain sizes
of the order of \SI{1}{mm}) for thermal insulation and light protection.
Each DOM combines a 10\,inch photo-multiplier tube (PMT) with
electronics for signal processing and readout~\cite{Abbasi:2008aa,Abbasi:2010vc}.

A \emph{discriminator trigger} occurs when the voltage in one of the DOMs in a tank
exceeds a predefined threshold and the capture of the PMT waveform and digitization are launched. Stations have two readout modes. A \emph{Hard Local Coincidence} (HLC) occurs when both tanks in a station have a discriminator trigger within a time window of \SI{1}{\micro s}. If there is a discriminator trigger in only one tank, it is called a \emph{Soft Local Coincidence} (SLC). 
The anode pulse is sampled in time bins, which are then digitized and baseline subtracted. In SLC mode, each launched DOM is read out but only coarse information, \emph{i.e.} time stamp and total charge of a DOM, is transmitted. In HLC mode, in addition to the SLC data, also the full waveform is transmitted. For HLCs, the total charge of a DOM is also calculated from the waveform offline, after a better estimate of the baseline becomes available.
In the case of SLCs, this is done by the DOM firmware, using the best estimate of the baseline at the time.
For the current analysis the total charge together with a time stamp constitutes the tank's \textit{signal}, $S$.
The calibrated charge is expressed in
units of \emph{Vertical Equivalent Muon} (VEM), which is the average charge
produced by a vertically through-going muon.

The shower direction, the intersection point of the shower axis with IceTop (the
\emph{shower core}), and the shower size are estimated by fitting the measured
signals with a \emph{Lateral Distribution Function} (LDF) and the times of the signals with a
phenomenological model of the shower front~\cite{IceCube:2012nn,Abbasi:2012wn}.
The LDF and the shower front model are functions of the lateral distance,
the distance of closest approach between the shower axis and the tanks.
The LDF includes an attenuation factor due to the snow cover on top of each
tank. To estimate the energy of the primary cosmic ray, we use the relationship
between the shower size $S_{125}$, defined as the signal at a lateral distance
of \SI{125}{m}, and the true primary energy, as obtained from simulations~\cite{Aartsen:2013wda}. This relation was derived using air shower simulations assuming a specific composition model commonly referred to as H3a.
This is a 4-component model based on the 5-component model by
Gaisser~\cite{Gaisser:2011cc}, with the nitrogen and aluminum components
merged into a single component with oxygen as the representative element. For $\cos \theta > 0.95$, where $\theta$ is the zenith angle of the arrival direction, the relation is:
\begin{equation}
\log_{10}(E) = 0.938 \log_{10} (S_{125}) + 6.018.
\end{equation}

\section{Datasets}
\label{section:datasets}

\subsection{Experimental dataset}

This analysis uses data collected between May 31, 2010 and May 2, 2013 in three yearly campaigns
(IC79.2010, IC86.2011 and IC86.2012). The detector grew from 73 stations in the 2010-2011 campaign to 81
stations thereafter. More than 18 million events with reconstructed energy $E_\mathrm{reco}>\SI{1}{PeV}$ and passing the selection criteria were collected during this 3-year period, totaling around 947 days of data acquisition.

To improve the general quality of reconstructions and to
stay within the simulated zenith range, the following cuts
were applied to the simulated and the experimental data. These cuts are the same
as in previous IceTop analyses~\cite{Aartsen:2013wda,IceCube:2019}:
\begin{itemize}
\item Events must trigger at least five stations and the reconstruction fit must succeed.
\item Reconstructed cores must be within the geometric boundary of the array.
\item The station with the largest signal must not be at the edge of the detector.
\item There must be at least one station with signal greater than \SI{6}{VEM}.
\end{itemize}

After cuts, the energy resolution, defined as the standard deviation
of the distribution of $\log_{10}(E_\mathrm{reco}/E_\mathrm{true})$, is better than 0.1 at \SI{1}{PeV} and around 0.05 at \SI{100}{PeV}. For further analysis, only events with reconstructed energy $E_\mathrm{reco}\geq \SI{2.5}{PeV}$ are considered, an energy above which IceTop reaches a detection efficiency close to $ 100\%$ for all cosmic ray masses, from hydrogen up to iron~\cite{IceCube:2019}. This analysis is further restricted to events with zenith angles $\theta < \SI{18}{\degree}$, in contrast
with previous IceTop analyses~\cite{Abbasi:2012wn,Aartsen:2013wda,IceCube:2019}. This is
because vertical muons form a distinctive peak around \SI{1}{VEM} in the signal distribution, a feature that is used in this analysis to determine the muon density in air showers.

Another difference with previous IceTop analyses is the addition of SLCs.
In previous publications, only HLC signals were used. The
SLC signals are not used to reconstruct the energy and arrival
direction of the air showers, but they are
essential for the measurement of the muon density, since they occur at large lateral
distances, where the muon component dominates over the electromagnetic
component and there is a high probability that only one tank of a station is hit. For HLC signals used in the muon determination, only the information also available for SLC signals is used (\emph{i.e.} total charge and time stamp).

While precipitation at the South Pole only amounts to about \SI{2}{cm} per year,
there is significant snow accumulation due to wind-driven drift. The snow
height on top of the detectors increases on average by \SI{20}{cm} every
year. The snow cover over each tank is measured twice a year and is accounted for during the air shower reconstruction process (see Section~\ref{section:snow}).

\subsection{Simulated datasets}
\label{section:simulation}

The simulation process
starts with the production of air showers using \corsika~\cite{corsika}.
The resulting air showers are used as input to a detailed simulation of the
detector response. As in previous IceTop analyses~\cite{Aartsen:2013wda,IceCube:2019}, at the input to the detector response, each \corsika{} shower is resampled 100 times
to increase statistics. Shower cores are uniformly distributed over areas
larger than the detector area, with an energy-dependent resampling radius.
Resampling radii are chosen at the largest distance possible for the shower to
trigger the array. The detector response is simulated using custom software
which simulates the entire hardware and data acquisition chain. The
interactions of particles with the IceTop tanks are simulated using the
\geant{} package~\cite{geant4}, which handles the treatment of particles starting well above the detector, including their interactions in the snow. The simulated shower energies are distributed according to an
$E^{-1}$ differential spectrum between \SI{0.1}{} and \SI{100}{PeV}, and
their zenith angles are distributed uniformly in
$\sin^2\theta$ between \SI{0}{} and \SI{65}{\degree}. This angular distribution accounts for the
projection of the detector area on a plane perpendicular to the air
shower direction. Showers induced by four primary types (H, He, O, Fe) are
simulated.

This particular analysis is sensitive to \textit{accidental}
coincidences.  These are signals uncorrelated with the event, but
coincident in time. They are produced by lower-energy showers.
This was not an issue in previous analyses because they used only HLC
signals, and HLC signals are rarely produced by lower-energy showers.
For this reason, we enhanced the standard simulation chain to include a background
consisting of accidental coincidences as well as a potential fluctuations in the VEM calibration values over time.
These are described in Section~\ref{section:simulation}.

Several simulated datasets are combined in this study in order to
reproduce the experimental conditions over multiple years. The accumulation of snow on the
IceTop stations increases the energy threshold of the trigger. To
account for this effect, we simulate the detector in three \emph{epochs}
using different values of snow coverage in each epoch. For
convenience, the epochs are chosen to correspond to each campaign
year. The amount of snow specified for each
epoch corresponds to in-situ measurements made roughly half way through each
campaign, in October 2010, November 2011 and October 2012 respectively. The simulations use moderately different software versions and two models for the South Pole atmosphere for an atmospheric mass overburden of \SI{692.9}{g/cm^2} (\SI{680}{hPa}).

The IC79.2010 epoch simulated datasets, each with 60,000 showers per primary, are the same
ones used in a previous publication~\cite{Aartsen:2013wda}. They were produced
with \corsika{} v69900, using atmosphere 12, which is based on the July 1, 1997
South Pole atmosphere. The datasets in IC86.2011 and IC86.2012 epochs, each with 20,000 showers per
primary, were produced with \corsika{} v73700 using an atmospheric profile for April, 
obtained from data of the atmospheric conditions between 2007 and 2011~\cite{sderidder_thesis}. 
In all cases, the hadronic model is Sibyll 2.1~\cite{Ahn:2009wx} for interactions with energies greater
than \SI{80}{GeV} and \fluka{} 2011.2c~\cite{fluka,Ferrari:2005zk} below \SI{80}{GeV}. The differences of the atmospheric models and software versions used for different epochs are negligible in this analysis.

The \textit{true} muon densities at ground are determined directly from the \corsika{} simulations,
without simulating the detector. These densities are compared to the measured muon densities
in order to estimate and correct various effects.

The main dataset based on Sibyll~2.1 is used to develop the general analysis pipeline. However, in order to study and compare predictions from different hadronic models, smaller datasets were produced with \corsika{} v73700 according to the IC86.2012 epoch, using the post-LHC hadronic interaction models EPOS-LHC~\cite{Pierog:2013ria} and QGSJet-II.04~\cite{Ostapchenko:2013pia}. These datasets are used to study systematic shifts in this analysis due to different hadronic models and to compare the final results to their predictions. Simulated datasets based on the Sibyll 2.3 hadronic model~\cite{Riehn:2015oba,Engel:2019dsg} were not available at the time of this analysis and will not be considered in this work.

\subsubsection{Simulating accidental coincidence background}
\label{section:background_sim}

Accidental coincidences are simulated at the end of the standard simulation
chain, at which point the signals in each DOM have already been
calculated in the absence of low-energy background. Accidental coincidence signals
are then added to the event at a fixed rate of 1470 Hz.
The accidental rate of 1470 Hz and its signal probability distribution function is determined
from experimental data using an off-time window as described in Section~\ref{section:background}.
If an accidental signal is generated within \SI{427}{ns} of an existing signal, the two are replaced by
a single signal equal to their sum. The analysis
of measured events shows that the smallest possible time interval between two
successive signals from the same DOM is about \SI{1.65}{\micro s}. We implement
this dead time in the simulation by sweeping over the
time-ordered signals of each DOM, after the background has been added. If
two signals are found within an interval of \SI{1.65}{\micro s}, the
second signal is discarded.

The last step in the simulation of accidental background is to recalculate
the local coincidences. A signal could be added to a tank in
a station where the neighboring tank already had a signal within the HLC
coincidence window of \SI{1}{\micro s}. In this case, the background signal and
the existing SLC are reclassified as HLC. Conversely, due to the dead time
simulation, it can happen that one HLC signal is discarded after an
accidental background signal is added, causing the remaining HLC signal,
in the neighboring tank, to be reclassified as SLC. Because of this, we
sweep over all tanks after the background simulation and re-assign HLC
and SLC trigger bits according to the HLC criterion of having a station
neighbor within \SI{1}{\micro s}.

\subsubsection{Simulating VEM fluctuations between calibration runs}
\label{section:vemdrift}

Calibration of tank signals is performed in regular bi-weekly intervals,
and the position of the peak representing one VEM can fluctuate within 3\% over time,
smearing the features in the signal histogram.
The SLC signals are not calibrated with the
same accuracy as the HLC signals, because SLCs are only calibrated online by
the DOM firmware, while HLCs are calibrated offline using more
sophisticated algorithms. This results in different smearing for SLC and HLC
signal histograms. To approximately include both effects in the simulations,
we apply a Gaussian smearing to simulated
signals. The smearing width is \SI{0.05}{VEM} for HLC and \SI{0.1}{VEM} for SLC.
If the resulting smeared signal is negative, the random sampling is
repeated until a positive signal is produced. The sampling widths
are chosen to match the width of signal distributions in data.

\section{Analysis}
\label{section:analysis}

\subsection{Method}
\label{section:analysis_method}

\begin{figure}[b]
  \vspace{-1.5em}\includegraphics[width=0.5\textwidth]{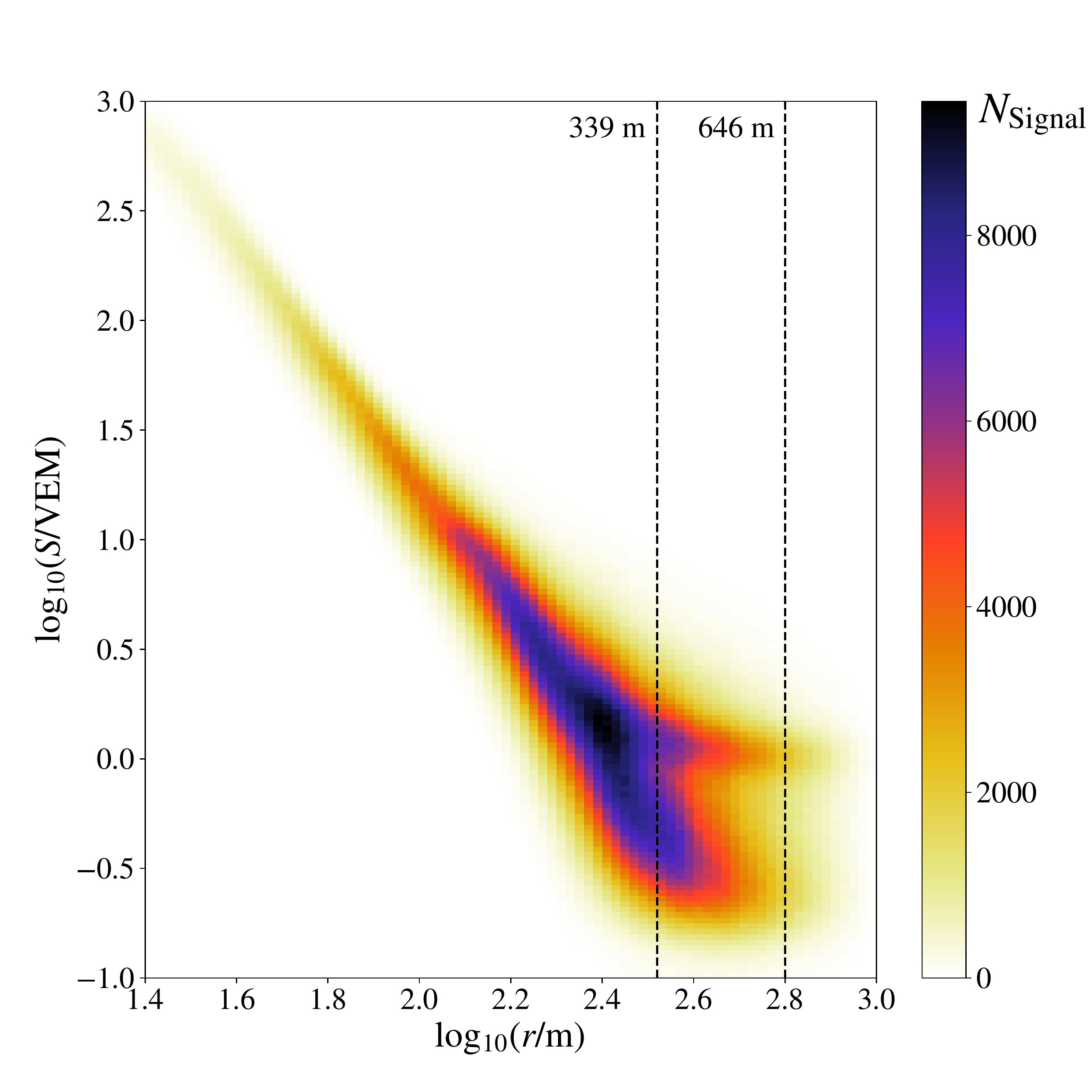}%
  \caption{Distribution of HLC and SLC tank signals for near-vertical events ($\theta < \SI{18}{\degree}$) with reconstructed energies between \SI{10}{PeV} and \SI{12.5}{PeV} as a function of lateral distance and charge. A structure produced by muons in IceTop tanks at large distances, the so-called \emph{muon thumb}, is visible at signals around 1 VEM.}
  \label{fig:thumb_plot}%
\end{figure}

\begin{figure*}[t]
  \centering
  \vspace{-2em}
  
  \subfloat[]{%
    \includegraphics[width=0.49\textwidth]{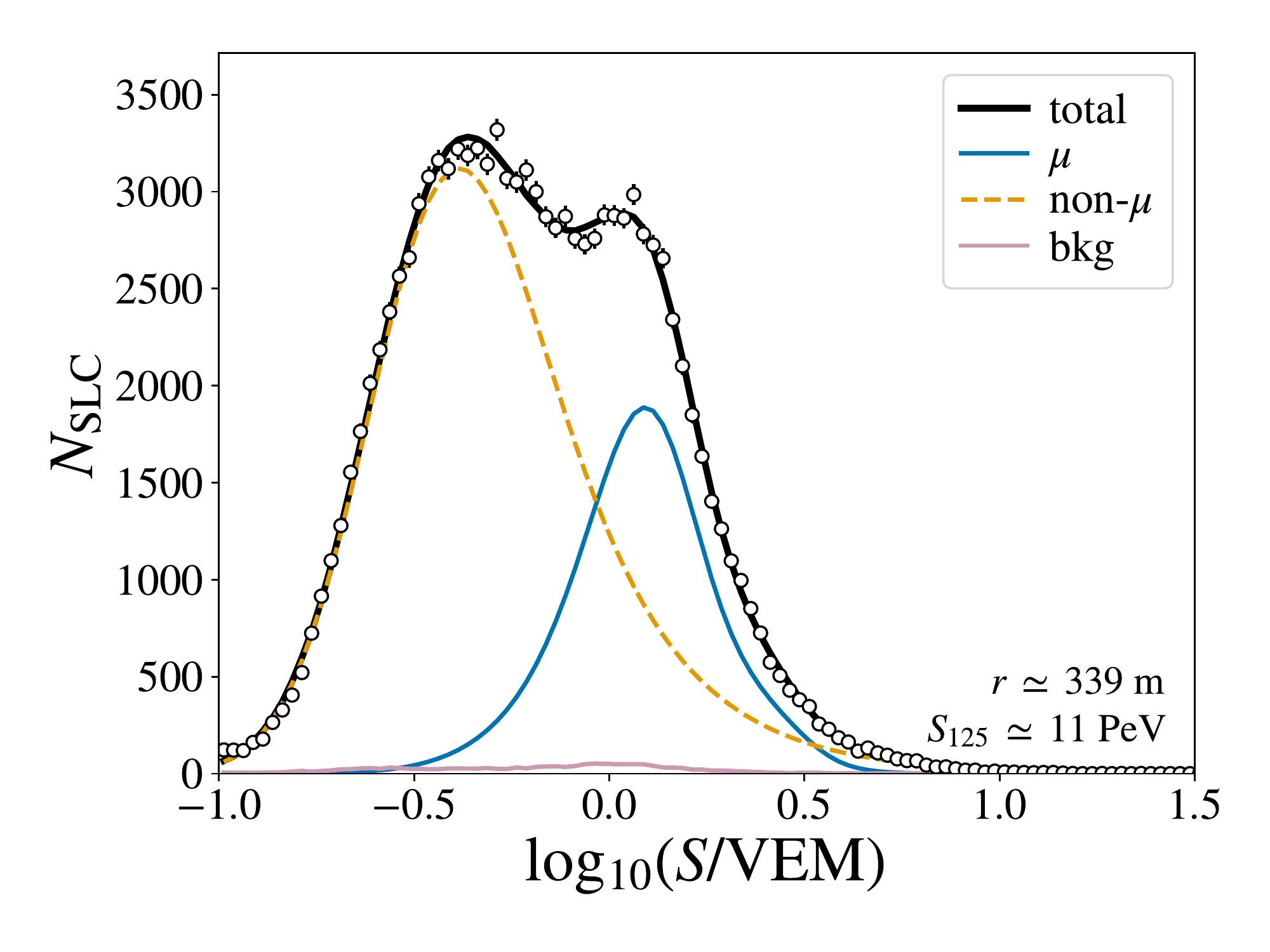}%
    \label{fig:slice_257m}%
  }\;\;
  \subfloat[]{%
    \includegraphics[width=0.49\textwidth]{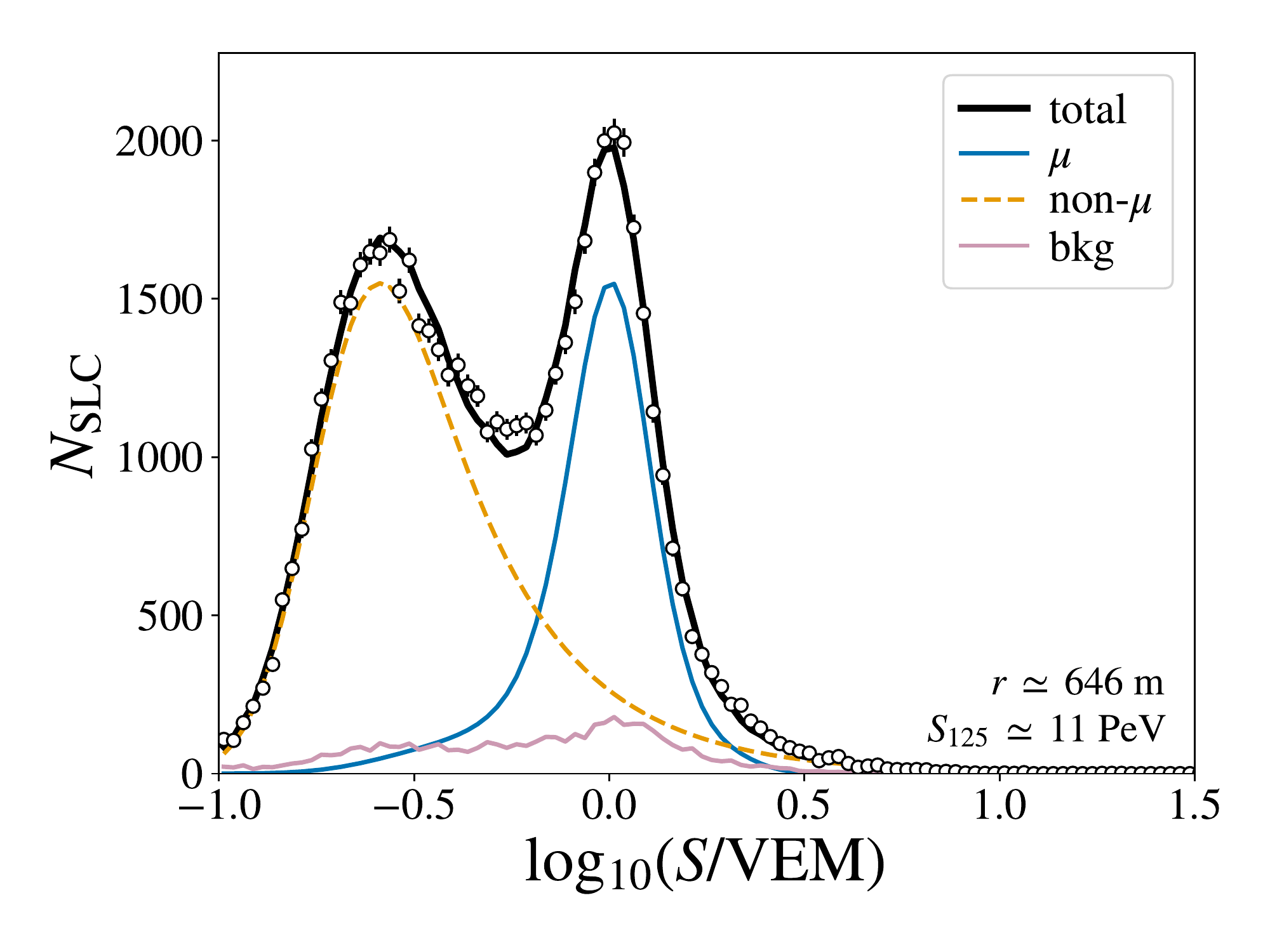}%
    \label{fig:slice_646m}%
  }
  \vspace{-.5em}
  
  \caption{Signal distribution (HLC+SLC) at lateral distances of \SI{339}{m} (a) and \SI{646}{m} (b), with fits to the signal model. The figures correspond to vertical slices in Fig.~\ref{fig:thumb_plot}. The lines show the muon signal model (blue solid line), the distribution of signals with no muons (dashed orange line) and the distribution of accidental signals (pink solid line).}
  \label{fig:r_signal_slices}%
\end{figure*}

The basis of this analysis is the different response of the detector to electrons and muons.
Muons produce a characteristic signal that can be identified at large lateral distances,
as shown in Figs.~\ref{fig:thumb_plot} and~\ref{fig:r_signal_slices}.
Figure~\ref{fig:thumb_plot} is a 2-dimensional histogram of lateral distance in
the horizontal axis and tank signal in the vertical axis. It includes signals from air
showers with energy $\SI{10}{PeV} < E < \SI{12.5}{PeV}$ and zenith angle $\theta < \SI{18}{\degree}$.  At large
distances, there are two distinct populations of signals. One population is the
continuation of the main distribution which approximately follows a power law over the
entire lateral distance range. The other population, with signals, $S$, around 1 VEM ($\log_{10}(S/\rm{VEM}) = 0$),
is visible only at large lateral distances ($\log_{10}(R/\rm{m}) \gtrsim 2.5$ in the figure). This population comprises mostly tanks hit by one muon, with a much smaller contribution from tanks hit by
more than one muon. The main step in the analysis, which will be described
in greater detail below, is to estimate the number of muons by measuring this
muon-induced population. The two populations are clearly seen in Fig.~\ref{fig:r_signal_slices},
which histograms the signals at selected fixed lateral distances.

\begin{figure}[b]
\vspace{-1em}

  \mbox{\hspace{-0.8em}\includegraphics[width=0.5\textwidth]{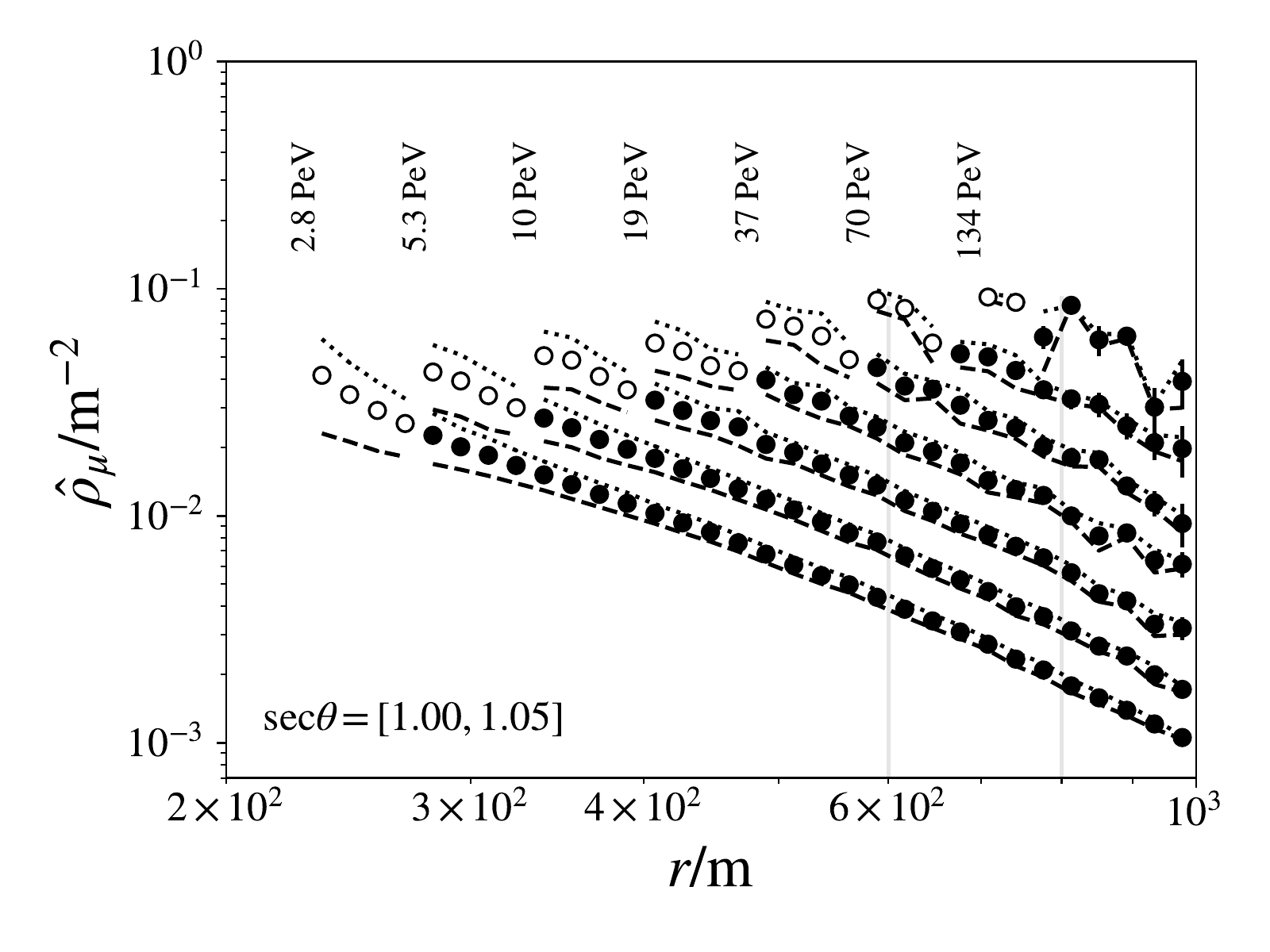}}%
  \vspace{-1.5em}
  
  \caption{The raw reconstructed muon densities measured in IceTop, $\hat\rho_\mu$, as a function of lateral distance for seven energy bins. The lines indicate the systematic uncertainty associated to the function used to model signals with no muons. Filled markers correspond to lateral distances where more than 80\% of signals only have SLC information.}
  \label{fig:data_rho_ldf}%
\end{figure}

The number of muons is determined using a log-likelihood method to fit the
signal distributions of all events at fixed energy, zenith, and lateral distance, using
a multi-component model as illustrated in
Fig.~\ref{fig:r_signal_slices}. The figures show the muon signal distribution,
the empirically-determined distribution of signals with no muons, and the distribution of
accidental signals. Each of these distributions is described in the following subsections.
The model has a total of eight parameters: six in the muon
response model described in Section~\ref{section:muon_model}, and two in the electromagnetic
(EM) model presented in Section~\ref{section:em_model}. This empirically-driven fit determines the muon signal parameters from the data independent of air shower simulations. It finds the mean number of muons per event per radial and energy bin, $\langle N_{\mu}\rangle$, which is then divided by the cross-sectional
area of all IceTop tanks within that bin projected onto a plane perpendicular to the zenith angle direction to yield the muon density. This assumes that the
direction of motion of the muons coincides with the direction of the reconstructed air
shower. The resulting \emph{raw} reconstructed muon densities, $\hat\rho_\mu$, are shown in 
Fig.~\ref{fig:data_rho_ldf}. These distributions are fit to interpolate the raw muon densities at radial distances of \SI{600}{m} and
\SI{800}{m}. With simulated datasets, a small deviation from the true muon number is observed.
This deviation is accounted for by multiplying a correction factor to the raw densities to determine the final muon densities, $\rho_\mu$, at lateral distances of \SI{600}{m} and \SI{800}{m}, as described in Section~\ref{section:mc_verification}.

\subsection{Detector response to muons}
\label{section:muon_model}

\begin{figure*}[t]
  \centering
   \vspace{-2em}
  
  \subfloat[]{%
 
   \hspace{-2em} \includegraphics[width=0.378\textwidth]{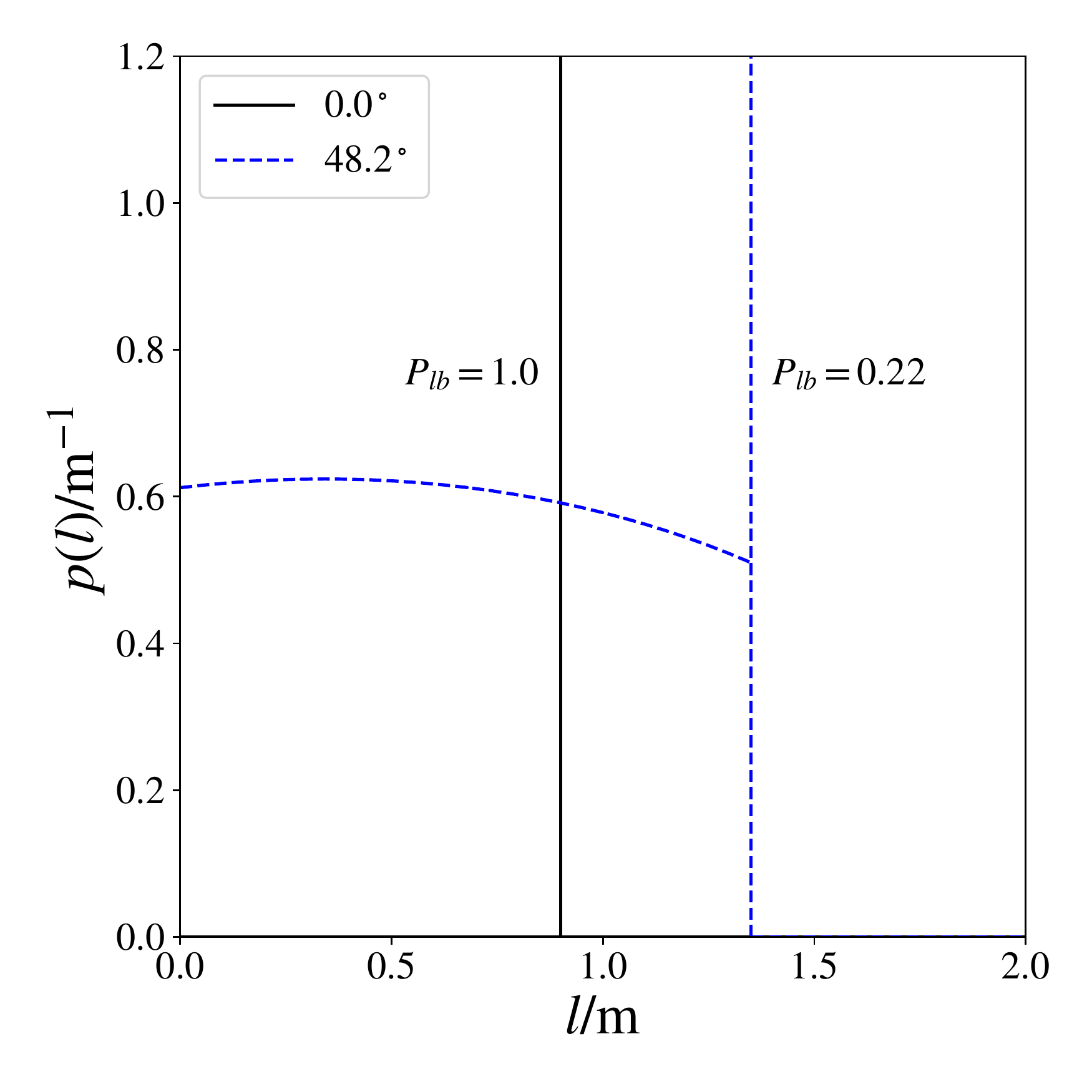}%
    \label{fig:length}%
  }
  \subfloat[]{%
    \includegraphics[width=0.378\textwidth]{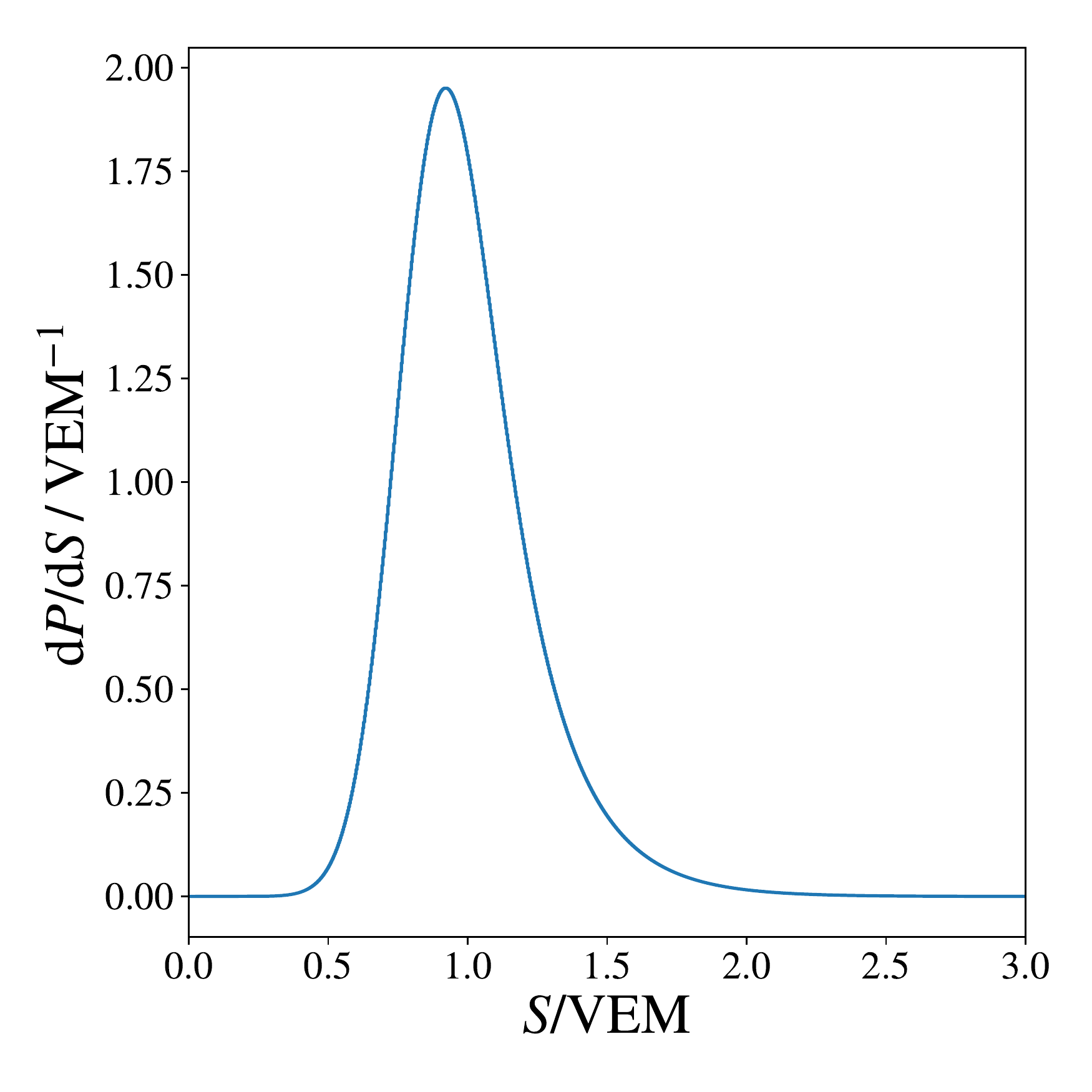}%
    \label{fig:kernel}%
  }
  \subfloat[]{%
  \includegraphics[width=0.24\textwidth]{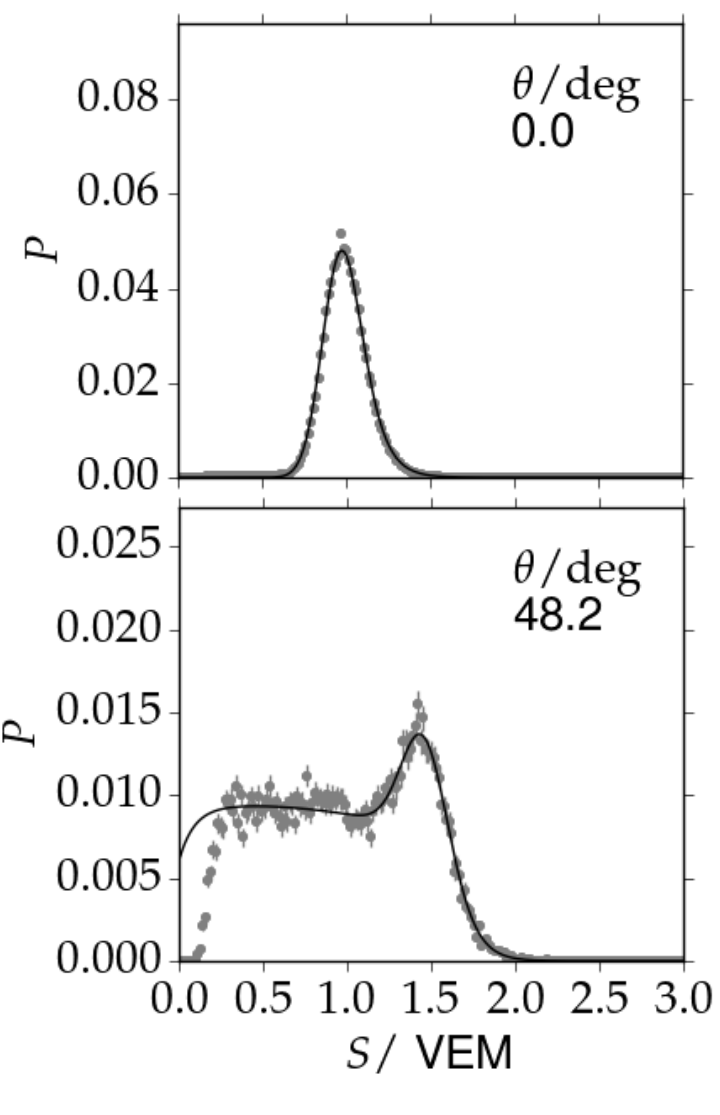}%
    \label{fig:response_both}
}
  \caption{Semi-analytical model of detector response. (a) Distribution of track lengths for muons crossing the lid to the bottom (lb) of a tank at two inclinations. The vertical lines represent $\delta$ functions. (b) Tank/PMT response kernel. This function describes the output signal for a given track length. (c) Comparisons of the semi-analytical model (solid line) with \geant{} simulations
(grey markers) at two zenith angles: $0^\circ$ (top) and $48.2^\circ$ (bottom).}
\end{figure*}

When a single particle enters the tank, the number of photoelectrons at the PMT is proportional to the particle's track length. The track length of a muon crossing the tank is determined entirely by geometry.
For a uniform beam of muons entering a tank, the
statistical distribution, $g(l)$, of track lengths, $l$,  can be calculated
analytically~\cite{single_mu_response}. Example distributions, for two arrival directions,
are shown in Fig.~\ref{fig:length}. In the case of vertical muons,
the distribution is a Dirac delta function, because all muons traverse
the tank from top to bottom. By definition, the signal recorded in this case
is \SI{1}{VEM}. For muons arriving at an angle $\theta$ with respect to the vertical, the
distribution is a sum of two terms: a Dirac delta function centered at
$l(\theta=0)\sec(\theta)$ produced by all muons that enter the tank
through the top and exit through the bottom, and a continuous
distribution that corresponds to muons that do not traverse the tank
from top to bottom. The latter are called \textit{corner
  clipping} muons. Given the track length of a muon, the
signal probability distribution is given by an exponentially modified Gaussian
kernel function, $K(S|l)$, such as the one shown in
Fig.~\ref{fig:kernel}~\cite{exp_gaus}. The kernel function is specified by three
parameters: a width $\sigma$, an exponent $\lambda$ and its mean, $\mu$. This function takes care of
effects such as the photon sampling and PMT collection efficiency.

\begin{figure*}
  \centering
  \vspace{-1em}\subfloat[]{%
    \includegraphics[width=0.471\textwidth]{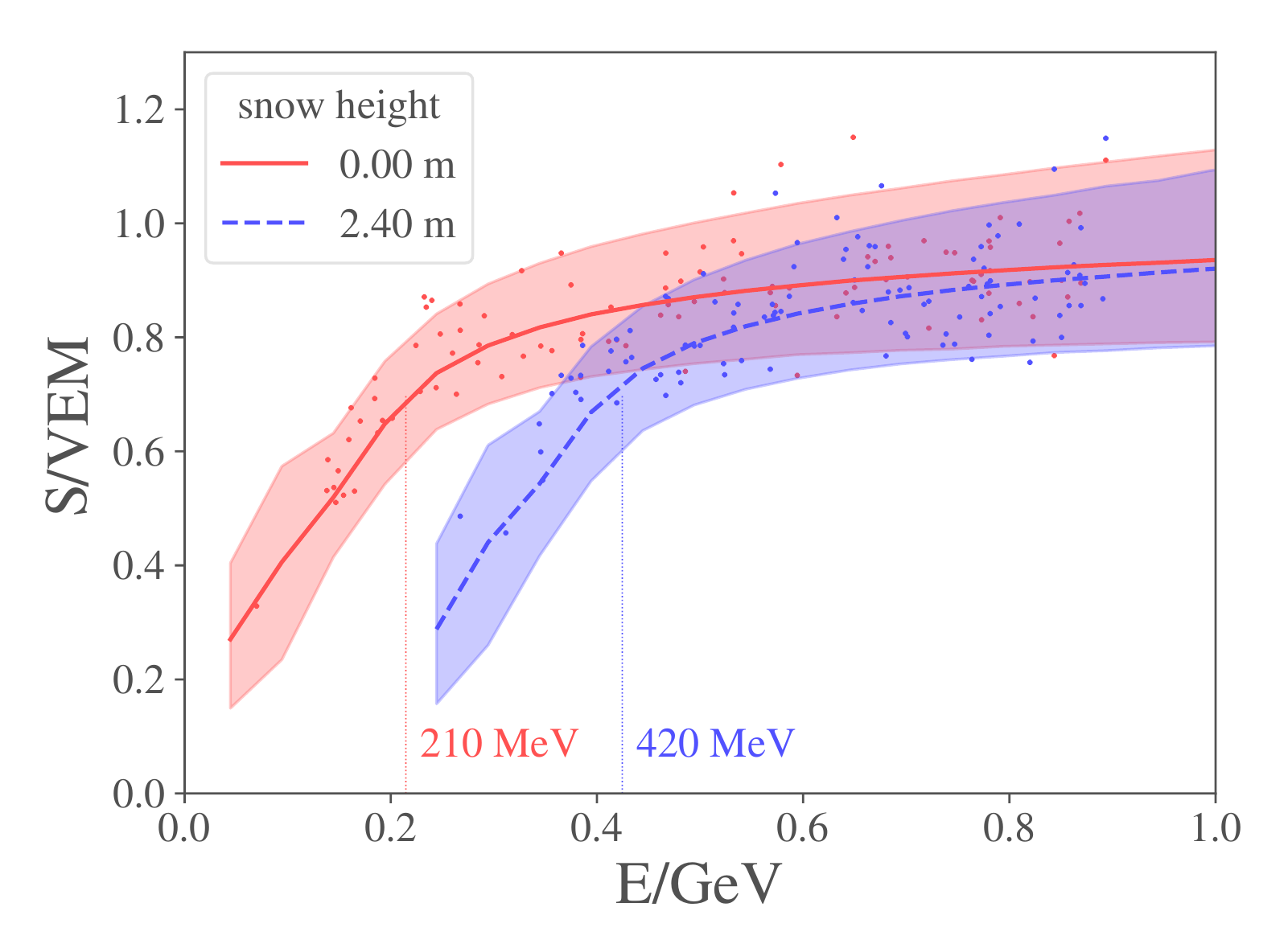}%
    \label{fig:muon_response_v_snow}%
  }\;\;
  \subfloat[]{%
    \includegraphics[width=0.468\textwidth]{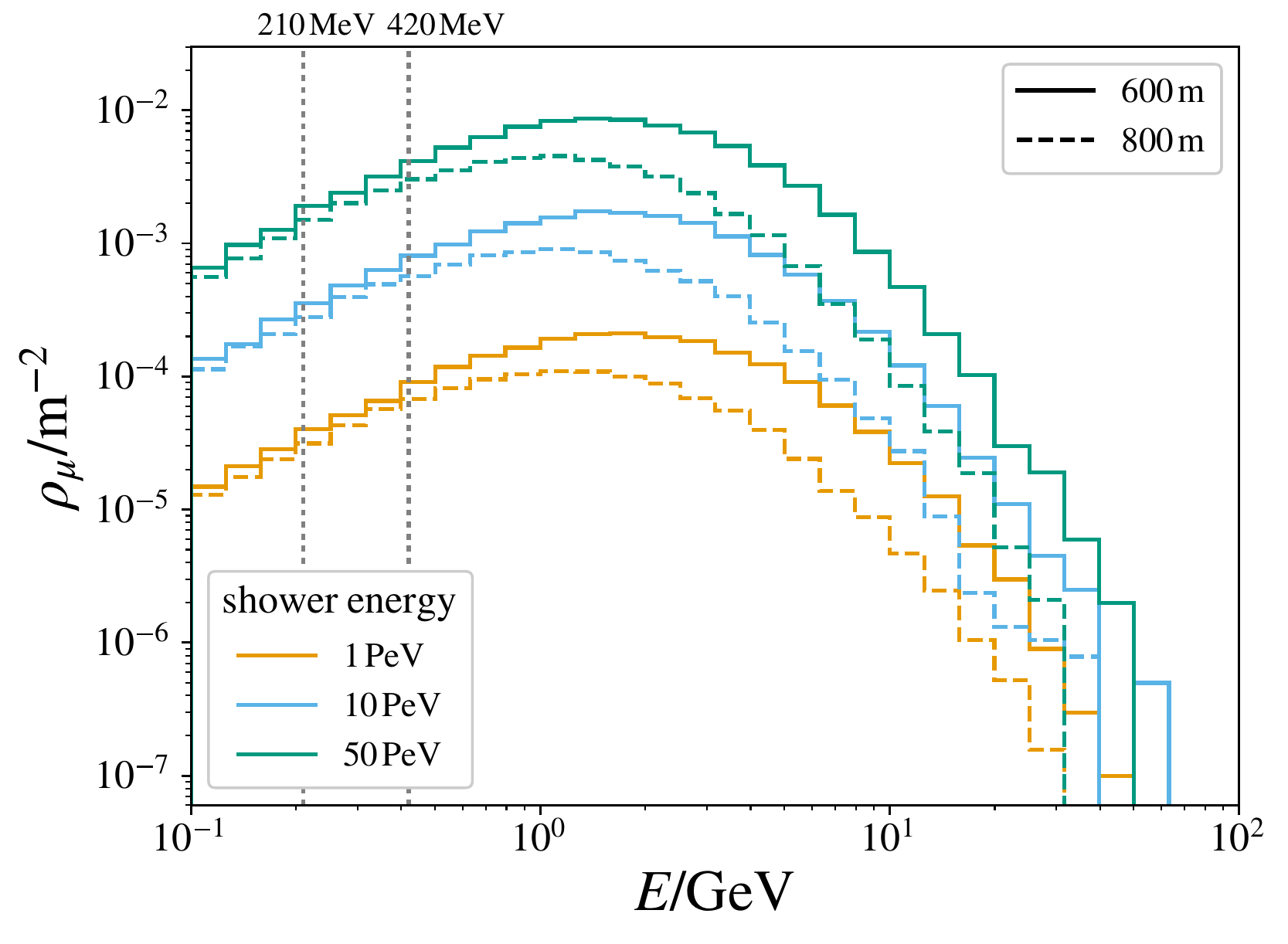}%
    \label{fig:muon_spectrum}%
  }
  \vspace{-0.2em}
  \caption{(a) Simulated signal produced in a PMT by muons entering the tank
    vertically after passing through different amounts of snow. The
    signal is measured in units of VEM and is plotted as a function of
    the kinetic energy of the incoming muon. Vertical lines show the threshold estimated as the
    energy at which the median signal becomes larger than \SI{0.7}{VEM}. (b) The muon density spectra in vertical proton showers obtained from \corsika{} simulations using the Sibyll model. The fraction of muons below \SI{210}{MeV} and \SI{420}{MeV} is 2\% and 10\%, respectively.
  }
\end{figure*}

The signal distribution for a given number of muons per tank, $\langle N_{\mu} \rangle_\mathrm{tank}$, is found by adding the
contributions of an integer number of muons, weighted by the Poisson probability.
The signal distribution for an integer number of muons is given by
the signal distribution of a single particle convolved multiple times with itself.
The resulting distribution is multiplied by a function that
models the reduced efficiency for detecting low signals, caused by
the discriminator trigger in each detector. This function takes
the form of a Gaussian cumulative density function in the logarithm of the
signal, $S$, with threshold $S_\mathrm{thr}$ and width $\sigma_\mathrm{thr}$. All this can be summarized in the
following equations:
\begin{eqnarray}
\label{eq:signal_model}
p\left(S \,|\, N_{\mu}=1\right) &=& \int_{0}^{l_\mathrm{max}} K(S\,|\,l, \mu, \sigma, \lambda) \nonumber  \\
&\times& g(l\,|\, \theta) \, dl\,,
\end{eqnarray}
\begin{eqnarray}
p\left(S \,|\, N_{\mu}=n \right) &=& \int_{0}^{S} p\left(t \,|\, N_{\mu}=1\right)  \nonumber\\
&\times& p\left(S - t \,|\, N_{\mu}=n-1\right) dt\,,
\label{eq:signal_model2}
\end{eqnarray}
\begin{eqnarray}
&p&\left(S \, \vert \, \langle N_{\mu} \rangle_\mathrm{tank} \right) = \frac{1}{2}\left( 1 + \text{erf}\left(\frac{\log (S/S_\mathrm{thr})}{\sigma_\mathrm{thr} \sqrt(2)}\right) \right)  \nonumber\\
&\times& \sum_{n=0}^{\infty} \frac{\langle N_{\mu} \rangle^n_\mathrm{tank}}{n!} e^{-\langle N_{\mu} \rangle_\mathrm{tank}} p\left(S \,|\, N_{\mu}=n \right),
\label{eq:signal_model3}
\end{eqnarray}
where $g(l\,|\, \theta)$ and $K(S\,|\,l, \mu, \sigma, \lambda)$ are the
track-length and tank-response functions shown in Figs.~\ref{fig:length} and~\ref{fig:kernel},
respectively. This model has six parameters: three in Eq.~\ref{eq:signal_model}
($\mu, \sigma, \lambda$), and three in Eq.~\ref{eq:signal_model3}
($\langle N_{\mu} \rangle_\mathrm{tank}, S_\mathrm{thr}, \sigma_\mathrm{thr}$). At large distances, the electromagnetic contribution is expected to be small compared to the muon signal in the same tank and can thus be neglected. The result of this semi-analytical model is displayed in Fig.~\ref{fig:response_both}. These figures show the signal distribution
obtained using \geant{}~\cite{geant4}
together with the prediction from the model.
The two panels (top and bottom) of Fig.~\ref{fig:response_both} correspond to two sets of parameters that differ only in their zenith angle (\SI{0}{\degree} and \SI{48.2}{\degree}).

\subsection{Electromagnetic signal distribution model}
\label{section:em_model}

The probability distribution for signals in air showers includes the model
described in Section~\ref{section:muon_model}, which describes signals produced by muons,
and a model for signals produced by gamma-rays, electrons and positrons
in the electromagnetic (EM) component of the air shower. Hadrons can also produce a signal in the
tanks, but they are so few in comparison to electrons, muons and gamma-rays, that their contribution
is negligible.

Most electrons and positrons lose all their energy inside the tank. Their track
lengths, and hence their signals, are typically smaller than those of muons.
As a result, the distribution of signals from the EM component roughly mimics their
energy distribution, with a mean signal that corresponds to a few tens of
centimeters of track length inside the tank, while the distribution of signals
from muons is determined by the geometry of the tank.

We assume that the energy distribution of the EM
component (and therefore their signal distribution) approximately follows a power law.
This is true at large lateral distances, where the mean expected EM signal is
much smaller than the threshold, $S_\mathrm{thr}$, leaving only the tail of the EM signal
distribution above threshold. With this assumption, the EM signal distribution for tanks with no muons is given by
\begin{eqnarray}
  \label{eq:em_distribution}
  p_\mathrm{EM}(S) &= \frac{1}{2}\left(
     1 + \text{erf}\left(\frac{\log (S/S_\mathrm{thr})}{\sigma_\mathrm{thr} \sqrt{2}}\right)\right) \nonumber\\
  &\times N_\mathrm{EM} S^{-\lambda_\mathrm{EM} + c \log(S)},
\end{eqnarray}
where we included the same efficiency function that appears in
Eq.~\ref{eq:signal_model3}. In Eq.~\ref{eq:em_distribution} we have added the
possibility of the $c\log{(s)}$ term to allow for small deviations
from a power law. We thus have two different models for the distribution of EM
signal, depending on whether or not we set the parameter $c$ to zero.
These two models will be referred to as \textit{EM1} and \textit{EM2}, respectively. The analysis is repeated
using both EM models in order to estimate the systematic uncertainty arising
from the choice of model.

\subsection{The effect of snow}
\label{section:snow}

The snow on top of each tank absorbs some of the energy carried by secondary
particles, sometimes stopping them before they enter the tank. This
raises the threshold energy above which the detection,
filter, and selection are fully efficient, and attenuates
the recorded signals. While the attenuation effect is
included in the reconstruction process, the loss in
trigger efficiency can not be corrected. This loss could
affect the muon density measurement, since it
depends on the air shower age, and thereby on the
mass of the primary cosmic ray.
Lower efficiency leads to a systematic decrease
in the measured muon density.

The mechanical structures of the IceTop tanks and snow accumulation can directly affect the measurement of the muon
density by reducing the number of muons that enter the tank
if their energy is low enough. A muon that enters the tank and produces a
signal smaller than about \SI{0.7}{VEM} will not contribute to the measured
muon density, as it would not contribute to the characteristic
peak around \SI{1}{VEM} in Fig.~\ref{fig:r_signal_slices}.

The results of detailed simulations of the signals produced by muons
crossing vertically through IceTop detectors using the
\geant{} package, which include the modeling of any detector effects, are shown in
Fig.~\ref{fig:muon_response_v_snow}. The (kinetic) energy threshold to produce a signal larger than \SI{0.7}{VEM}
for muons vertically entering a tank without snow is about \SI{210}{MeV}. However, assuming realistic snow depths on top of the tanks  of about 2 meters during the end of the data taking period, the energy threshold for muons in this analysis increases to about \SI{420}{MeV}.

\begin{figure*}[tb]
  \centering
   \vspace{-1em}\subfloat[]{%
   \hspace{-1.5em}\includegraphics[width=0.478\textwidth]{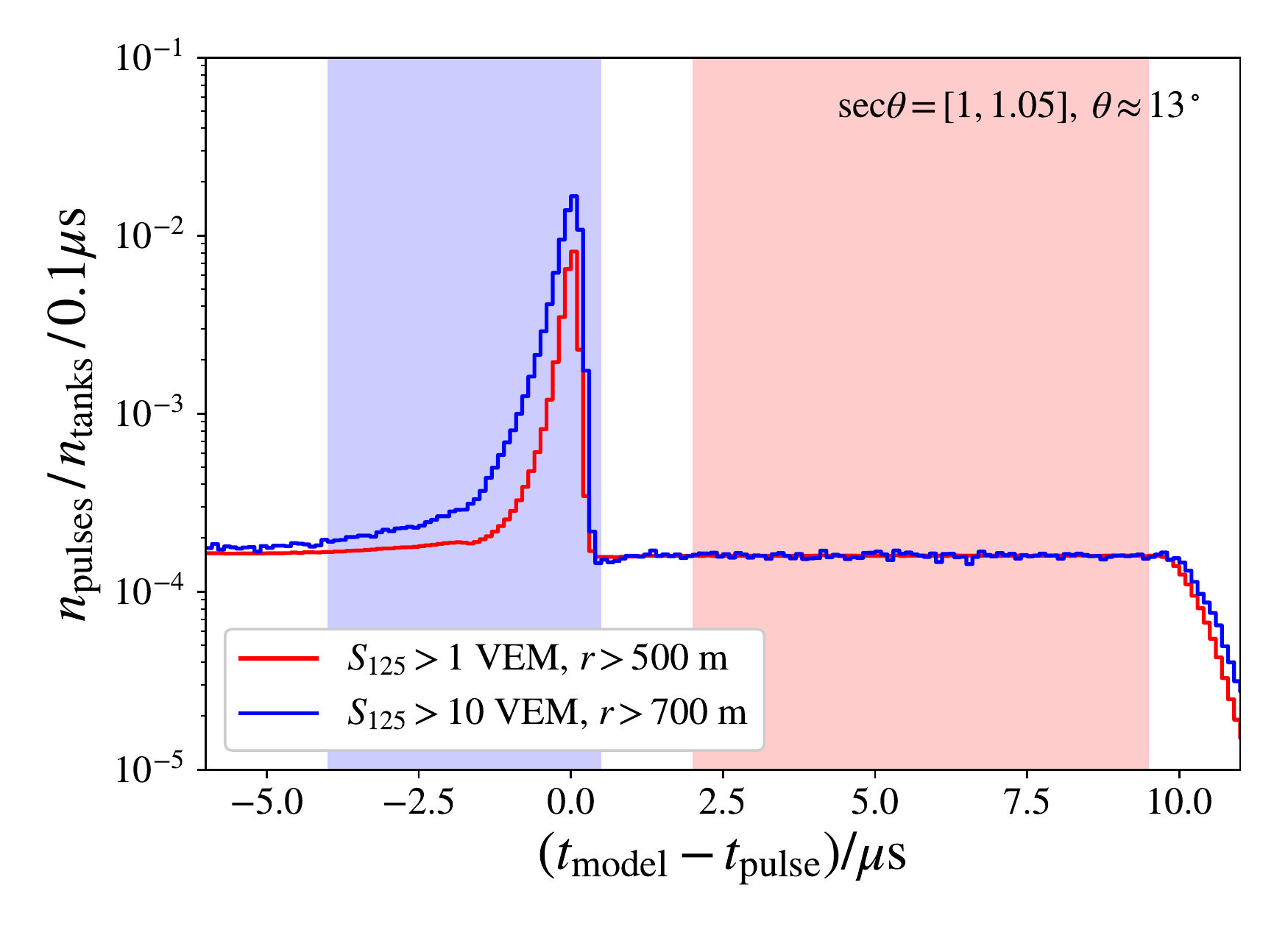}%
    \label{fig:on_off_regions}%
    }\;\;\;%
  \subfloat[]{%
    \includegraphics[width=0.46\textwidth]{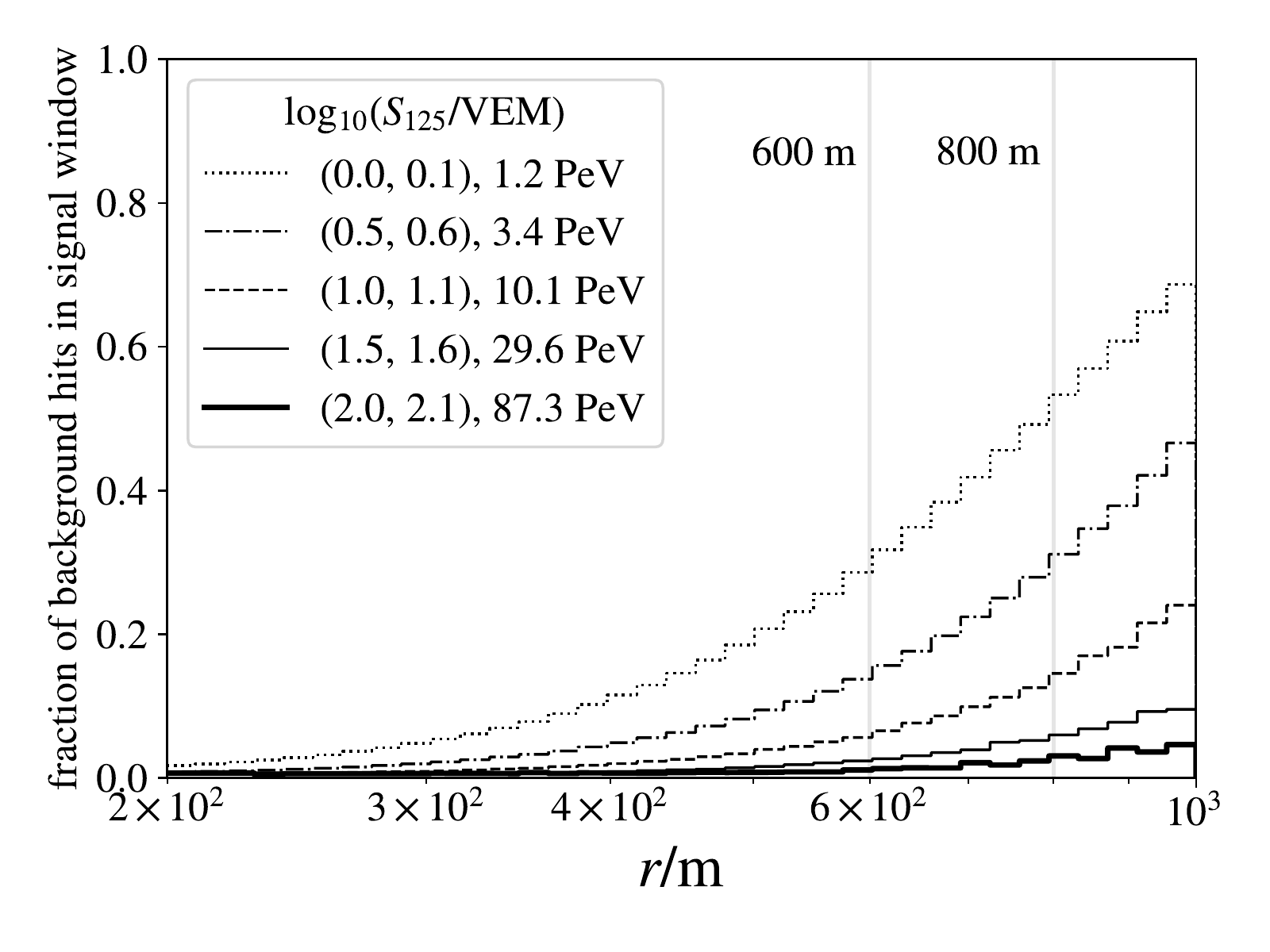}
    \label{fig:bkg_fraction}
    }
\caption{Accidental background. (a) Pulse rate as a function of time-offset with respect
    to the reconstructed shower front for near-vertical showers. Signal window
    in blue, background window in pink. Positive and negative values indicate signals arriving before and after the shower front respectively. (b) Fraction of hits in signal time window that originate from
    background events as a function of lateral distance to the shower axis. The different lines correspond
    to different ranges of $S_{125}$. The mean energy for each $S_{125}$ range is also shown.}
\end{figure*}

Typical muon density spectra from \corsika{}
simulations, at \SI{600}{m} and \SI{800}{m} from the shower axis, for different primary cosmic ray energies, are shown in
Fig.~\ref{fig:muon_spectrum}. The fractions of muons with energies below \SI{210}{MeV} and
\SI{420}{MeV} are about 2\% and 10\%, respectively. We therefore expect a
systematic shift of this order in the reconstructed muon density. The exact magnitude of the shift
depends on the primary energy, the arrival direction, and lateral
distance  because the muon spectrum depends on these variables.

A further complication is that the snow is spread unevenly over the
array. Between May 2010 and May 2011, the tanks were covered
by as little as a few centimeters and as much as \SI{1.5}{m} of
snow. Between May 2011 and May 2012 the snow coverage ranged from
\SI{0.4}{m} to \SI{2}{m} of snow. The attenuation in the snow is one of several systematic effects that are accounted for by
a correction factor that will be introduced in
Section~\ref{section:mc_verification}. 
As a final verification, the yearly variation in the muon density is used as an estimate for the uncertainty due to snow in the final results (Section~\ref{section:results}).

\subsection{Accidental coincidence background measurement}
\label{section:background}

The rate of single-DOM discriminator triggers is
\SI{1470}{Hz} per tank. Most of these triggers are produced by
low-energy showers which are independent of the main air shower event. They constitute a background
in the present analysis. If we consider all triggers within a time
window around an air shower event, there will certainly be some of these
\textit{accidental} signals, which will be distributed according to a different
probability distribution than the signals from the shower, potentially biasing the measured muon
density. 

To minimize the effect of accidental signals, we select the
signals based on their time difference with
respect to the reconstructed shower front. The time window must
be as small as possible to discard accidental signals, but large
enough to guarantee that all air shower signals are selected. This selection
is implemented assuming a plane shower front. Therefore, the time window must
be large enough to account for the curvature of the shower front, and
guarantee that air shower signals at
large lateral distances are selected. Figure~\ref{fig:on_off_regions} shows
histograms of the time difference using two different signal selection cuts.
The time difference distribution exhibits a constant floor and a
peak near zero, in coincidence with the reconstructed shower
front.
Based on this distribution, we define signal and background windows as follows:
\begin{itemize}
\item signal window (in blue): $\SI{-4.0}{\micro s} < dt < \SI{0.5}{\micro s}$
\item background window (in pink): $\SI{2}{\micro s} < dt < \SI{9.5}{\micro s}$
\end{itemize}

The expected accidental background rate per signal bin in the signal
window per tank is:
\begin{equation} \hat n_B = \frac{n_{\rm{pulses}}}{n_{\rm{tanks}}} \frac{\Delta t_S}{\Delta t_B}, \end{equation}
where $n_{\rm{pulses}}$ is the number of observed hits in each signal bin
in the background window, $n_{\rm{tanks}}$ is the number of detectors in
the lateral distance bin considered, independent of whether there was
a signal or not, $\Delta t_B$ is the size of the background time
window, $\Delta t_S$ is the size of the signal time window, and $\hat
n_B$ is the number of expected hits in each signal bin per tank in the
signal window.

\begin{figure}[tb]
\vspace{-.2em}

\mbox{\hspace{-0.2em}\includegraphics[width=0.5\textwidth]{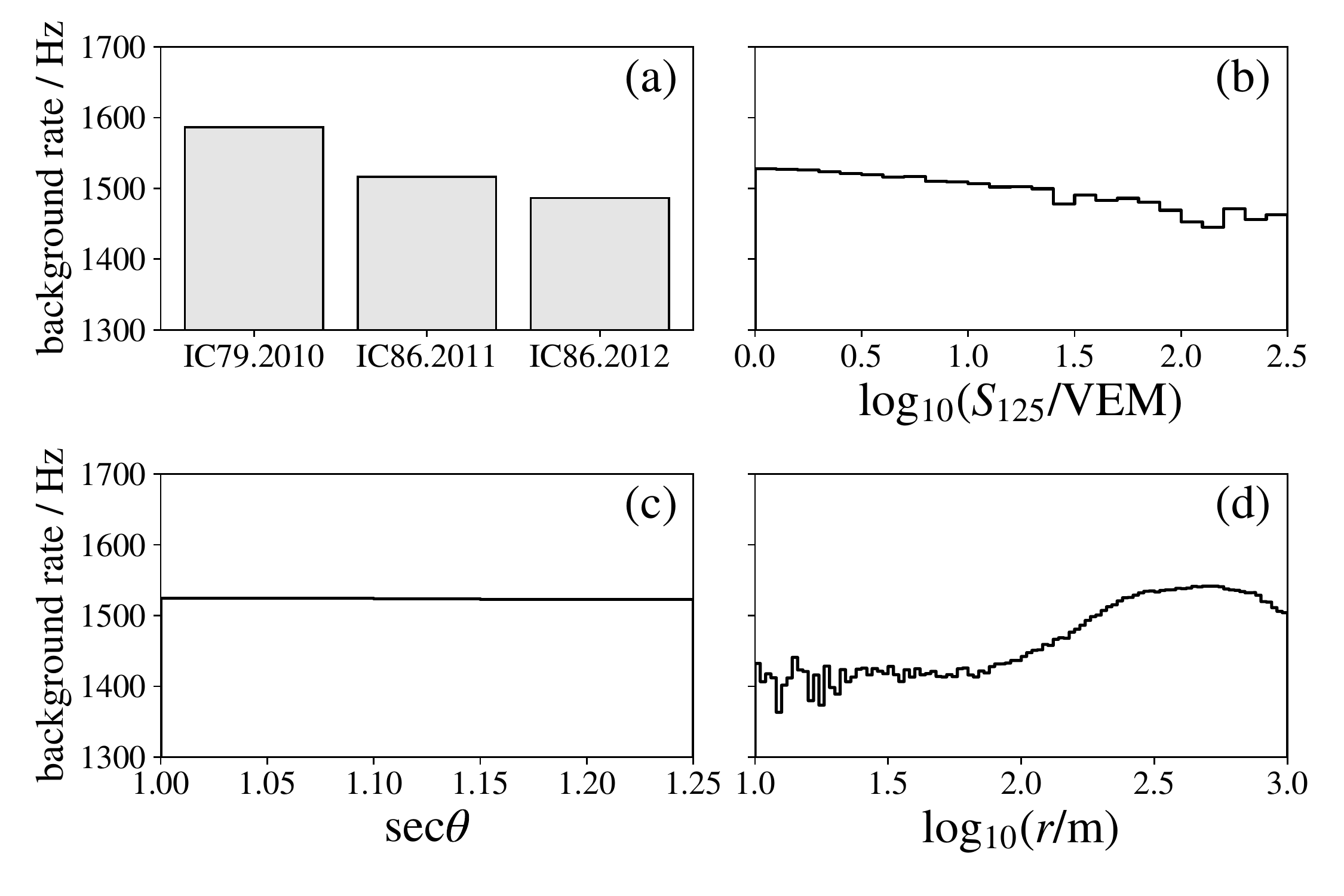}}
\caption{Background rate for tanks binned in different ways. (a) Binned by year, (b) binned by shower size $S_{125}$ of the event that was used to define the off-time window, (c) binned by $\sec\theta$ of the event, (d) binned by lateral distance $r$ from the shower axis of the event.}
\label{fig:bkg_uniformity_radius}
\vspace{-1em}

\end{figure}

The relative contribution of the accidental coincidence
background, in relation to all signals collected, is substantial at
large lateral distances or for low-energy showers. This becomes important at the lowest energies around \SI{2.5}{PeV}
 where showers are small and produce only
few signals. Figure~\ref{fig:bkg_fraction} is based on the background
data and shows the fraction of the signals in the signal window expected to originate from background, as a function of lateral distance from the
shower axis for several intervals in $\log_{10}(S_{125})$. The equivalent shower
energy for each bin of $S_{125}$ is given in the legend. For reference, the
lateral distances of \SI{600}{m} and \SI{800}{m} are indicated.
For example, in the signal window of a \SI{1.2}{PeV} shower the background amounts to 30\% at \SI{600}{m} lateral distance. 
If the background is not properly modeled, this would produce
an upward bias of 50\%. At 10 PeV, this fraction decreases to approximately 5\%.

The advantage of measuring the accidental background in the same data sample
used for the muon measurement is that it accounts for systematic
effects under the same conditions. The background present in the signal window is estimated by measuring
it in the background window. Both windows are equally affected by effects which can introduce systematic uncertainties, as discussed in Section~\ref{section:results}.
Some of these effects can be seen in Fig.~\ref{fig:bkg_uniformity_radius}.
The background rate decreases as snow accumulates on the  detectors, and therefore
the background rate decreases over the years. The
most notable effect is the dependence on lateral distance. This
dependence has its origin in the way snow is unevenly distributed over
the array. The detectors are buried under different amounts of
snow. Given a core location contained within the array and an arrival
direction of the air shower, the effective amount of snow on the tanks is
determined for in each radial bin.

\subsection{Monte-Carlo correction}
\label{section:mc_verification}

The accuracy of the muon density reconstruction depends on various assumptions used in this analysis. These
include the assumption that the muons travel in the same direction as the air
shower and the value of the mean zenith angle used in the signal model. There
is also the effect of the absorption in the snow, as previously discussed, and selection bias occurs if the
trigger or selection efficiency is not perfectly 100\%. Other effects can arise
from the finite resolution in the reconstructed parameters: energy, direction
and core location. 

The finite resolution in the reconstruction of the primary energy affects the muon density and becomes important in the
presence of steep spectra. For example, both the finite energy resolution and a systematic shift in energy can cause migrations in the energy bins. These effects produce an apparent shift in the muon density relative to the density at the true energy and thus need to be accounted for. A systematic shift in the reconstructed zenith angle would slightly alter
the signal response model. Any shift in core position or arrival direction
will cause migrations in the radial bins, which can increase the contributions
from the electromagnetic and muon components of the air showers, since they are
steeply falling functions of the lateral distance. To estimate the sizes of the various effects, the analysis is performed on
simulated datasets and the results are compared to
the \textit{true} muon density determined directly from the output of \corsika{}.

\begin{figure}[t]
    \centering
        \mbox{\hspace{-0.8em}\includegraphics[width=0.5\textwidth]{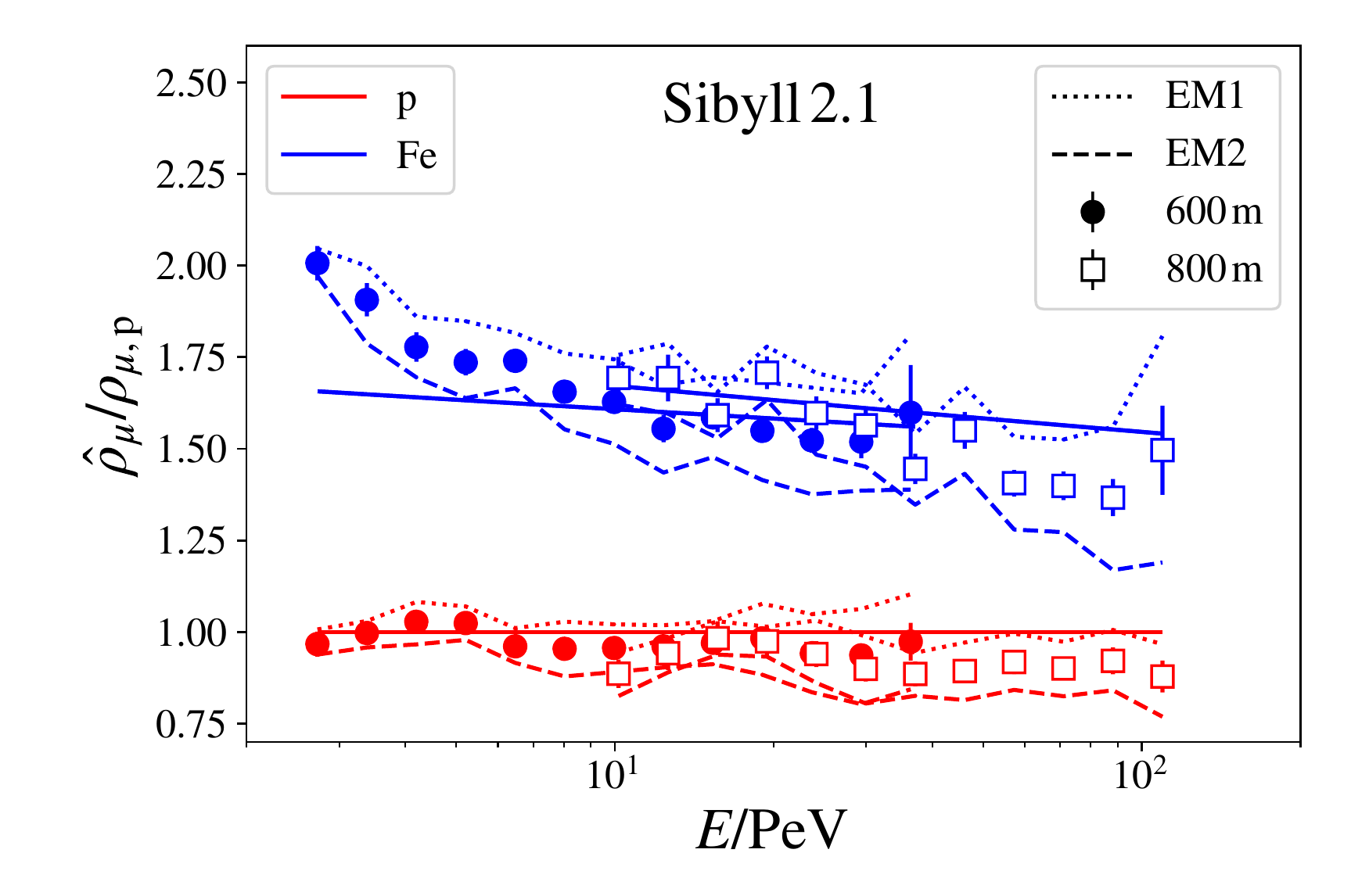}}
    \caption{Comparison of the raw reconstructed density of muons obtained from simulations, $\hat \rho_\mu$, (markers) and the true
muon density in proton and iron air showers simulated with Sibyll 2.1 (solid lines) at \SI{600}{m} and \SI{800}{m}. The distributions are normalized with respect to the muon density in simulated proton showers, $\rho_{\mu,\mathrm{p}}$. Dotted lines indicate the results obtained using the EM1 model and dashed lines those obtained with the EM2 model. Markers are placed at the midpoint.}
    \label{fig:mc_reco}
    
    \vspace{-1em}
\end{figure}

\begin{figure}
    \centering    
    \subfloat{
        \hspace{-0.8em}\includegraphics[width=0.5\textwidth]{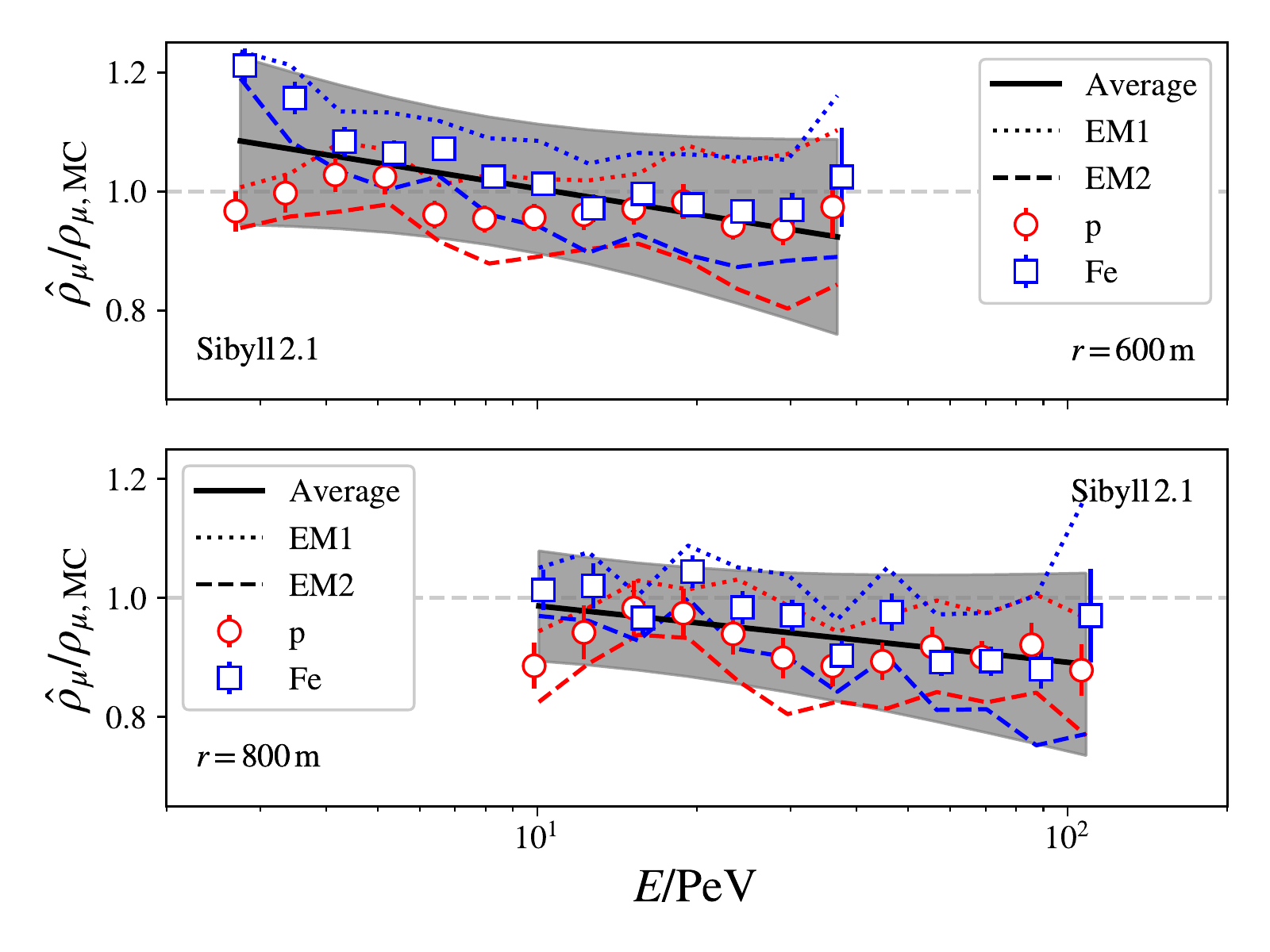}%
    }
   \vspace{-2em}
  
  \subfloat{
     \hspace{-0.8em}\includegraphics[width=0.5\textwidth]{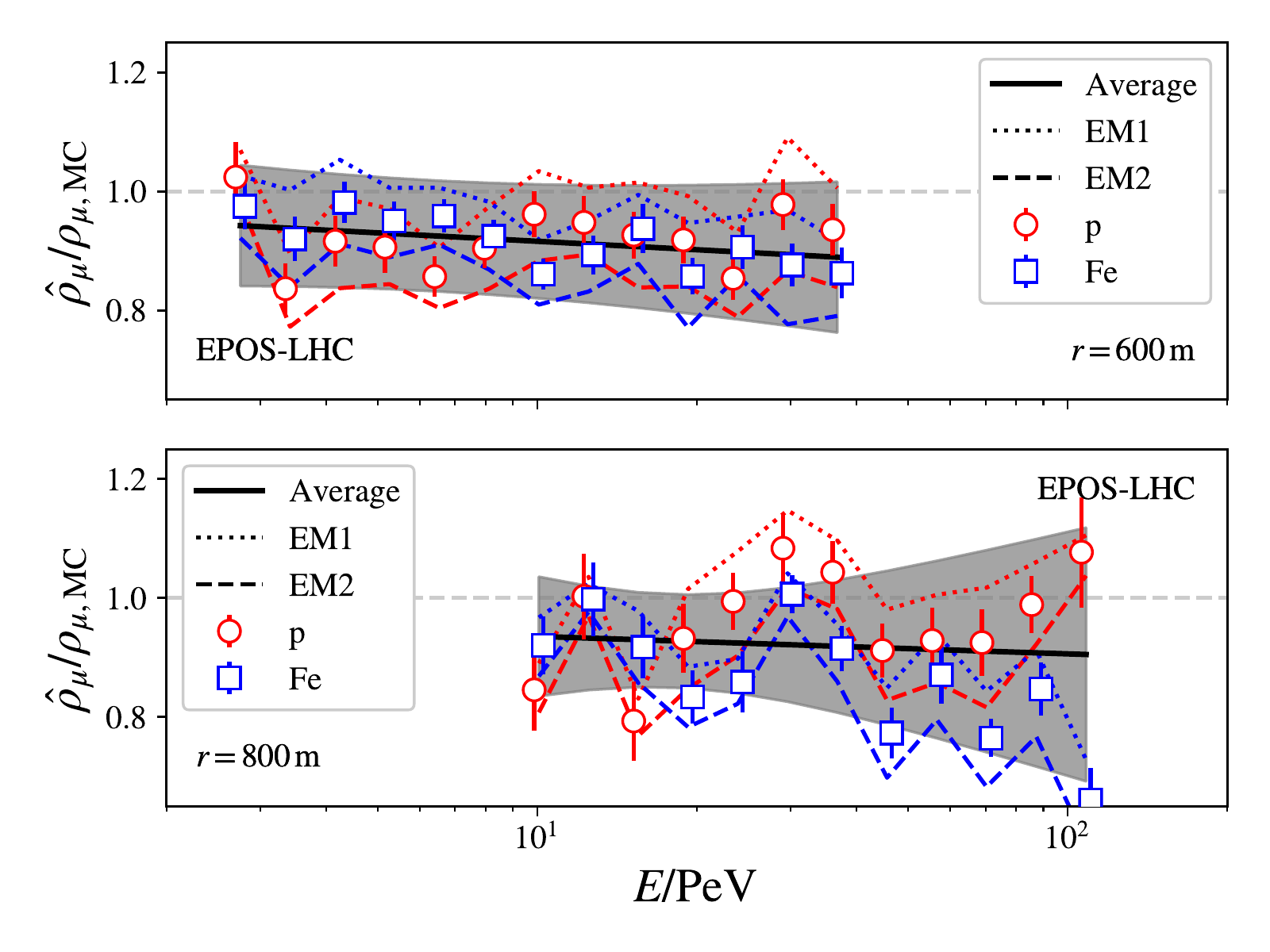}%
    }
    \vspace{-2em}
    
  \subfloat{
       \hspace{-0.8em}\includegraphics[width=0.5\textwidth]{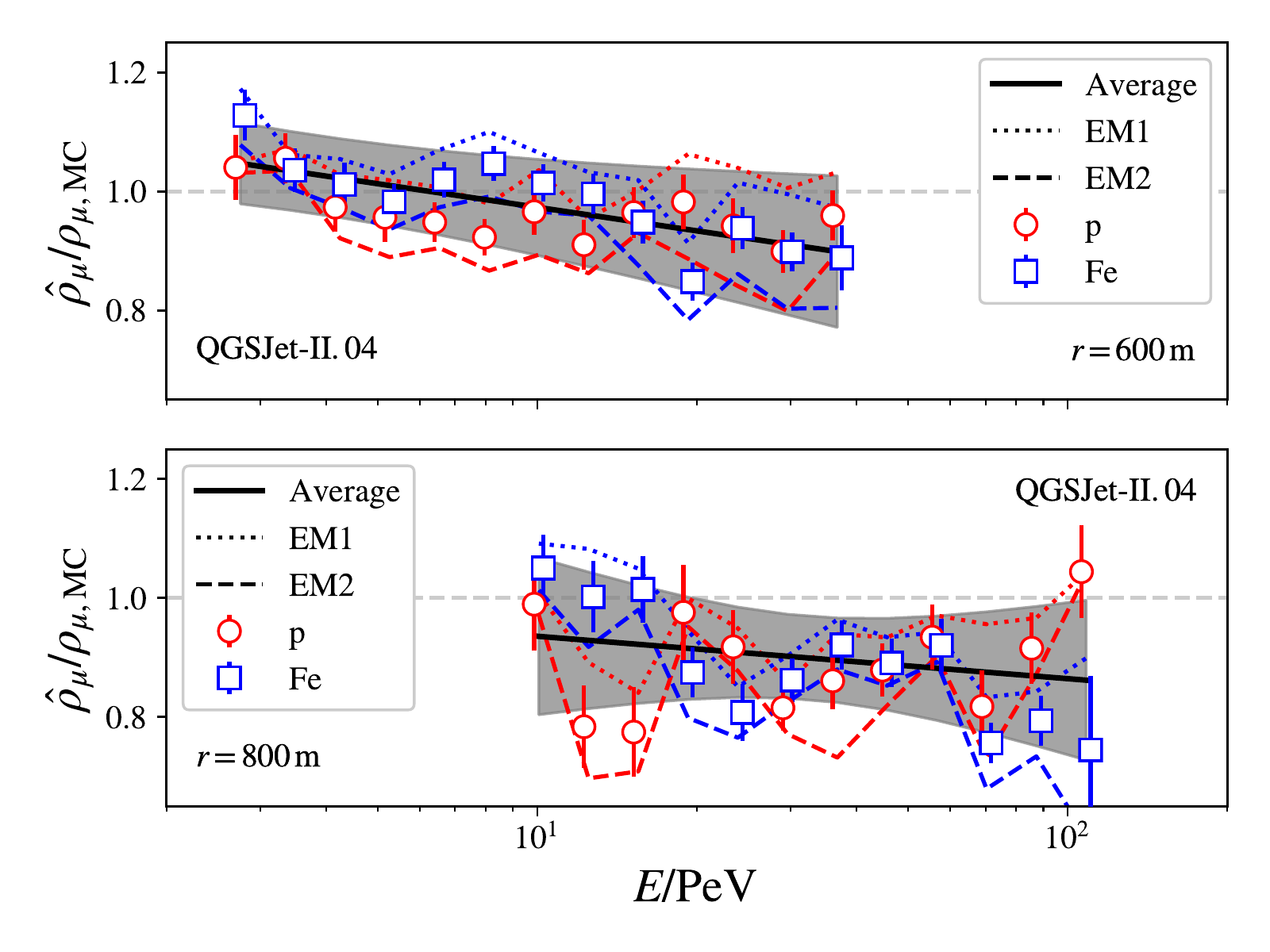}%
    }
    \vspace{-1em}
    
    \caption{Ratio of raw reconstructed over true muon density at \SI{600}{m} and \SI
    {800}{m} lateral distance, as a function of reconstructed energy, for simulated air showers using the hadronic interaction models Sibyll 2.1 (top), EPOS-LHC (center), and QGSJet-II.04 (bottom). Squares and circles correspond to iron and proton simulated showers, dotted and dashed lines indicate the ratios obtained using the two EM models. Corresponding fits are shown as solid lines where the grey band represents the corresponding uncertainties.}
    \label{fig:mc_correction_had_model}
\end{figure}

The raw reconstructed muon densities, $\hat\rho_\mu$, at lateral distances of \SI{600}{m} and \SI{800}{m}, obtained from simulations based on Sibyll 2.1, are shown in Fig.~\ref{fig:mc_reco} with respect to the true muon density in proton showers, $\rho_{\mu,\mathrm{p}}$, as a function of the reconstructed air shower energy. The dotted lines in Fig.~\ref{fig:mc_reco} indicate the results obtained using the EM1 model for the electromagnetic part, and the dashed lines represent the results obtained using the EM2 model. The markers are placed at the midpoint between the two. 

\begin{figure}
    \centering    
    \mbox{
     \hspace{-0.8em}\includegraphics[width=0.5\textwidth]{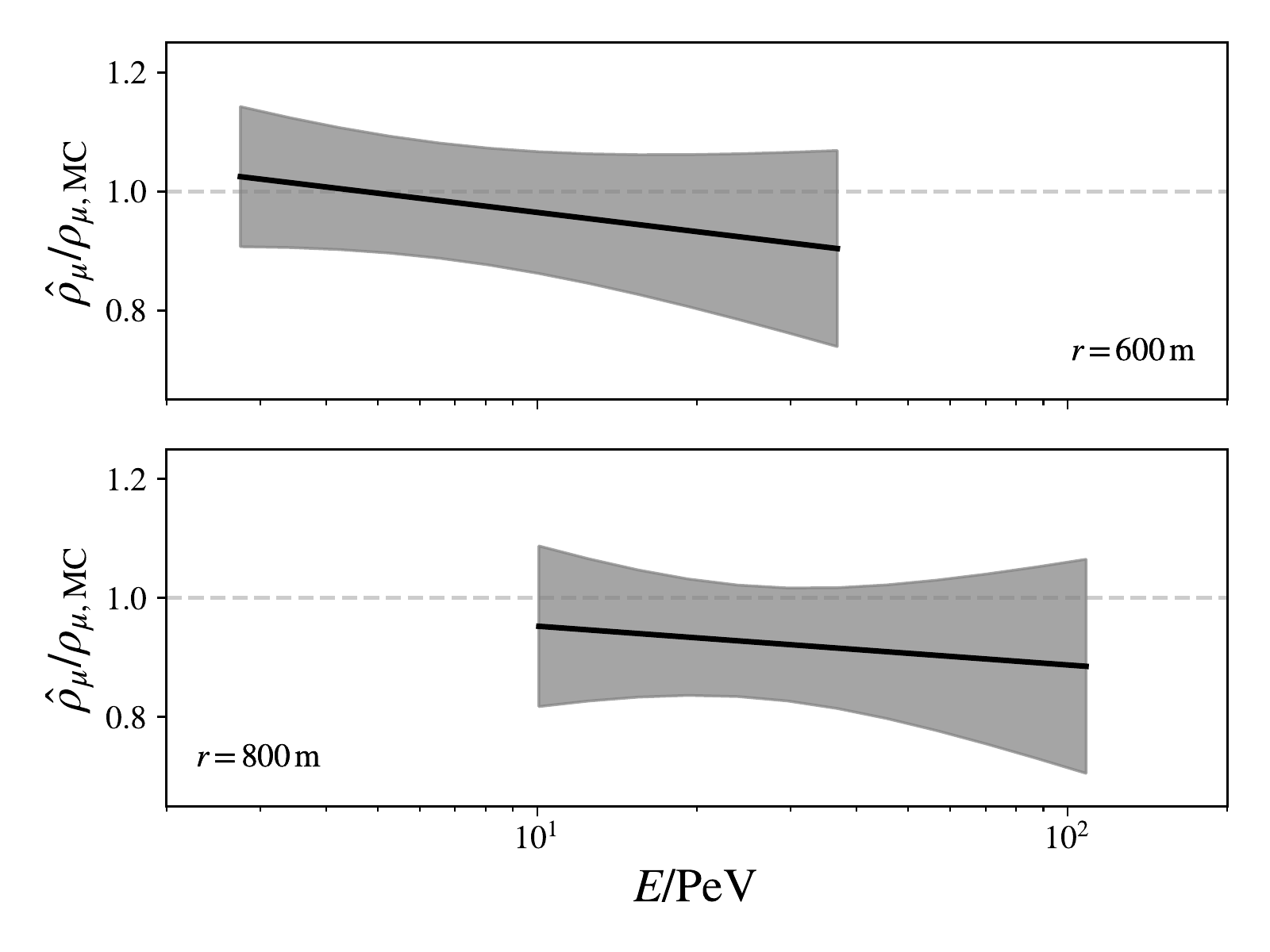}%
    }
    \vspace{-3em}
    
    \caption{Average ratio of raw reconstructed over true muon density at \SI{600}{m} and \SI
    {800}{m} lateral distance, derived from the individual ratios shown in Fig.~\ref{fig:mc_correction_had_model}. The grey band shows the resulting uncertainty which is accounted for in the final systematic uncertainties.}
    \label{fig:mc_correction_avg}
    \vspace{-1em}
    
\end{figure}

In the following, we use the difference
between the EM1 and EM2 result as an estimate of the systematic uncertainty
arising from the assumed functional form for the electromagnetic signal
distribution (see also Section~\ref{section:results}). The reconstructed muon distributions are corrected based on simulations which are evaluated in the same way as the experimental data.

The correction is determined by dividing the reconstructed muon
density obtained from simulations by the true muon density. The resulting ratios, using simulations based on the hadronic interaction models Sibyll 2.1, EPOS-LHC, and QGSJet-II.04, are shown separately in
Fig.~\ref{fig:mc_correction_had_model}. The inverse of this ratio is used as a
multiplicative factor to adjust the result in reconstructed experimental data. However, the correction factor clearly depends on
the mass composition of the sample, since iron primaries require a larger
correction. The actual composition is unknown, so the correction factor applied to the data will be the average of the proton and iron factors, with a systematic uncertainty of half of the difference. The actual composition is better known, but we are using this conservative approach for the uncertainty estimate. A linear fit to this average
yields the correction factors for each hadronic model, depicted as black lines in
Fig.~\ref{fig:mc_correction_had_model}. There are three contributions to
the uncertainty in the correction factor: the electromagnetic signal model used
(EM1/EM2), the assumed mass composition, and the statistical uncertainty in the
fit. All these uncertainties appear as systematic uncertainties in the muon density. 

The remaining uncertainty due to bin migration effects has been studied by comparing the corrected muon densities versus true and reconstructed observables in simulations. This has been done assuming three recent composition models in air shower simulations, commonly known as H3a~\cite{Gaisser:2011cc}, GST~\cite{Stanev:2014mla}, and GSF~\cite{Dembinski:2015xtn}. The remaining uncertainty has been estimated to be at the sub-percent level and can be neglected considering the uncertainties due to the electromagnetic model, the mass composition assumption, and the statistical uncertainty of the fit.

\begin{figure*}[t]
  \centering
  \vspace{-1em}
  
  \subfloat{%
    \hspace{-0.5em}\includegraphics[width=0.495\textwidth]{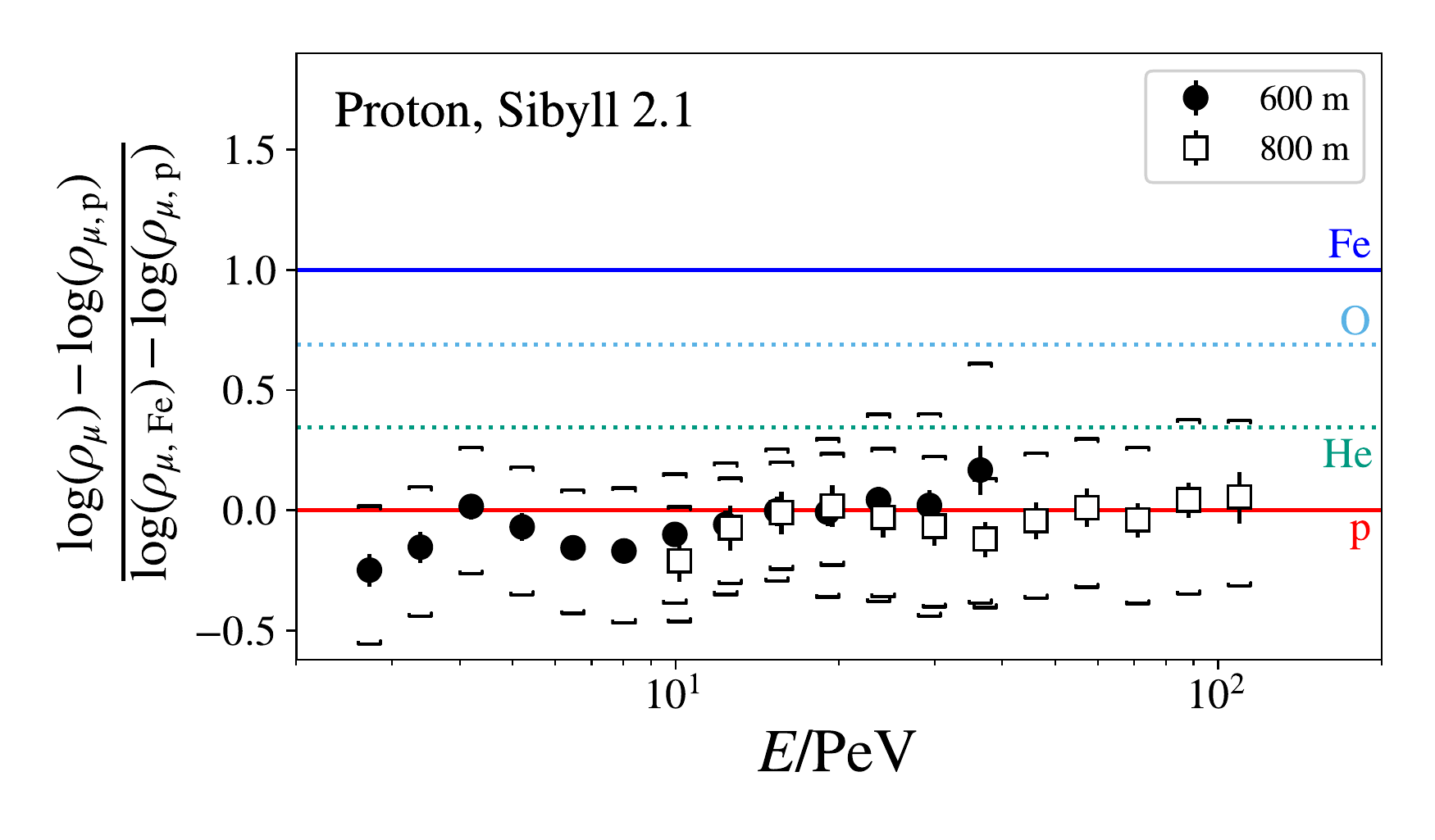}%
  }\;
  \subfloat{%
    \includegraphics[width=0.495\textwidth]{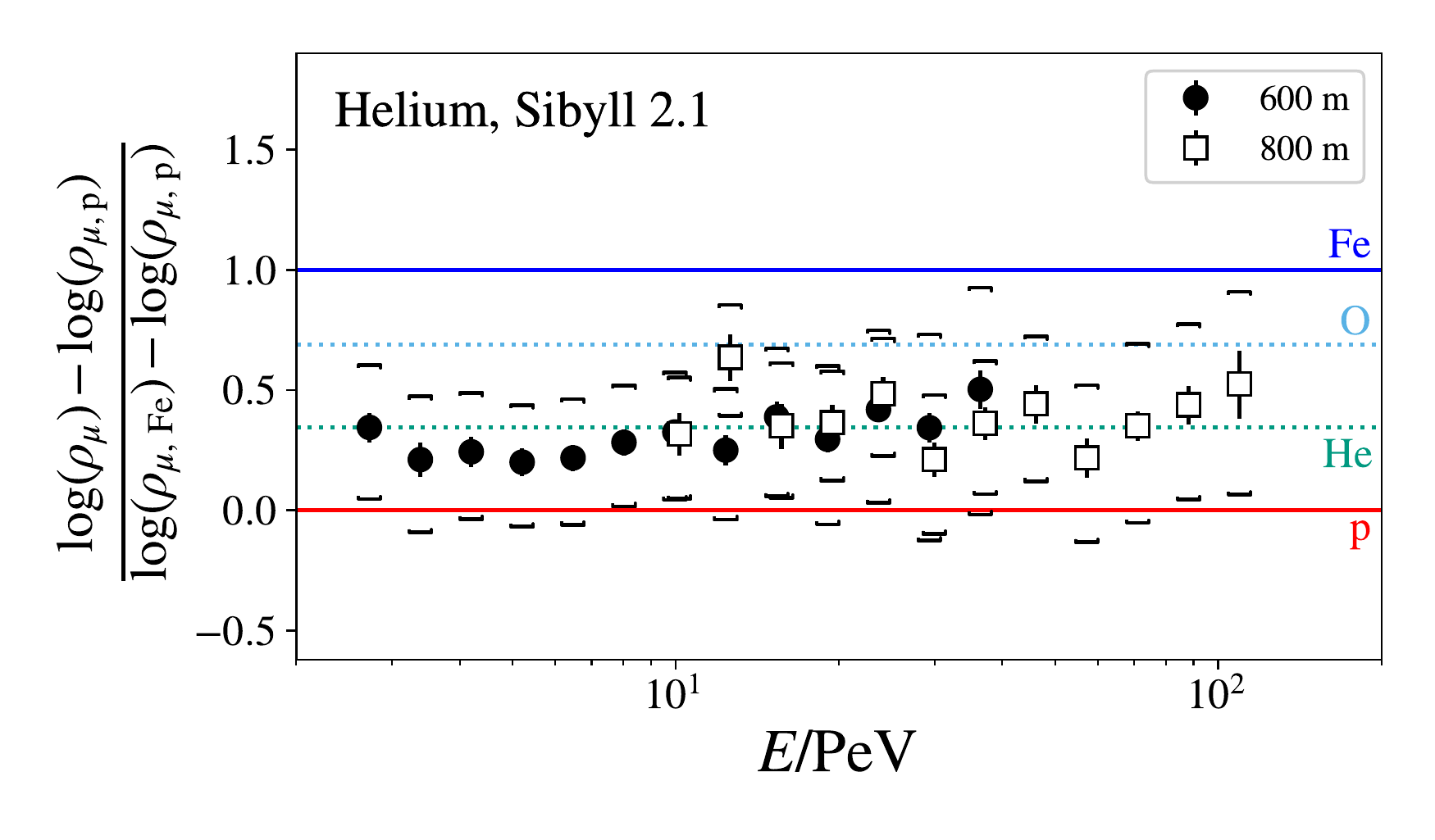}%
  }
  \vspace{-1.2em}
  
  \subfloat{%
    \hspace{-0.5em}\includegraphics[width=0.495\textwidth]{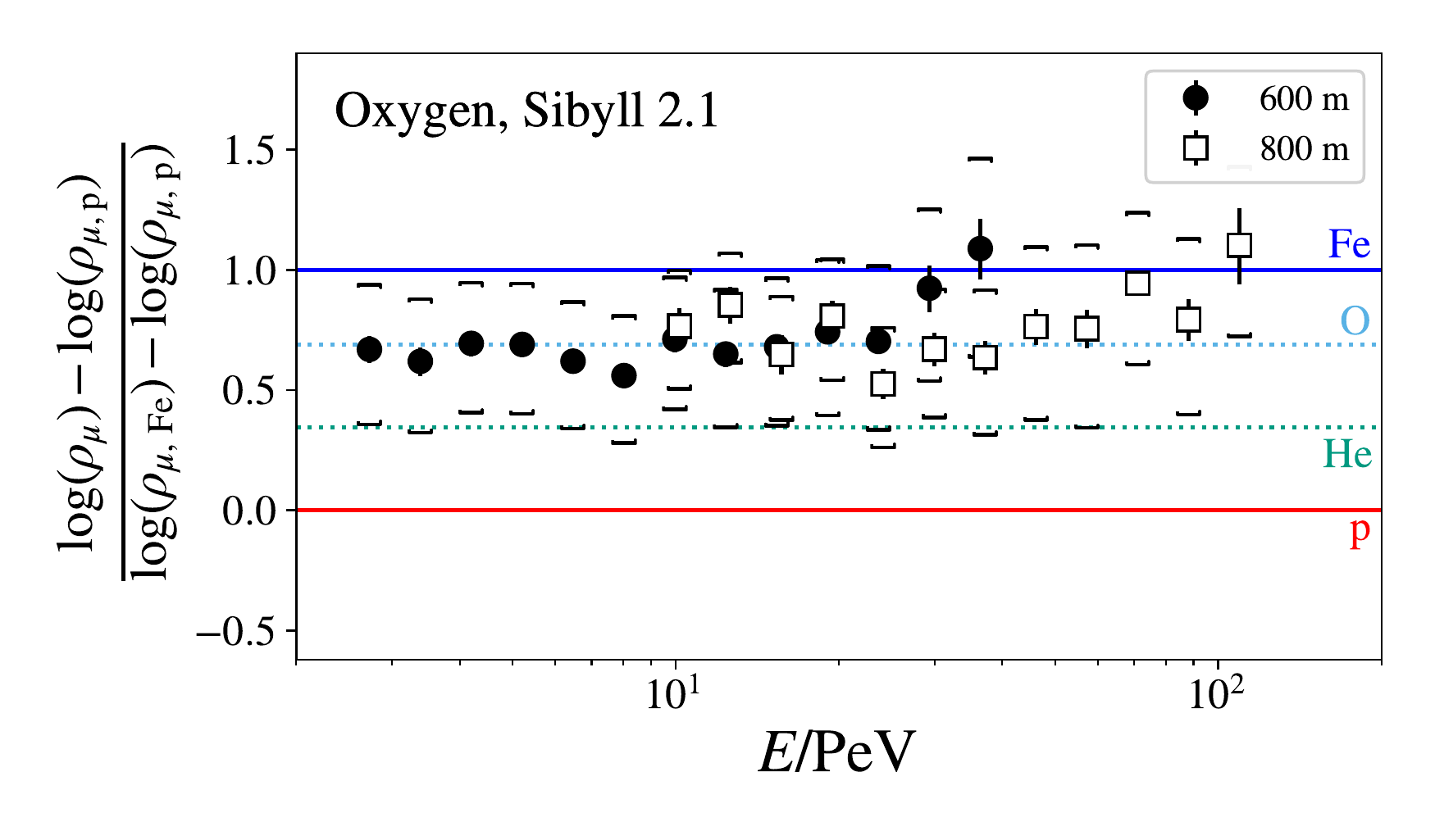}%
  }\;
  \subfloat{%
    \includegraphics[width=0.495\textwidth]{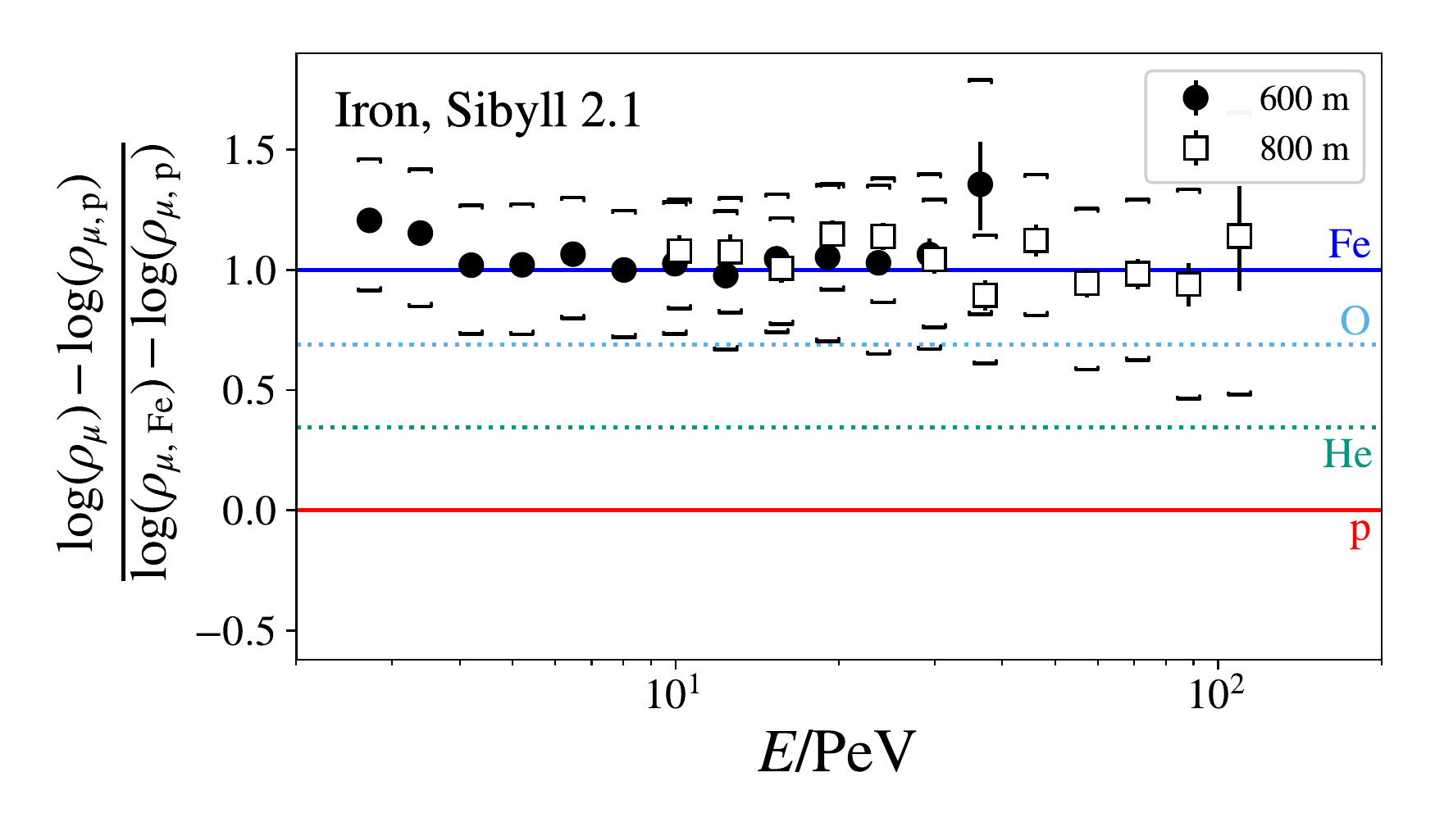}%
  }
  \vspace{-1.5em}
  
  \caption{Comparison between corrected muon density and true muon density obtained from simulated air showers using Sibyll 2.1. The logarithm of $\rho_{\mu}$ has been scaled such that the true value for proton is at zero and the true value for iron is at 1, as shown in Eq.~\ref{eq:z_value}. The oxygen and helium datasets are independent from the datasets used to develop the method. Their corresponding values correctly interpolate within the region of interest.}
  \label{fig:model_mc_comparison}
\end{figure*}

The hadronic interaction model can affect the reconstructed muon density in three ways. The first way is to influence
the functional form of the electromagnetic signal distribution. The
second way is to alter the fraction of muons below threshold. The
third way is to produce muons with an angular distribution that
differs significantly between the models. To account for these model dependencies of the correction factors, the muon densities are determined for each hadronic interaction model separately, as shown in Fig.~\ref{fig:mc_correction_had_model}.

In addition to the model-dependent results, the muon densities are also determined using an average correction factor, which is shown in Fig.~\ref{fig:mc_correction_avg}. This average correction is obtained from the three ratios of the muon densities shown in Fig.~\ref{fig:mc_correction_had_model} and a linear fit to this average
yields the average correction factors, depicted as a black line.

The correction factor is calculated using only proton and iron primaries and the
intermediate nuclei, helium and oxygen, can be used for validation purposes. It is expected that the muon density scales with the mass number, such that the logarithm of muon densities for helium and
oxygen primaries should lie in between those of proton and iron.
In Fig.~\ref{fig:model_mc_comparison}, the logarithm of the mean muon densities, obtained from simulations using Sibyll 2.1, are scaled such that the proton value is at zero and the iron value is at one on the vertical
axis for each model,
\begin{equation}
\label{eq:z_value}
z=\frac{\log(\rho_\mu)-\log(\rho_{\mu,\mathrm{p}})}{\log(\rho_{\mu,\mathrm{Fe}})-\log(\rho_{\mu,\mathrm{p}})}
\end{equation}
which is often referred to as the \emph{z-value} in the context of muon measurements. As expected, the logarithm of the reconstructed average muon densities for intermediate masses correctly interpolates linearly with $\log(A)$ between proton and iron.

\section{Results}
\label{section:results}

\begin{figure}
  \centering
  \subfloat{%
    \hspace{-0.8em}\includegraphics[width=0.5\textwidth]{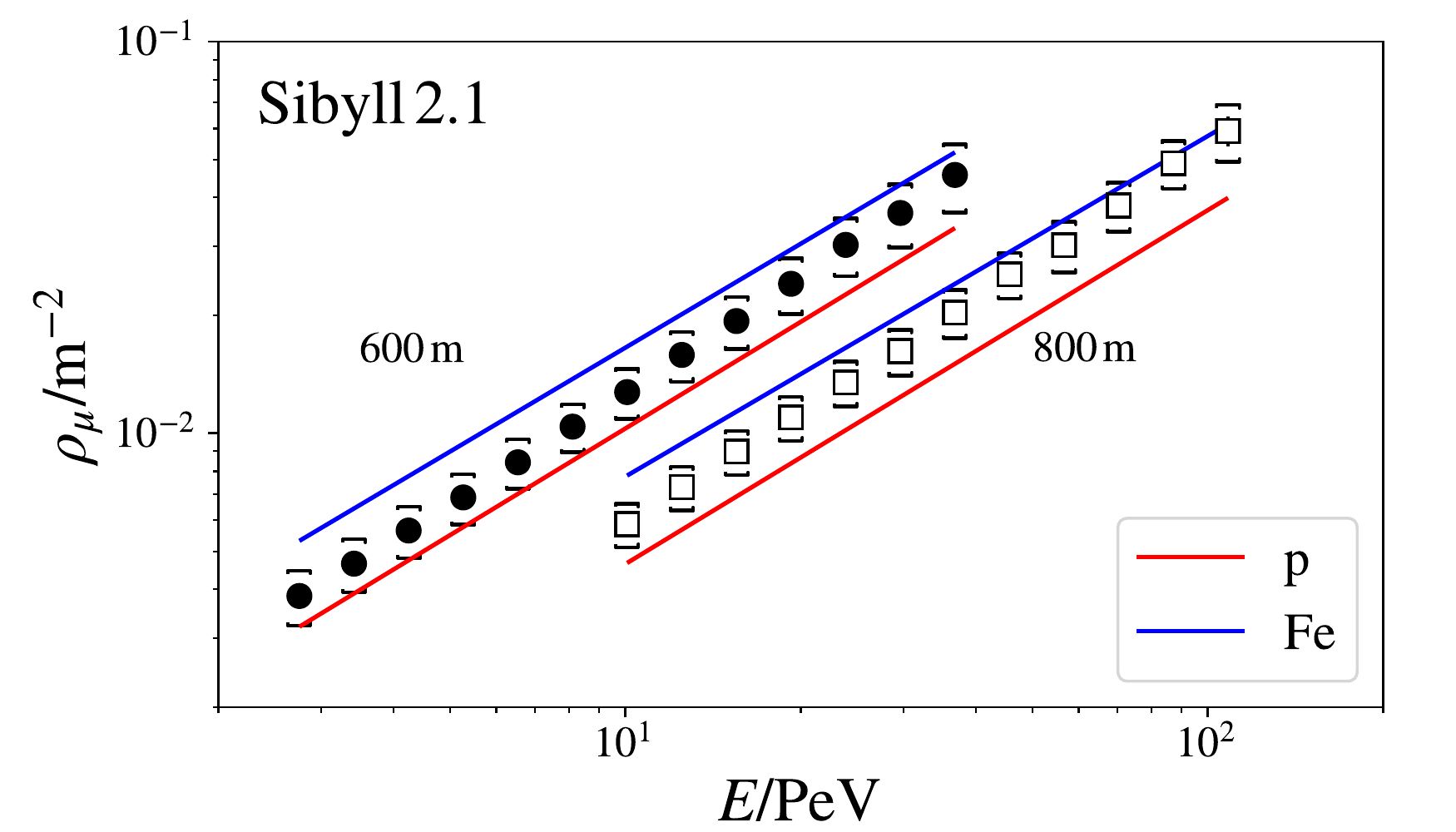}%
  }
  \vspace{-1.2em}
  
  \subfloat{%
    \hspace{-0.8em}\includegraphics[width=0.5\textwidth]{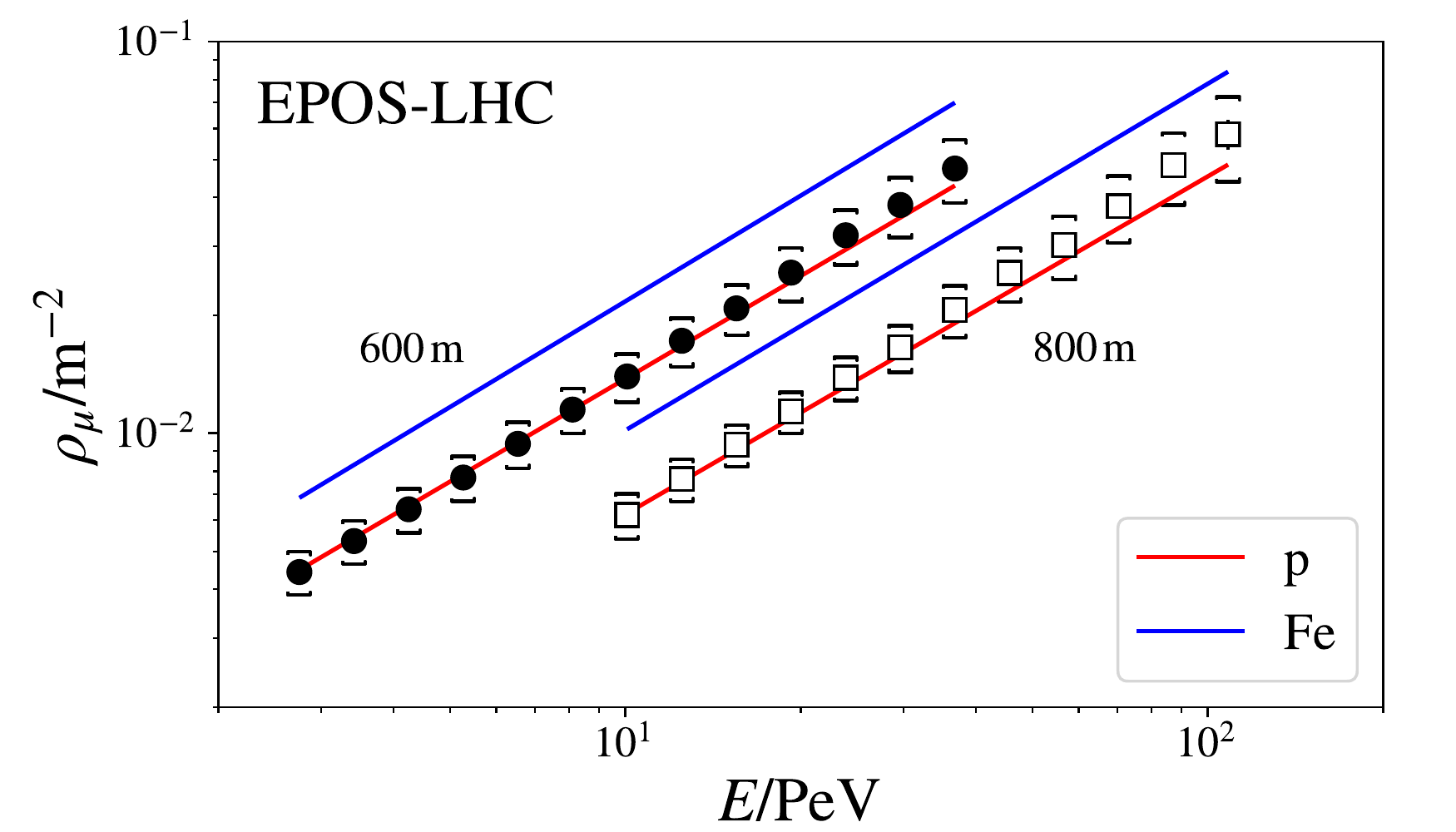}%
  }
  \vspace{-1.2em}

  \subfloat{%
     \hspace{-0.8em}\includegraphics[width=0.5\textwidth]{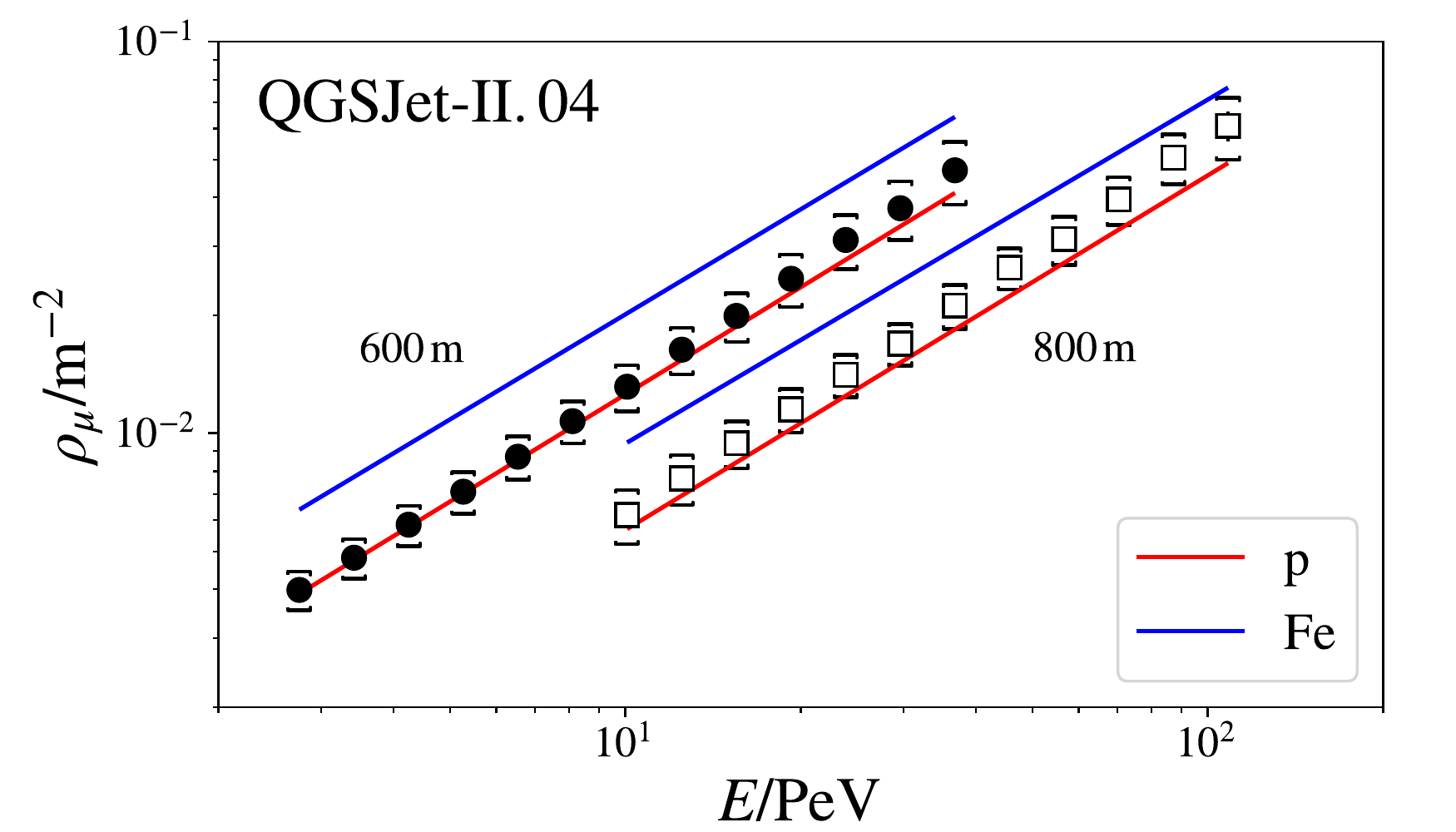}%
  }
  \caption{Measured muon densities at \SI{600}{m} (solid circles) and \SI{800}{m} (white squares) lateral distance after applying the corrections from Fig.~\ref{fig:mc_correction_had_model}, for the hadronic interaction models Sibyll~2.1, EPOS-LHC, and QGSJet-II.04. Error bars indicate the statistical uncertainty, brackets the systematic uncertainty. Shown for comparison are the corresponding simulated densities for proton and iron (red and blue lines). Tables of these data are available in a separate public data release~\cite{DataRelease}.}
  \label{fig:data_rho_final_model}
    \vspace{-1em}
    
\end{figure}

\begin{figure}[t]
  \hspace{-1.2em}\includegraphics[width=0.5\textwidth]{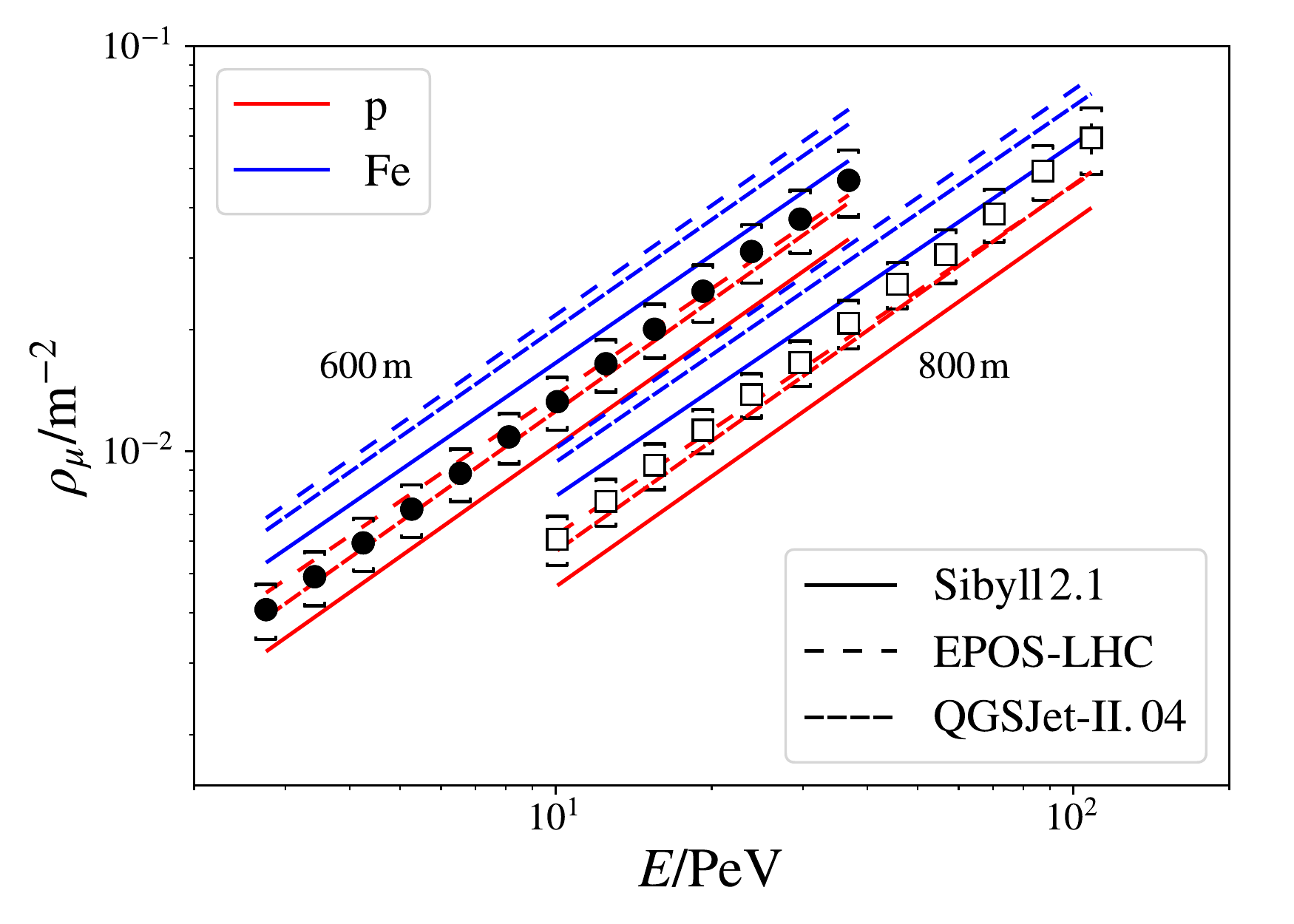}
  \vspace{-2em}
  
  \caption{Measured muon density at \SI{600}{m} (solid circles) and \SI{800}{m} (white squares) lateral distance after applying the average correction from Fig.~\ref{fig:mc_correction_avg}. Error bars indicate the statistical uncertainty, brackets the systematic uncertainty. Shown for comparison are the corresponding simulated densities for proton and iron (red and blue lines). Tables of these data are available in a separate public data release~\cite{DataRelease}.}
  \label{fig:data_rho_final}
   \vspace{-1em}
   
\end{figure}

Figure~\ref{fig:data_rho_final_model} shows the mean muon density at \SI{600}{m} from the shower
axis for air showers with energies between \SI{2.5}{PeV} and \SI{40}{PeV},
and the mean muon density \SI{800}{m} from the shower axis for air
showers with energies between \SI{9}{PeV} and \SI{120}{PeV}. To produce these model-dependent results, the individual corrections shown in Fig.~\ref{fig:mc_correction_had_model} have been used. Also shown are predictions of the mean muon density in proton and iron showers (solid lines), simulated with \corsika{} using the Sibyll~2.1, EPOS-LHC, and QGSJet-II.04 hadronic interaction models. In addition, the resulting model-independent muon density, using the average correction from Fig.~\ref{fig:mc_correction_avg}, is shown in Fig.~\ref{fig:data_rho_final}. The brackets represent the systematic uncertainties, while statistical uncertainties are not visible in these figures. The systematic uncertainties are larger than the statistical uncertainties over the entire energy range. The distributions qualitatively agree with the na\"ive expectation that the mean mass of the primary cosmic rays becomes larger as the primary energy increases~\cite{Kampert:2012mx}.

\begin{figure*}[tb]
    \centering
   \vspace{-1.5em}
    
    \hspace{-1em}
    \subfloat[]{%
       \includegraphics[width=0.48\textwidth]{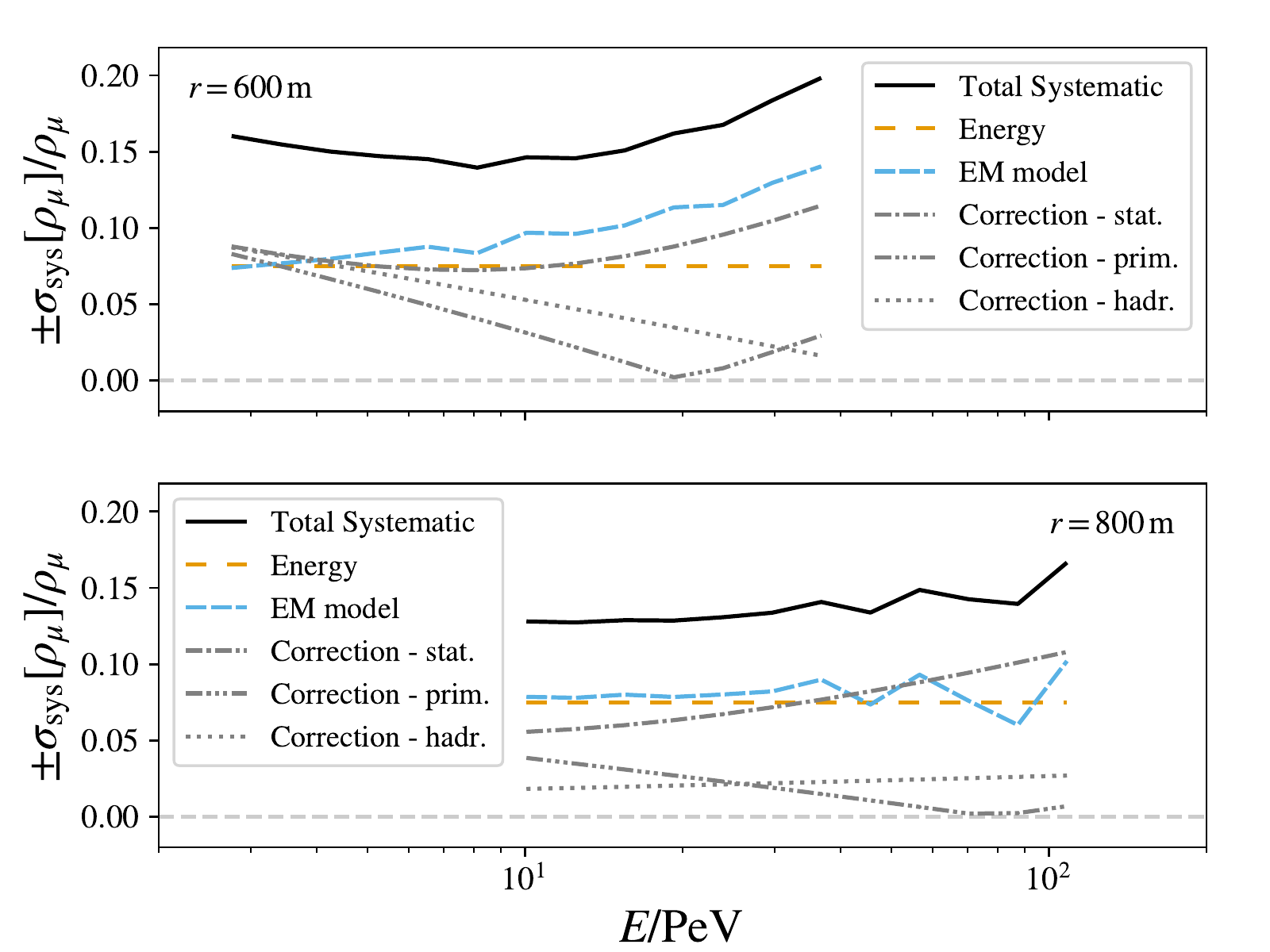}%
        \label{fig:data_sys}%
    }
    \subfloat[]{%
       \includegraphics[width=0.48\textwidth]{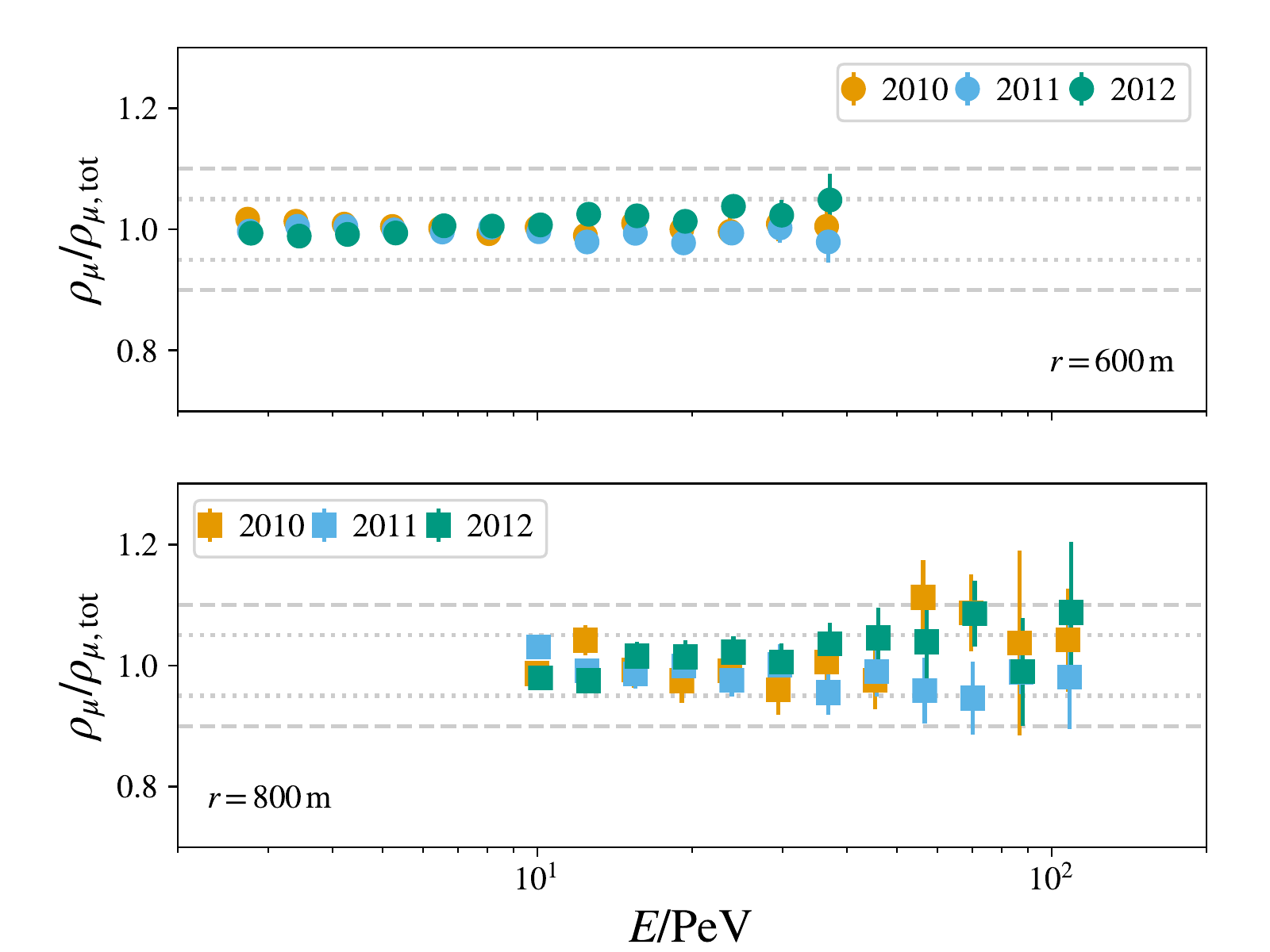}%
        \label{fig:rho_comp_year}%
    }
    \caption{(a) Relative systematic uncertainties of the muon density measurement at \SI{600}{m} and \SI{800}{m} due to energy scale and the electromagnetic model used, as well as the uncertainties from the average correction factor shown in Fig.~\ref{fig:mc_correction_avg} due to limited MC statistics (stat.), the primary mass assumptions (prim.), and hadronic models (hadr.). The individual contributions are added in quadrature to obtain the total systematic uncertainty (solid black line). (b) Relative deviation from the average muon density estimates at \SI{600}{m} and \SI{800}{m} lateral distance with the data split by year. The points are horizontally displaced slightly for better visibility, dotted and dashed lines indicate the 5\,\% and 10\,\% deviation thresholds, respectively. }
  \vspace{-.5em}
    
\end{figure*}

The systematic uncertainty included in these distributions arises from four main sources which are depicted in Fig.~\ref{fig:data_sys} individually.  The uncertainty in the energy determination causes a systematic uncertainty in the muon density because of the correlation between shower energy and muon number. While the effect of bin migrations due to the finite resolution of the reconstructions is covered by the Monte-Carlo correction described in Section~\ref{section:mc_verification}, a systematic shift of the absolute energy scale can also cause an apparent discrepancy in muon density and needs to be considered. The associated uncertainty in the muon density has been determined based on the uncertainties of the cosmic ray flux reported by IceTop in Ref.~\cite{Aartsen:2013wda}. The individual flux uncertainties are added in quadrature and the corresponding energy uncertainty is calculated assuming a power law with spectral index 2.7 for the cosmic ray flux. Assuming the uncertainty in muon density is directly proportional to the uncertainty in the energy scale\footnote{The Heitler-Matthews model, \emph{i.e.} Eq.~\ref{eq:muon_scaling}, yields a sub-linear scaling between muon number and energy. Thus, assuming linearity produces conservative uncertainties.}, the resulting uncertainty in muon density has been determined to be approximately 7.5\%, as shown in Fig.~\ref{fig:data_sys}.

The other three sources were already discussed, namely the functional
form in the electromagnetic signal distribution described in
Section~\ref{section:em_model} (orange lines in
Fig.~\ref{fig:r_signal_slices}), the mass composition assumed to derive the
correction shown in Fig.~\ref{fig:mc_correction_had_model}, and the statistical uncertainty when fitting the correction, both
discussed in Section~\ref{section:mc_verification}. The effect of the EM model
assumption enters in two different places, when
reconstructing the muon density in data, and when reconstructing the muon
density in simulations in order to derive the correction factor. However, these
two effects are correlated. The difference for the two EM models in the reconstructed muon density are
the same size in simulations and in experimental data.

The snow is corrected for in a time-dependent way by dividing 
the data sample in subsamples
corresponding to campaign years. The mean muon density in each
subsample is compared to the main result. The difference between the muon
density from yearly samples and the main result is shown in
Fig.~\ref{fig:rho_comp_year}. At a lateral distance of
\SI{600}{m} we find differences between -2\% and +4\% with respect to the average muon densities.  At \SI{800}{m},
no systematic effect is detectable, because the statistical scatter is
much larger than any inter-year effect. 
Thus, the results for the muon densities are consistent with being independent of the detector configuration.

\begin{figure*}[bt]
    \centering
    \vspace{-1.em}
    
    \subfloat[]{%
        \hspace{-1em}\includegraphics[width=0.48\textwidth]{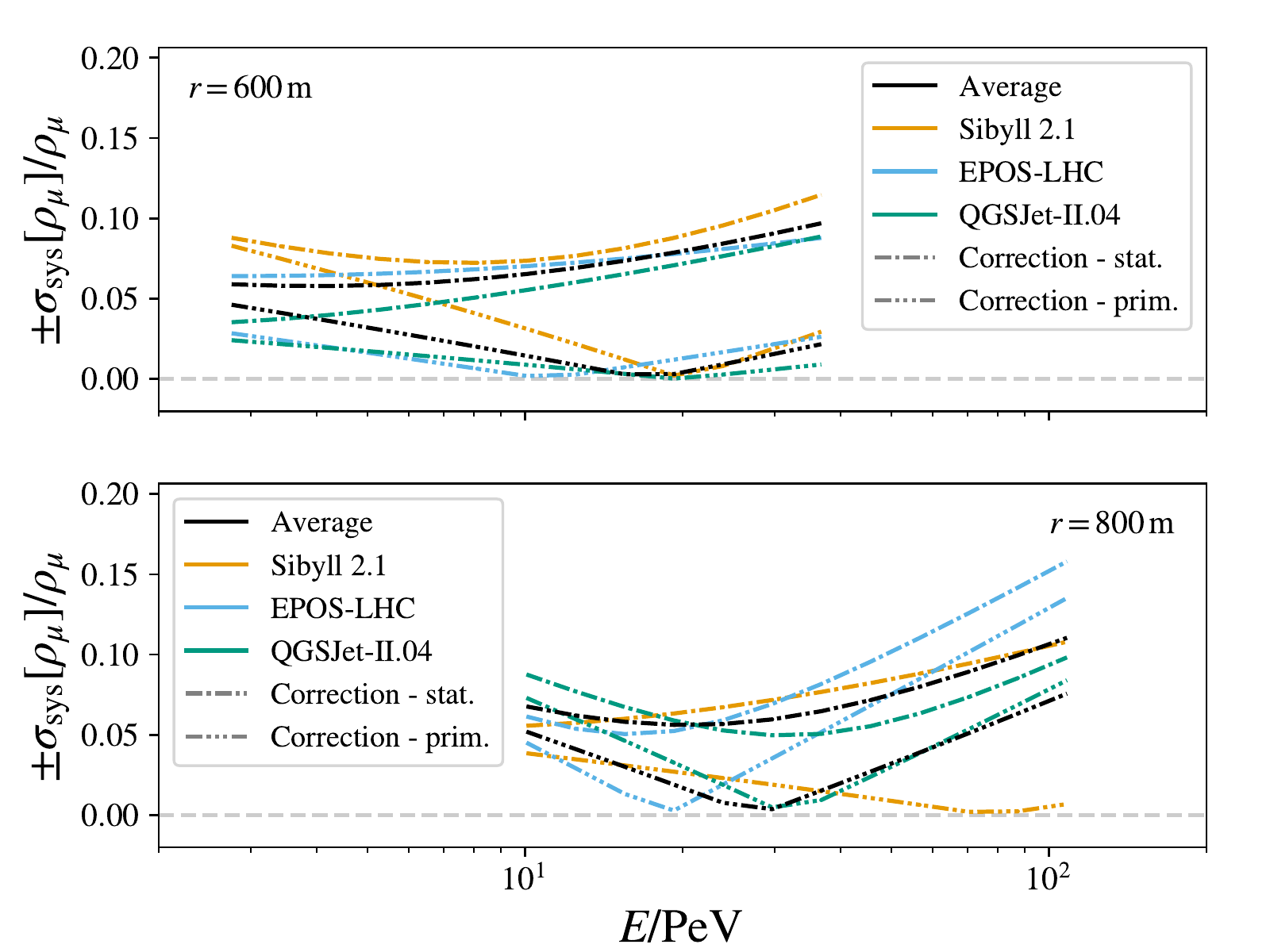}%
         \label{fig:data_sys_model}%
    }
    \subfloat[]{%
        \hspace{-1em}\includegraphics[width=0.48\textwidth]{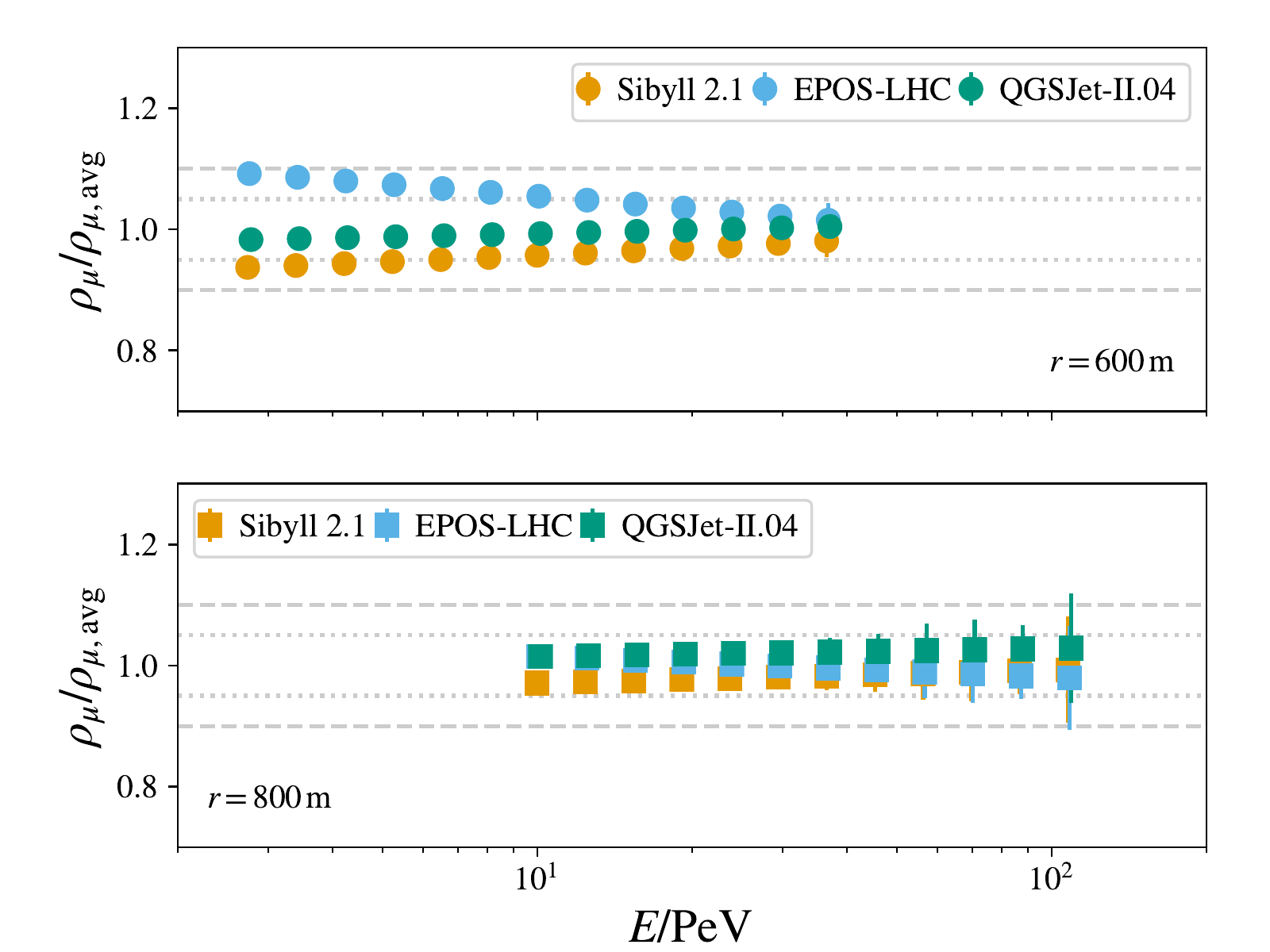}%
         \label{fig:rho_comp_model}%
    }\;\;\;
    \caption{(a) Relative systematic uncertainty due to hadronic interaction models assumed to determine the correction factors. The uncertainties due to limited statistics (stat.) and the primary mass assumption (prim.) are shown separately. These hadronic model uncertainties are added in quadrature to the uncertainties from Fig.~\ref{fig:data_sys} and included in the results shown in Figs.~\ref{fig:data_rho_final_model} and \ref{fig:data_rho_final}. (b) Relative deviation of the muon density estimates at \SI{600}{m} and \SI{800}{m} lateral distance obtained from different hadronic interaction models w.r.t. the average result shown in Fig.~\ref{fig:data_rho_final}. The points are horizontally displaced slightly for better visibility, dotted and dashed lines indicate the 5\,\% and 10\,\% deviation thresholds, respectively.}
   \vspace{-.5em}
    
\end{figure*}

\begin{figure}
  \centering
  \vspace{-.5em}
  
  \subfloat{%
  \hspace{-0.7em}\includegraphics[width=0.482\textwidth]{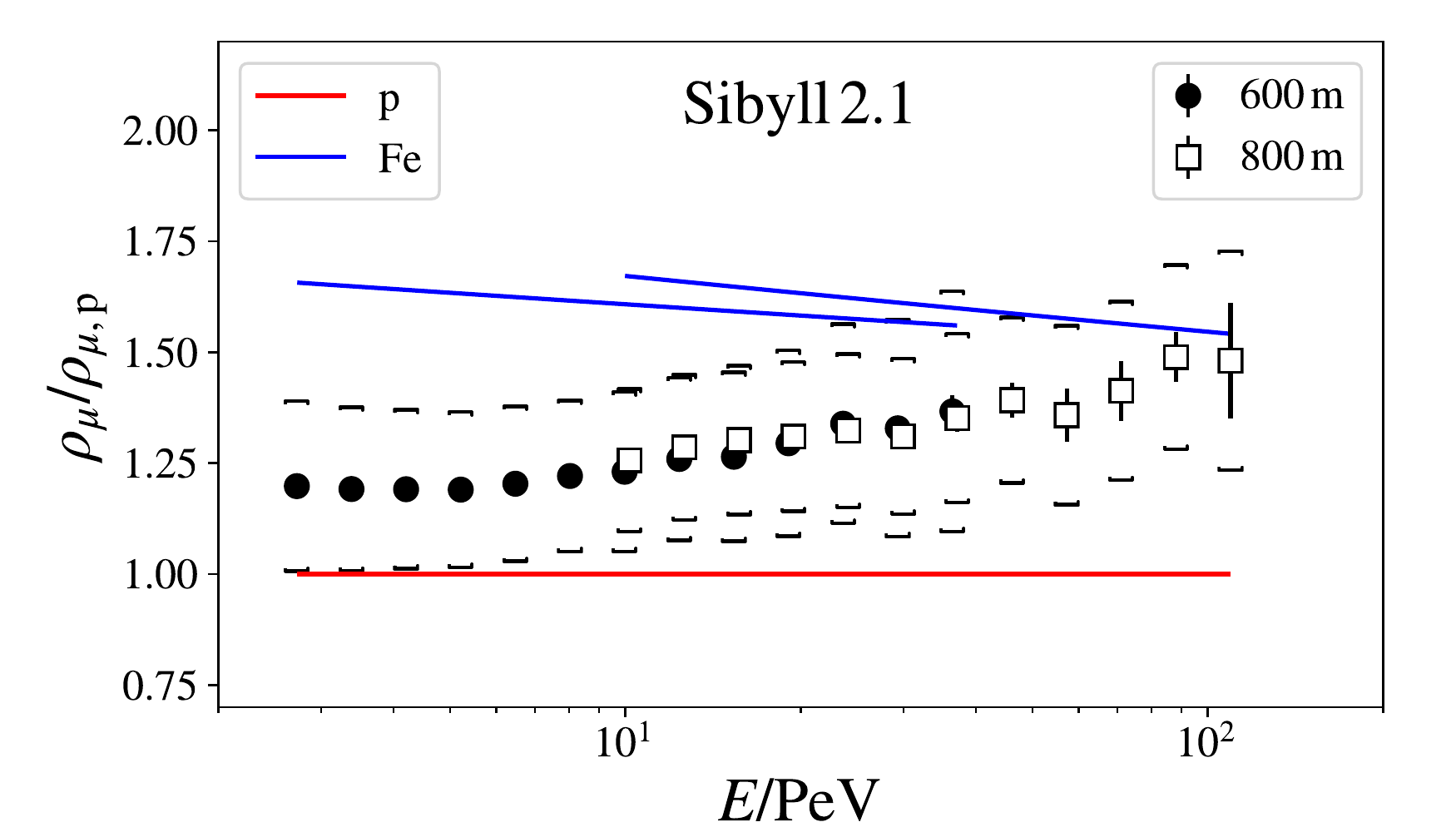}%
  }
  \vspace{-1.1em}
  
  \subfloat{%
    \hspace{-0.7em}\includegraphics[width=0.482\textwidth]{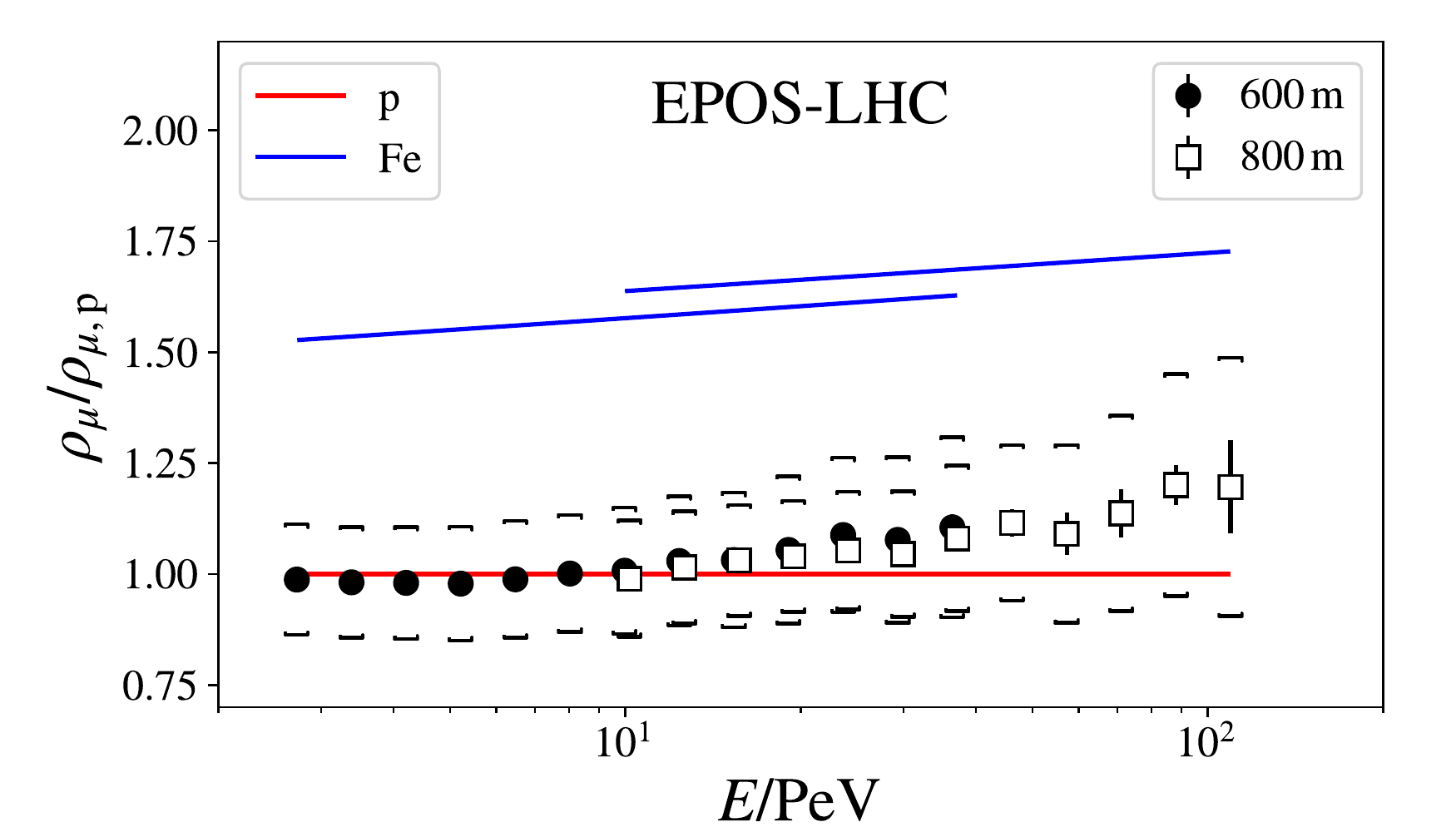}%
  }
  \vspace{-1.1em}

  \subfloat{%
     \hspace{-0.7em}\includegraphics[width=0.482\textwidth]{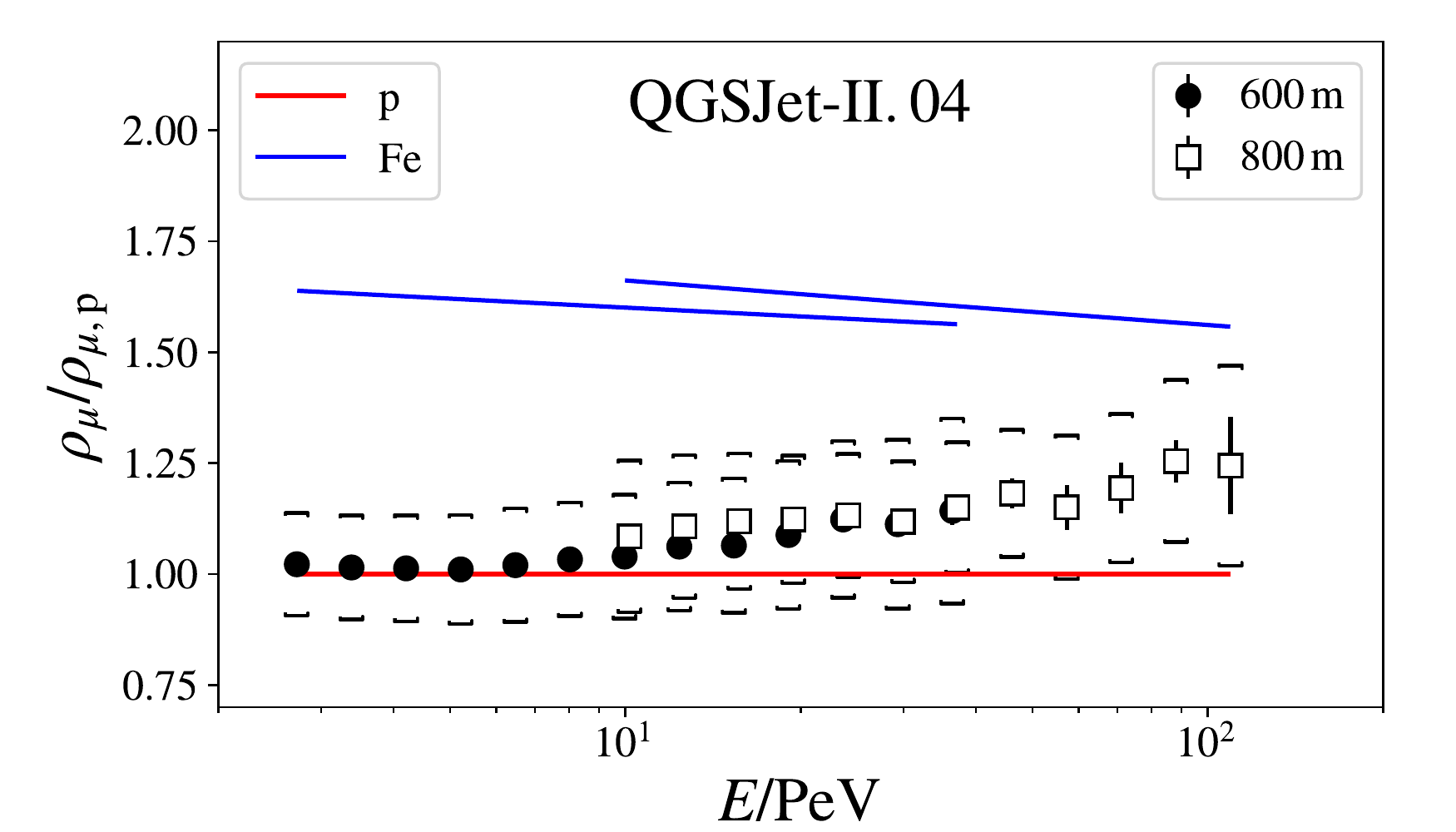}%
  }
  
  \caption{Measured muon densities at \SI{600}{m} (solid circles) and \SI{800}{m} (white squares) lateral distance after applying the corrections from Fig.~\ref{fig:mc_correction_had_model}, normalized to the muon density obtained from proton simulations. Error bars indicate the statistical uncertainty, brackets the systematic uncertainty. The points are horizontally displaced slightly for better visibility. Tables of these data are available in a separate public data release~\cite{DataRelease}.}
  \label{fig:data_rho_final_scaled}
    \vspace{-1.5em}
    
\end{figure}

The detection efficiencies for proton and iron showers can be different, in particular near 
the threshold at low energies around \SI{2.5}{PeV}. This selection bias would in turn introduce a 
systematic shift of the measured muon densities: a selection that prefers protons, for example, 
lowers the muon density at a given true primary energy. This effect, however, would be visible as a 
time-dependent shift of the muon densities in the threshold region because the snow 
correction applied during reconstruction accounts for the accumulation effect but not for a loss in efficiency. Since no significant time-dependent trend in the muon densities from the subsamples 
for each year is observed the potential bias due to mass-dependent detection efficiencies can be neglected.

\begin{figure}[!h]
  \centering
  \vspace{-.7em}
  
  \subfloat{%
    \hspace{-2em}\includegraphics[width=0.519\textwidth]{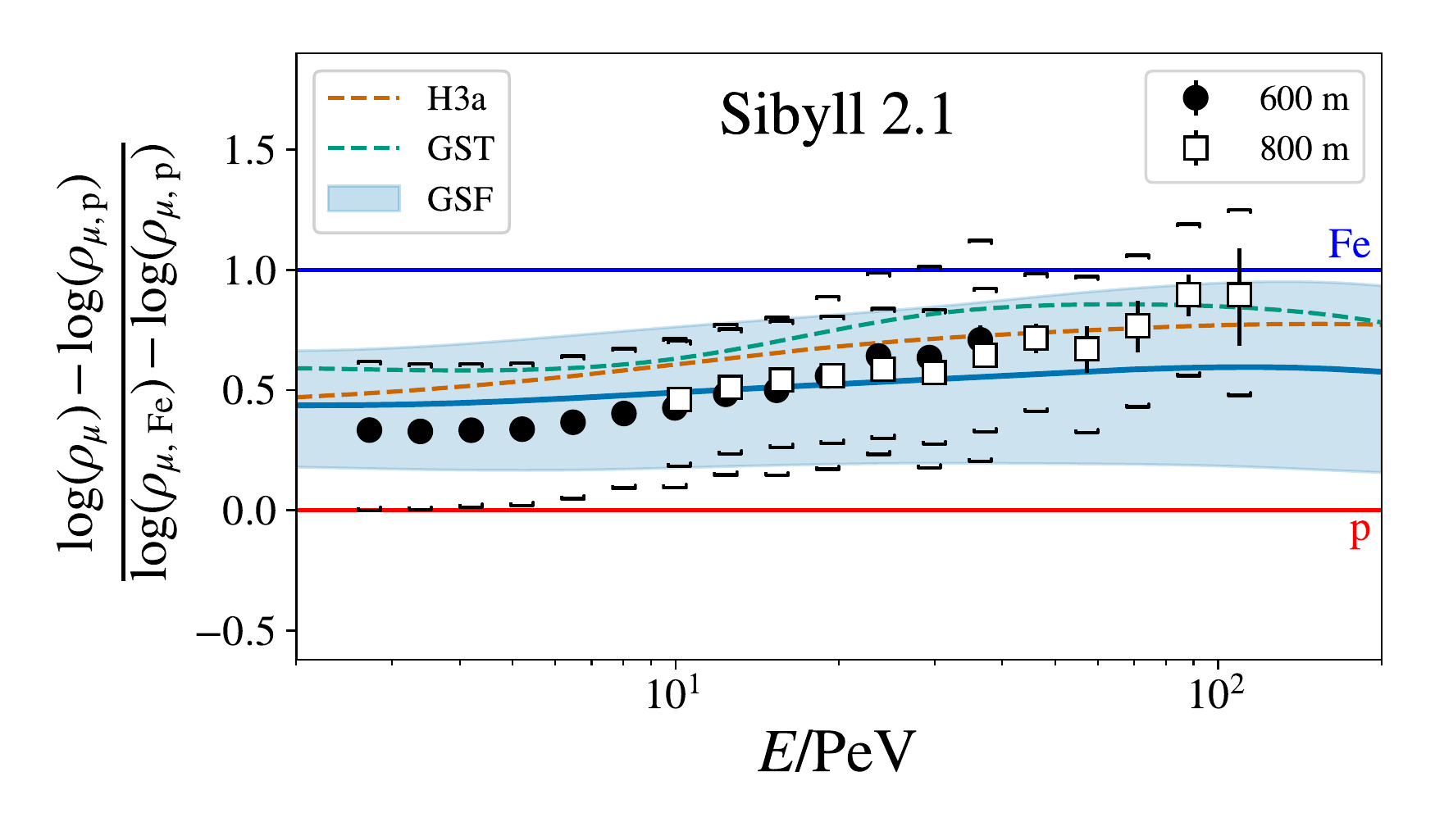}%
  }
  \vspace{-2.5em}
  
  \subfloat{%
    \hspace{-2em}\includegraphics[width=0.519\textwidth]{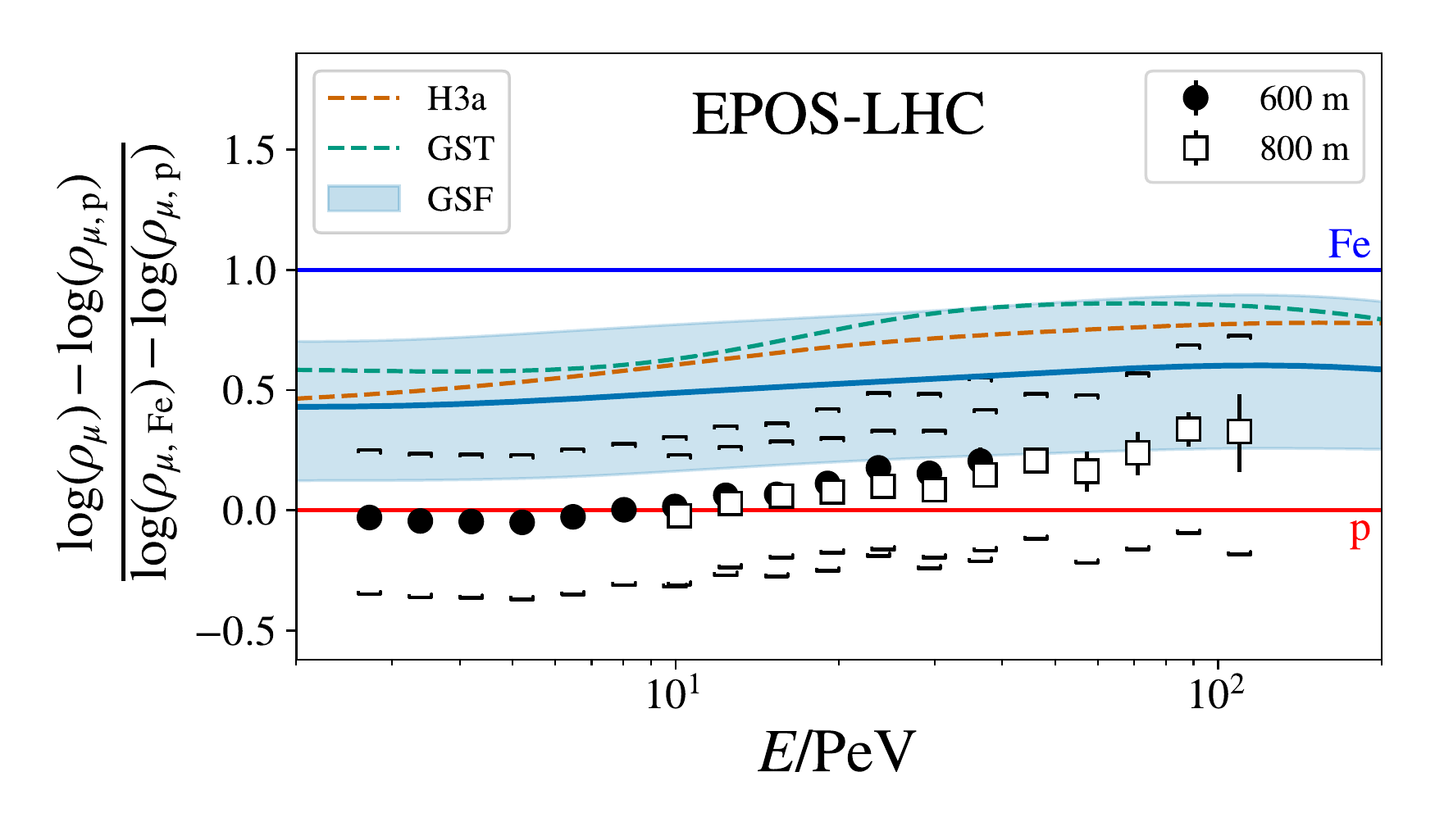}%
  }
  \vspace{-2.5em}

  \subfloat{%
     \hspace{-2em}\includegraphics[width=0.519\textwidth]{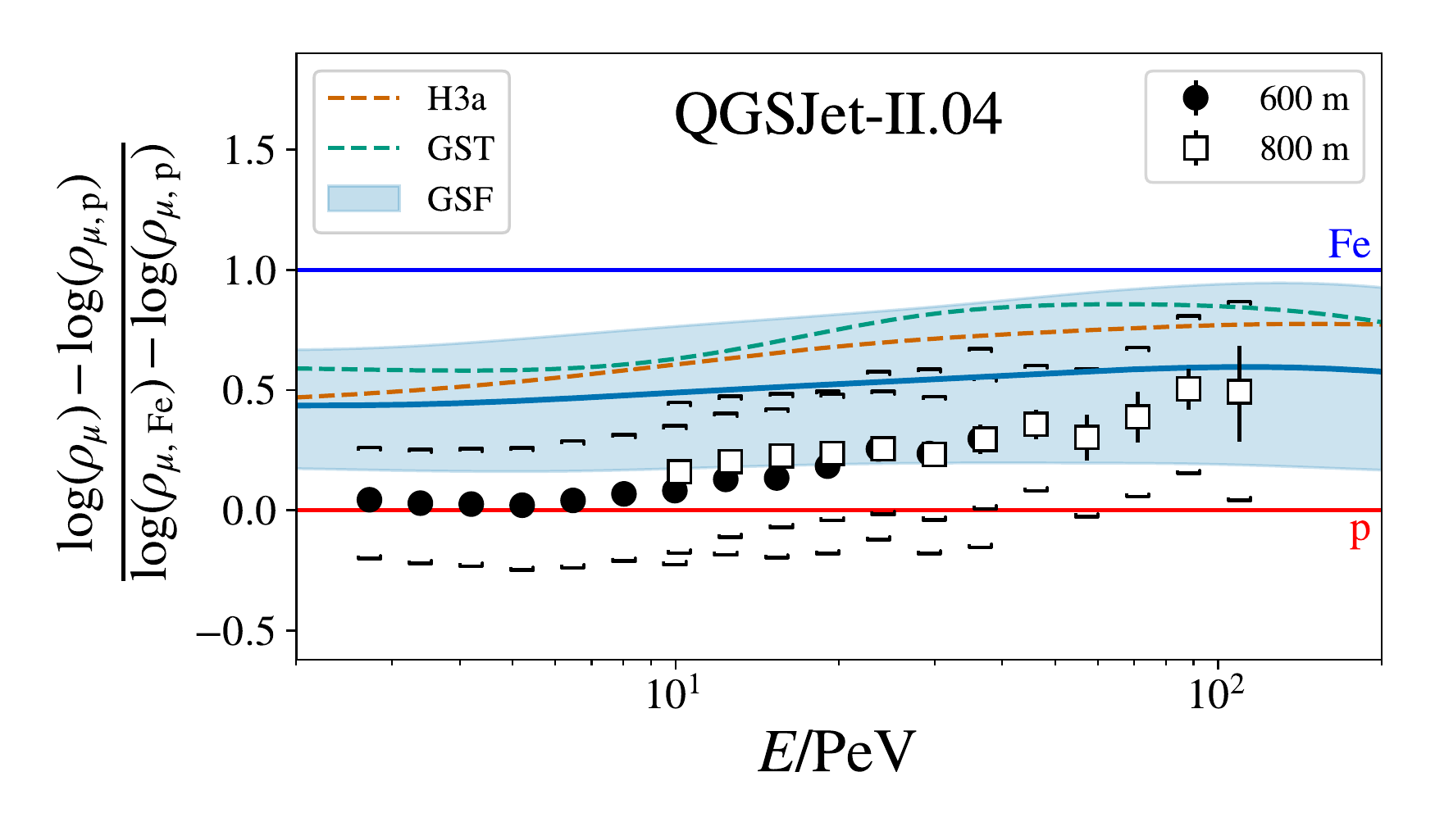}%
  }
  
  \vspace{-1.4em}
  
  \caption{Measured muon densities at \SI{600}{m} (solid circles) and \SI{800}{m} (white squares) compared to predictions from different hadronic models. These figures show the muon density scaled such that $\log \langle \rho_{\mu} \rangle$ for proton is 0 and for iron is 1. The lines display the mean muon density according to the three cosmic ray flux models H3a~\cite{Gaisser:2011cc}, GST~\cite{Stanev:2014mla}, and GSF~\cite{Dembinski:2015xtn}. The grey bands show the uncertainity of the GSF model. The differences between models and data indicate a discrepancy in the simulated muon densities in post-LHC models. The points are horizontally displaced slightly for better visibility. Tables of these data are available in a separate public data release~\cite{DataRelease}.}
  \label{fig:model_comparison}
  \vspace{-1.5em}
  
\end{figure}

As the Monte-Carlo correction depends on the underlying hadronic interaction model, the uncertainty due to the mass composition assumed to derive the correction, as well as the statistical uncertainty when fitting the correction, need to be considered for each model separately. The corresponding uncertainties for the average correction factor shown in Fig.~\ref{fig:mc_correction_avg}, which are used to derive the result in Fig.~\ref{fig:data_rho_final}, are shown in Fig.~\ref{fig:data_sys} (Correction - stat./prim.). The corresponding uncertainties due to the individual correction depicted in Fig.~\ref{fig:mc_correction_had_model}, which are used to derive the results in Fig.~\ref{fig:data_rho_final_model}, are shown in Fig.~\ref{fig:data_sys_model}. Also shown again are the uncertainties for the average correction (black lines). For the average correction factor, we also derive a systematic uncertainty due to the hadronic model assumption. Figure~\ref{fig:rho_comp_model} shows the variation of the muon density derived from the model-dependent correction with respect to the average result from Fig.~\ref{fig:data_rho_final}. In the post-LHC models EPOS-LHC and QGSJet-II.04, the muon density is larger than in Sibyll 2.1 and at a lateral distance of
\SI{600}{m} we find differences up to about 10\%. The largest difference between the average result and the individual models is accounted for as a systematic uncertainty, also shown in Fig.~\ref{fig:data_sys} (Correction - hadr.). All uncertainties are added in quadrature and are shown as brackets in the main results.

A comparison between model predictions and experimental data is displayed in
Fig.~\ref{fig:data_rho_final_scaled}, showing the ratio of reconstructed muon
density to expected muon density in simulated proton showers. The consistency of the hadronic interaction models can be examined by checking if
the data are bracketed by the expectations given by a cosmic ray flux with 100\% proton (red line) and
100\% iron (blue lines). In this sense QGSJet-II.04 and Sibyll 2.1 perform fairly well, since the corresponding
proton and iron lines bracket the data. For QGSJet-II.04,
the data between \SI{2.5}{PeV} and about \SI{10}{PeV} is very close to the expectation for a 100\% proton flux.
The EPOS-LHC lines do not bracket the data at the lowest energies, requiring an average composition slightly lighter than proton below approximately \SI{6}{PeV}. The cosmic ray composition has been previously measured in this energy range \cite{IceCube:2012vv,Aartsen:2013wda,IceCube:2019} and various cosmic ray flux models consistent with existing experimental data are available~\cite{Gaisser:2011cc,Stanev:2014mla,Dembinski:2015xtn}. 
The comparison between the measured muon densities and predictions based on different cosmic ray flux models is displayed in Fig.~\ref{fig:model_comparison}.
For this comparison, the logarithms of the muon densities are again scaled such that the proton
value is at zero and the iron value is at one on the vertical axis,
the same as in Fig.~\ref{fig:model_mc_comparison} and as described in Eq.~\ref{eq:z_value} (z-value).
The expected z-values according to three recent
composition models commonly known as H3a~\cite{Gaisser:2011cc},
GST~\cite{Stanev:2014mla}, and GSF~\cite{Dembinski:2015xtn} are shown as lines. While for H3a
and GST only the best fit curve is provided, for the GSF model also the entire
covariance matrix of the fit is available. This allows the calculation of the uncertainty in the
expected muon density, which is shown as a grey band in Fig.~\ref{fig:model_comparison}. While the shape of the muon density distributions agrees fairly well within uncertainties, the main difference between data and model expectations is the normalization of the muon densities. The mean model expectation according to Sibyll 2.1, assuming the GSF flux model and its uncertainty band (which covers H3a and GST predictions), is close to the data over the entire energy range. However, the post-LHC models EPOS-LHC and QGSJet-II.04 yield higher muon densities that are in tension with the measurement, assuming any realistic cosmic ray flux model.

\section{Conclusions}
\label{section:conclusion}

We have presented measurements of the muon density in air showers at lateral distances of \SI{600}{m}
and \SI{800}{m} for cosmic ray energies from \SI{2.5}{PeV} to \SI{40}{PeV} and
\SI{9}{PeV} to \SI{120}{PeV}, respectively, which are shown in Fig.~\ref{fig:data_rho_final}. We have compared these results
with expectations assuming a realistic primary composition and using the hadronic interaction models Sibyll 2.1 (pre-LHC), EPOS-LHC, and QGSJet-II.04 (both post-LHC).

For low-energy muons ($\sim\SI{1}{GeV}$) from air showers with energies above about $\SI{1}{EeV}$ (HiRes-MIA, Pierre Auger Observatory, NEVOD-DECOR), 
the measured muon densities are always higher than the predictions from simulations and could not be
accounted for even by LHC-tuning the models. In contrast, for the muon densities reported in this work, the post-LHC models
predict too large average muon densities over the entire energy range up to about \SI{100}{PeV}, assuming realistic cosmic ray flux models consistent with existing experimental data. 
The large muon density in the models correspondingly yields very light mass compositions when the experimental data are interpreted using these models. 
At around \SI{100}{PeV}, the muon density is consistent with that of a mixed composition. 
This is in agreement with the measurement done with the
EAS-MSU surface detector which is consistent with
a 57\% proton fraction in a na\"ive proton+iron model interpreted with QGSJet-II.04~\cite{Fomin:2016kul}. 

Within the pre-LHC baseline model Sibyll~2.1, the expectations based on realistic flux models are close to the measurement of low-energy muon densities reported here. This observation is consistent 
with recent measurements of high-energy muons in IceCube's deep ice detector ($E_\mu\gtrsim\SI{300}{GeV}$)
in the primary energy region between \SI{2.5}{PeV} and \SI{1}{EeV}~\cite{DeRidder:2017alk,IceCube:2019,IceCube:2021ixw}. 
The post-LHC models, however, yield lower muon densities for any given mass for the high-energy muons measured in IceCube, 
in contrast to the comparison for low-energy muon densities. This suggests that 
the muon densities in post-LHC models are not correct for primary energies below approximately \SI{100}{PeV}.

However, a systematic shift in the reconstructed primary energy can also cause an apparent
discrepancy in muon density.  A 20\% shift in the energy scale, for example, translates to an apparent shift of about $0.5$ in the z-value, defined in Eq.~\ref{eq:z_value} and shown in Fig.~\ref{fig:model_comparison}. For this reason, a detailed comparison of results from multiple observatories across all energies is needed to understand the production of GeV muons in air showers~\cite{Dembinski:2019uta,Soldin:2021wyv}. 
This comparison is beyond the scope of this work, since it needs to account for the
systematic effects of each detector, including zenith angle range, lateral distance,
and systematic uncertainty in the absolute energy scale~\cite{Cazon:2020zhx}.

The method of measuring the density of muons in air showers as described here can be extended to a larger phase space.
The current version focuses on mostly vertical events and by
extending it to a larger zenith angle range we could
compare with the measurements of air shower attenuation
done with KASCADE-Grande~\cite{Arteaga-Velazquez:2015jpa}. As the KASCADE-Grande experiment has measured the muon fluxes at sea level while IceTop is at an atmospheric depth of about $690\,\mathrm{g/cm}^2$, this measurement would allow a dedicated comparison of the muon attenuation in dependence of the mass overburden. 
In addition, by studying  a larger range of lateral distances, instead of restricting it to
the interpolated values at \SI{600}{} and \SI{800}{m}, it would be possible to study deviations
from the muon lateral distribution function given by simulations. This
can shed light on systematic differences between detectors that measure
muons at different lateral distances. Systematic studies of the low-energy interaction model and the transition to the high-energy regime will provide information about muon production at low energies.

This work creates more opportunities for further studies. 
The signal probability distribution models used
to estimate the mean muon density in this work can be used on an
event-by-event basis, as a likelihood function, to estimate the muon
content in individual air showers. In addition, further studies
of the mean muon density and its correlation with the muon bundle
multiplicity in the deep portion of IceCube can provide unique information and strong
constraints on hadronic interaction models. These coincident measurements would relate
to the energy spectrum of the muons in air showers since the deep part of
IceCube has an energy threshold of several hundred GeV, as opposed to
IceTop's threshold of a few hundred MeV. These analysis improvements and future 
measurements will be crucial to further understand the production of GeV muons in air showers in the primary energy range from \SI{2.5}{PeV} to \SI{100}{PeV} and higher.

\begin{acknowledgments}

The IceCube collaboration acknowledges the significant contributions to this manuscript from the Bartol Research Institute at the University of Delaware. This paper also profited enormously from the contributions of our departed colleague Thomas~K.~Gaisser (1940-2022).

USA {\textendash} U.S. National Science Foundation-Office of Polar Programs,
U.S. National Science Foundation-Physics Division,
U.S. National Science Foundation-EPSCoR,
Wisconsin Alumni Research Foundation,
Center for High Throughput Computing (CHTC) at the University of Wisconsin{\textendash}Madison,
Open Science Grid (OSG),
Extreme Science and Engineering Discovery Environment (XSEDE),
Frontera computing project at the Texas Advanced Computing Center,
U.S. Department of Energy-National Energy Research Scientific Computing Center,
Particle astrophysics research computing center at the University of Maryland,
Institute for Cyber-Enabled Research at Michigan State University,
and Astroparticle physics computational facility at Marquette University;
Belgium {\textendash} Funds for Scientific Research (FRS-FNRS and FWO),
FWO Odysseus and Big Science programmes,
and Belgian Federal Science Policy Office (Belspo);
Germany {\textendash} Bundesministerium f{\"u}r Bildung und Forschung (BMBF),
Deutsche Forschungsgemeinschaft (DFG),
Helmholtz Alliance for Astroparticle Physics (HAP),
Initiative and Networking Fund of the Helmholtz Association,
Deutsches Elektronen Synchrotron (DESY),
and High Performance Computing cluster of the RWTH Aachen;
Sweden {\textendash} Swedish Research Council,
Swedish Polar Research Secretariat,
Swedish National Infrastructure for Computing (SNIC),
and Knut and Alice Wallenberg Foundation;
Australia {\textendash} Australian Research Council;
Canada {\textendash} Natural Sciences and Engineering Research Council of Canada,
Calcul Qu{\'e}bec, Compute Ontario, Canada Foundation for Innovation, WestGrid, and Compute Canada;
Denmark {\textendash} Villum Fonden and Carlsberg Foundation;
New Zealand {\textendash} Marsden Fund;
Japan {\textendash} Japan Society for Promotion of Science (JSPS)
and Institute for Global Prominent Research (IGPR) of Chiba University;
Korea {\textendash} National Research Foundation of Korea (NRF);
Switzerland {\textendash} Swiss National Science Foundation (SNSF);
United Kingdom {\textendash} Department of Physics, University of Oxford.

\end{acknowledgments}

\bibliographystyle{apsrev4-1}
\bibliography{main}

\begin{thebibliography}{61}%
\makeatletter
\providecommand \@ifxundefined [1]{%
 \@ifx{#1\undefined}
}%
\providecommand \@ifnum [1]{%
 \ifnum #1\expandafter \@firstoftwo
 \else \expandafter \@secondoftwo
 \fi
}%
\providecommand \@ifx [1]{%
 \ifx #1\expandafter \@firstoftwo
 \else \expandafter \@secondoftwo
 \fi
}%
\providecommand \natexlab [1]{#1}%
\providecommand \enquote  [1]{``#1''}%
\providecommand \bibnamefont  [1]{#1}%
\providecommand \bibfnamefont [1]{#1}%
\providecommand \citenamefont [1]{#1}%
\providecommand \href@noop [0]{\@secondoftwo}%
\providecommand \href [0]{\begingroup \@sanitize@url \@href}%
\providecommand \@href[1]{\@@startlink{#1}\@@href}%
\providecommand \@@href[1]{\endgroup#1\@@endlink}%
\providecommand \@sanitize@url [0]{\catcode `\\12\catcode `\$12\catcode
  `\&12\catcode `\#12\catcode `\^12\catcode `\_12\catcode `\%12\relax}%
\providecommand \@@startlink[1]{}%
\providecommand \@@endlink[0]{}%
\providecommand \url  [0]{\begingroup\@sanitize@url \@url }%
\providecommand \@url [1]{\endgroup\@href {#1}{\urlprefix }}%
\providecommand \urlprefix  [0]{URL }%
\providecommand \Eprint [0]{\href }%
\providecommand \doibase [0]{http://dx.doi.org/}%
\providecommand \selectlanguage [0]{\@gobble}%
\providecommand \bibinfo  [0]{\@secondoftwo}%
\providecommand \bibfield  [0]{\@secondoftwo}%
\providecommand \translation [1]{[#1]}%
\providecommand \BibitemOpen [0]{}%
\providecommand \bibitemStop [0]{}%
\providecommand \bibitemNoStop [0]{.\EOS\space}%
\providecommand \EOS [0]{\spacefactor3000\relax}%
\providecommand \BibitemShut  [1]{\csname bibitem#1\endcsname}%
\let\auto@bib@innerbib\@empty
\bibitem [{\citenamefont {Kampert}\ and\ \citenamefont
  {Unger}(2012)}]{Kampert:2012mx}%
  \BibitemOpen
  \bibfield  {author} {\bibinfo {author} {\bibfnamefont {K.-H.}\ \bibnamefont
  {Kampert}}\ and\ \bibinfo {author} {\bibfnamefont {M.}~\bibnamefont
  {Unger}},\ }\href {\doibase 10.1016/j.astropartphys.2012.02.004} {\bibfield
  {journal} {\bibinfo  {journal} {Astropart. Phys.}\ }\textbf {\bibinfo
  {volume} {35}},\ \bibinfo {pages} {660} (\bibinfo {year} {2012})}\BibitemShut
  {NoStop}%
\bibitem [{\citenamefont {Engel}\ \emph {et~al.}(2011)\citenamefont {Engel},
  \citenamefont {Heck},\ and\ \citenamefont {Pierog}}]{Engel:2011zzb}%
  \BibitemOpen
  \bibfield  {author} {\bibinfo {author} {\bibfnamefont {R.}~\bibnamefont
  {Engel}}, \bibinfo {author} {\bibfnamefont {D.}~\bibnamefont {Heck}}, \ and\
  \bibinfo {author} {\bibfnamefont {T.}~\bibnamefont {Pierog}},\ }\href
  {\doibase 10.1146/annurev.nucl.012809.104544} {\bibfield  {journal} {\bibinfo
   {journal} {Ann. Rev. Nucl. Part. Sci.}\ }\textbf {\bibinfo {volume} {61}},\
  \bibinfo {pages} {467} (\bibinfo {year} {2011})}\BibitemShut {NoStop}%
\bibitem [{\citenamefont {Albrecht}\ \emph {et~al.}(2022)\citenamefont
  {Albrecht} \emph {et~al.}}]{Albrecht:2021yla}%
  \BibitemOpen
  \bibfield  {author} {\bibinfo {author} {\bibfnamefont {J.}~\bibnamefont
  {Albrecht}} \emph {et~al.},\ }\href {\doibase 10.1007/s10509-022-04054-5}
  {\bibfield  {journal} {\bibinfo  {journal} {Astrophys. Space Sci.}\ }\textbf
  {\bibinfo {volume} {367}},\ \bibinfo {pages} {27} (\bibinfo {year}
  {2022})}\BibitemShut {NoStop}%
\bibitem [{\citenamefont {Riehn}\ \emph {et~al.}(2016)\citenamefont {Riehn},
  \citenamefont {Engel}, \citenamefont {Fedynitch}, \citenamefont {Gaisser},\
  and\ \citenamefont {Stanev}}]{Riehn:2015oba}%
  \BibitemOpen
  \bibfield  {author} {\bibinfo {author} {\bibfnamefont {F.}~\bibnamefont
  {Riehn}}, \bibinfo {author} {\bibfnamefont {R.}~\bibnamefont {Engel}},
  \bibinfo {author} {\bibfnamefont {A.}~\bibnamefont {Fedynitch}}, \bibinfo
  {author} {\bibfnamefont {T.~K.}\ \bibnamefont {Gaisser}}, \ and\ \bibinfo
  {author} {\bibfnamefont {T.}~\bibnamefont {Stanev}},\ }\bibfield  {booktitle}
  {\emph {\bibinfo {booktitle} {{Proceedings, 34th International Cosmic Ray
  Conference (ICRC 2015): The Hague, The Netherlands, July 30-August 6,
  2015}}},\ }\href {\doibase 10.22323/1.236.0558} {\bibfield  {journal}
  {\bibinfo  {journal} {PoS}\ }\textbf {\bibinfo {volume} {ICRC2015}},\
  \bibinfo {pages} {558} (\bibinfo {year} {2016})}\BibitemShut {NoStop}%
\bibitem [{\citenamefont {Riehn}\ \emph {et~al.}(2020)\citenamefont {Riehn},
  \citenamefont {Engel}, \citenamefont {Fedynitch}, \citenamefont {Gaisser},\
  and\ \citenamefont {Stanev}}]{Engel:2019dsg}%
  \BibitemOpen
  \bibfield  {author} {\bibinfo {author} {\bibfnamefont {F.}~\bibnamefont
  {Riehn}}, \bibinfo {author} {\bibfnamefont {R.}~\bibnamefont {Engel}},
  \bibinfo {author} {\bibfnamefont {A.}~\bibnamefont {Fedynitch}}, \bibinfo
  {author} {\bibfnamefont {T.~K.}\ \bibnamefont {Gaisser}}, \ and\ \bibinfo
  {author} {\bibfnamefont {T.}~\bibnamefont {Stanev}},\ }\href {\doibase
  10.1103/PhysRevD.102.063002} {\bibfield  {journal} {\bibinfo  {journal}
  {Phys. Rev. D}\ }\textbf {\bibinfo {volume} {102}},\ \bibinfo {pages}
  {063002} (\bibinfo {year} {2020})}\BibitemShut {NoStop}%
\bibitem [{\citenamefont {Pierog}\ \emph {et~al.}(2015)\citenamefont {Pierog},
  \citenamefont {Karpenko}, \citenamefont {Katzy}, \citenamefont {Yatsenko},\
  and\ \citenamefont {Werner}}]{Pierog:2013ria}%
  \BibitemOpen
  \bibfield  {author} {\bibinfo {author} {\bibfnamefont {T.}~\bibnamefont
  {Pierog}}, \bibinfo {author} {\bibfnamefont {I.}~\bibnamefont {Karpenko}},
  \bibinfo {author} {\bibfnamefont {J.~M.}\ \bibnamefont {Katzy}}, \bibinfo
  {author} {\bibfnamefont {E.}~\bibnamefont {Yatsenko}}, \ and\ \bibinfo
  {author} {\bibfnamefont {K.}~\bibnamefont {Werner}},\ }\href {\doibase
  10.1103/PhysRevC.92.034906} {\bibfield  {journal} {\bibinfo  {journal} {Phys.
  Rev.}\ }\textbf {\bibinfo {volume} {C92}},\ \bibinfo {pages} {034906}
  (\bibinfo {year} {2015})}\BibitemShut {NoStop}%
\bibitem [{\citenamefont {Ostapchenko}(2013)}]{Ostapchenko:2013pia}%
  \BibitemOpen
  \bibfield  {author} {\bibinfo {author} {\bibfnamefont {S.}~\bibnamefont
  {Ostapchenko}},\ }\bibfield  {booktitle} {\emph {\bibinfo {booktitle}
  {{Proceedings, 17th International Symposium on Very High Energy Cosmic Ray
  Interactions (ISVHECRI 2012): Berlin, Germany, August 10-15, 2012}}},\ }\href
  {\doibase 10.1051/epjconf/20125202001} {\bibfield  {journal} {\bibinfo
  {journal} {EPJ Web Conf.}\ }\textbf {\bibinfo {volume} {52}},\ \bibinfo
  {pages} {02001} (\bibinfo {year} {2013})}\BibitemShut {NoStop}%
\bibitem [{\citenamefont {Ahn}\ \emph {et~al.}(2009)\citenamefont {Ahn},
  \citenamefont {Engel}, \citenamefont {Gaisser}, \citenamefont {Lipari},\ and\
  \citenamefont {Stanev}}]{Ahn:2009wx}%
  \BibitemOpen
  \bibfield  {author} {\bibinfo {author} {\bibfnamefont {E.-J.}\ \bibnamefont
  {Ahn}}, \bibinfo {author} {\bibfnamefont {R.}~\bibnamefont {Engel}}, \bibinfo
  {author} {\bibfnamefont {T.~K.}\ \bibnamefont {Gaisser}}, \bibinfo {author}
  {\bibfnamefont {P.}~\bibnamefont {Lipari}}, \ and\ \bibinfo {author}
  {\bibfnamefont {T.}~\bibnamefont {Stanev}},\ }\href {\doibase
  10.1103/PhysRevD.80.094003} {\bibfield  {journal} {\bibinfo  {journal} {Phys.
  Rev.}\ }\textbf {\bibinfo {volume} {D80}},\ \bibinfo {pages} {094003}
  (\bibinfo {year} {2009})}\BibitemShut {NoStop}%
\bibitem [{\citenamefont {{Heitler}}(1954)}]{Heitler}%
  \BibitemOpen
  \bibfield  {author} {\bibinfo {author} {\bibfnamefont {W.}~\bibnamefont
  {{Heitler}}},\ }\href@noop {} {\emph {\bibinfo {title} {{Quantum theory of
  radiation}}}},\ \bibinfo {edition} {3rd}\ ed.\ (\bibinfo  {publisher} {Oxford
  University Press},\ \bibinfo {year} {1954})\ \bibinfo {note} {international
  Series of Monographs on Physics}\BibitemShut {NoStop}%
\bibitem [{\citenamefont {Matthews}(2005)}]{Matthews:2005sd}%
  \BibitemOpen
  \bibfield  {author} {\bibinfo {author} {\bibfnamefont {J.}~\bibnamefont
  {Matthews}},\ }\href {\doibase 10.1016/j.astropartphys.2004.09.003}
  {\bibfield  {journal} {\bibinfo  {journal} {Astropart. Phys.}\ }\textbf
  {\bibinfo {volume} {22}},\ \bibinfo {pages} {387} (\bibinfo {year}
  {2005})}\BibitemShut {NoStop}%
\bibitem [{\citenamefont {Ulrich}\ \emph {et~al.}(2011)\citenamefont {Ulrich},
  \citenamefont {Engel},\ and\ \citenamefont {Unger}}]{Ulrich:2010rg}%
  \BibitemOpen
  \bibfield  {author} {\bibinfo {author} {\bibfnamefont {R.}~\bibnamefont
  {Ulrich}}, \bibinfo {author} {\bibfnamefont {R.}~\bibnamefont {Engel}}, \
  and\ \bibinfo {author} {\bibfnamefont {M.}~\bibnamefont {Unger}},\ }\href
  {\doibase 10.1103/PhysRevD.83.054026} {\bibfield  {journal} {\bibinfo
  {journal} {Phys. Rev.}\ }\textbf {\bibinfo {volume} {D83}},\ \bibinfo {pages}
  {054026} (\bibinfo {year} {2011})}\BibitemShut {NoStop}%
\bibitem [{\citenamefont {Greisen}(1960)}]{Greisen:1960wc}%
  \BibitemOpen
  \bibfield  {author} {\bibinfo {author} {\bibfnamefont {K.}~\bibnamefont
  {Greisen}},\ }\href {\doibase 10.1146/annurev.ns.10.120160.000431} {\bibfield
   {journal} {\bibinfo  {journal} {Ann. Rev. Nucl. Part. Sci.}\ }\textbf
  {\bibinfo {volume} {10}},\ \bibinfo {pages} {63} (\bibinfo {year}
  {1960})}\BibitemShut {NoStop}%
\bibitem [{\citenamefont {Bennett}\ and\ \citenamefont
  {Greisen}(1961)}]{PhysRev.124.1982}%
  \BibitemOpen
  \bibfield  {author} {\bibinfo {author} {\bibfnamefont {S.}~\bibnamefont
  {Bennett}}\ and\ \bibinfo {author} {\bibfnamefont {K.}~\bibnamefont
  {Greisen}},\ }\href {\doibase 10.1103/PhysRev.124.1982} {\bibfield  {journal}
  {\bibinfo  {journal} {Phys. Rev.}\ }\textbf {\bibinfo {volume} {124}},\
  \bibinfo {pages} {1982} (\bibinfo {year} {1961})}\BibitemShut {NoStop}%
\bibitem [{\citenamefont {{Rossi, B.}}(1960)}]{RossiICRC1960}%
  \BibitemOpen
  \bibfield  {author} {\bibinfo {author} {\bibnamefont {{Rossi, B.}}},\
  }\href@noop {} {\bibfield  {journal} {\bibinfo  {journal} {{Proceedings, 6th
  International Cosmic Ray Conference (ICRC 1959): Moscow, USSR, 1959}}\
  }\textbf {\bibinfo {volume} {4}},\ \bibinfo {pages} {18} (\bibinfo {year}
  {1960})}\BibitemShut {NoStop}%
\bibitem [{\citenamefont {Nagano}\ \emph {et~al.}(1984)\citenamefont {Nagano}
  \emph {et~al.}}]{Akeno_1984}%
  \BibitemOpen
  \bibfield  {author} {\bibinfo {author} {\bibfnamefont {M.}~\bibnamefont
  {Nagano}} \emph {et~al.},\ }\href
  {http://stacks.iop.org/0305-4616/10/i=9/a=016} {\bibfield  {journal}
  {\bibinfo  {journal} {J. Phys. G: Nucl. Phys.}\ }\textbf {\bibinfo {volume}
  {10}},\ \bibinfo {pages} {1295} (\bibinfo {year} {1984})}\BibitemShut
  {NoStop}%
\bibitem [{\citenamefont {Hayashida}\ \emph {et~al.}(1995)\citenamefont
  {Hayashida} \emph {et~al.}}]{Akeno_muons_1995}%
  \BibitemOpen
  \bibfield  {author} {\bibinfo {author} {\bibfnamefont {N.}~\bibnamefont
  {Hayashida}} \emph {et~al.},\ }\href
  {http://stacks.iop.org/0954-3899/21/i=8/a=008} {\bibfield  {journal}
  {\bibinfo  {journal} {J. Phys. G: Nucl. Phys.}\ }\textbf {\bibinfo {volume}
  {21}},\ \bibinfo {pages} {1101} (\bibinfo {year} {1995})}\BibitemShut
  {NoStop}%
\bibitem [{\citenamefont {Apel}\ \emph {et~al.}(2013)\citenamefont {Apel} \emph
  {et~al.}}]{Apel:2013dga}%
  \BibitemOpen
  \bibfield  {author} {\bibinfo {author} {\bibfnamefont {W.~D.}\ \bibnamefont
  {Apel}} \emph {et~al.} (\bibinfo {collaboration} {KASCADE-Grande
  Collaboration}),\ }\href {\doibase 10.1016/j.astropartphys.2013.06.004}
  {\bibfield  {journal} {\bibinfo  {journal} {Astropart. Phys.}\ }\textbf
  {\bibinfo {volume} {47}},\ \bibinfo {pages} {54} (\bibinfo {year}
  {2013})}\BibitemShut {NoStop}%
\bibitem [{\citenamefont {Apel}\ \emph {et~al.}(2015)\citenamefont {Apel} \emph
  {et~al.}}]{Apel:2014qqa}%
  \BibitemOpen
  \bibfield  {author} {\bibinfo {author} {\bibfnamefont {W.~D.}\ \bibnamefont
  {Apel}} \emph {et~al.} (\bibinfo {collaboration} {KASCADE-Grande
  Collaboration}),\ }\href {\doibase 10.1016/j.astropartphys.2014.12.001}
  {\bibfield  {journal} {\bibinfo  {journal} {Astropart. Phys.}\ }\textbf
  {\bibinfo {volume} {65}},\ \bibinfo {pages} {55} (\bibinfo {year}
  {2015})}\BibitemShut {NoStop}%
\bibitem [{\citenamefont {A.~N.~Bunner}\ and\ \citenamefont
  {Landecker}(1968)}]{CornellFlourescence}%
  \BibitemOpen
  \bibfield  {author} {\bibinfo {author} {\bibfnamefont {K.~G.}\ \bibnamefont
  {A.~N.~Bunner}}\ and\ \bibinfo {author} {\bibfnamefont {P.~B.}\ \bibnamefont
  {Landecker}},\ }\href@noop {} {\bibfield  {journal} {\bibinfo  {journal}
  {Can. J. Phys.}\ }\textbf {\bibinfo {volume} {46}},\ \bibinfo {pages} {S266}
  (\bibinfo {year} {1968})}\BibitemShut {NoStop}%
\bibitem [{\citenamefont {Baltrusaitis}\ \emph {et~al.}(1985)\citenamefont
  {Baltrusaitis} \emph {et~al.}}]{Baltrusaitis:1985mx}%
  \BibitemOpen
  \bibfield  {author} {\bibinfo {author} {\bibfnamefont {R.~M.}\ \bibnamefont
  {Baltrusaitis}} \emph {et~al.},\ }\href {\doibase
  10.1016/0168-9002(85)90658-8} {\bibfield  {journal} {\bibinfo  {journal}
  {Nucl. Instrum. Meth.}\ }\textbf {\bibinfo {volume} {A240}},\ \bibinfo
  {pages} {410} (\bibinfo {year} {1985})}\BibitemShut {NoStop}%
\bibitem [{\citenamefont {Abu-Zayyad}\ \emph {et~al.}(2000)\citenamefont
  {Abu-Zayyad} \emph {et~al.}}]{PhysRevLett.84.4276}%
  \BibitemOpen
  \bibfield  {author} {\bibinfo {author} {\bibfnamefont {T.}~\bibnamefont
  {Abu-Zayyad}} \emph {et~al.} (\bibinfo {collaboration} {HiRes-MIA
  Collaboration}),\ }\href {\doibase 10.1103/PhysRevLett.84.4276} {\bibfield
  {journal} {\bibinfo  {journal} {Phys. Rev. Lett.}\ }\textbf {\bibinfo
  {volume} {84}},\ \bibinfo {pages} {4276} (\bibinfo {year}
  {2000})}\BibitemShut {NoStop}%
\bibitem [{\citenamefont {Kalmykov}\ \emph {et~al.}(1997)\citenamefont
  {Kalmykov}, \citenamefont {Ostapchenko},\ and\ \citenamefont
  {Pavlov}}]{Kalmykov:1997te}%
  \BibitemOpen
  \bibfield  {author} {\bibinfo {author} {\bibfnamefont {N.~N.}\ \bibnamefont
  {Kalmykov}}, \bibinfo {author} {\bibfnamefont {S.~S.}\ \bibnamefont
  {Ostapchenko}}, \ and\ \bibinfo {author} {\bibfnamefont {A.~I.}\ \bibnamefont
  {Pavlov}},\ }\bibfield  {booktitle} {\emph {\bibinfo {booktitle}
  {{Proceedings, 9th International Symposium on Very High Energy Cosmic Ray
  Interactions (ISVHECRI 1996): Karlsruhe, Germany, August 19-23, 1996}}},\
  }\href {\doibase 10.1016/S0920-5632(96)00846-8} {\bibfield  {journal}
  {\bibinfo  {journal} {Nucl. Phys. Proc. Suppl.}\ }\textbf {\bibinfo {volume}
  {52}},\ \bibinfo {pages} {17} (\bibinfo {year} {1997})}\BibitemShut {NoStop}%
\bibitem [{\citenamefont {Fletcher}\ \emph {et~al.}(1994)\citenamefont
  {Fletcher}, \citenamefont {Gaisser}, \citenamefont {Lipari},\ and\
  \citenamefont {Stanev}}]{Fletcher:1994bd}%
  \BibitemOpen
  \bibfield  {author} {\bibinfo {author} {\bibfnamefont {R.~S.}\ \bibnamefont
  {Fletcher}}, \bibinfo {author} {\bibfnamefont {T.~K.}\ \bibnamefont
  {Gaisser}}, \bibinfo {author} {\bibfnamefont {P.}~\bibnamefont {Lipari}}, \
  and\ \bibinfo {author} {\bibfnamefont {T.}~\bibnamefont {Stanev}},\ }\href
  {\doibase 10.1103/PhysRevD.50.5710} {\bibfield  {journal} {\bibinfo
  {journal} {Phys. Rev.}\ }\textbf {\bibinfo {volume} {D50}},\ \bibinfo {pages}
  {5710} (\bibinfo {year} {1994})}\BibitemShut {NoStop}%
\bibitem [{\citenamefont {Aab}\ \emph {et~al.}(2015)\citenamefont {Aab} \emph
  {et~al.}}]{Aab:2014pza}%
  \BibitemOpen
  \bibfield  {author} {\bibinfo {author} {\bibfnamefont {A.}~\bibnamefont
  {Aab}} \emph {et~al.} (\bibinfo {collaboration} {Pierre Auger
  Collaboration}),\ }\href {\doibase 10.1103/PhysRevD.91.032003} {\bibfield
  {journal} {\bibinfo  {journal} {Phys. Rev. D}\ }\textbf {\bibinfo {volume}
  {91}},\ \bibinfo {pages} {032003} (\bibinfo {year} {2015})}\BibitemShut
  {NoStop}%
\bibitem [{\citenamefont {Aab}\ \emph {et~al.}(2016)\citenamefont {Aab} \emph
  {et~al.}}]{Aab:2016hkv}%
  \BibitemOpen
  \bibfield  {author} {\bibinfo {author} {\bibfnamefont {A.}~\bibnamefont
  {Aab}} \emph {et~al.} (\bibinfo {collaboration} {Pierre Auger
  Collaboration}),\ }\href {\doibase 10.1103/PhysRevLett.117.192001} {\bibfield
   {journal} {\bibinfo  {journal} {Phys. Rev. Lett.}\ }\textbf {\bibinfo
  {volume} {117}},\ \bibinfo {pages} {192001} (\bibinfo {year}
  {2016})}\BibitemShut {NoStop}%
\bibitem [{\citenamefont {Aab}\ \emph {et~al.}(2020)\citenamefont {Aab} \emph
  {et~al.}}]{Aab:2020frk}%
  \BibitemOpen
  \bibfield  {author} {\bibinfo {author} {\bibfnamefont {A.}~\bibnamefont
  {Aab}} \emph {et~al.} (\bibinfo {collaboration} {Pierre Auger
  Collaboration}),\ }\href {\doibase 10.1140/epjc/s10052-020-8055-y} {\bibfield
   {journal} {\bibinfo  {journal} {Eur. Phys. J. C}\ }\textbf {\bibinfo
  {volume} {80}},\ \bibinfo {pages} {751} (\bibinfo {year} {2020})}\BibitemShut
  {NoStop}%
\bibitem [{\citenamefont {Kokoulin}\ \emph {et~al.}(2009)\citenamefont
  {Kokoulin}, \citenamefont {Bogdanov}, \citenamefont {Mannocchi},
  \citenamefont {Petrukhin}, \citenamefont {Saavedra}, \citenamefont
  {Shutenko}, \citenamefont {Trinchero},\ and\ \citenamefont
  {Yashin}}]{Kokoulin:2009zz}%
  \BibitemOpen
  \bibfield  {author} {\bibinfo {author} {\bibfnamefont {R.~P.}\ \bibnamefont
  {Kokoulin}}, \bibinfo {author} {\bibfnamefont {A.~G.}\ \bibnamefont
  {Bogdanov}}, \bibinfo {author} {\bibfnamefont {G.}~\bibnamefont {Mannocchi}},
  \bibinfo {author} {\bibfnamefont {A.~A.}\ \bibnamefont {Petrukhin}}, \bibinfo
  {author} {\bibfnamefont {O.}~\bibnamefont {Saavedra}}, \bibinfo {author}
  {\bibfnamefont {V.~V.}\ \bibnamefont {Shutenko}}, \bibinfo {author}
  {\bibfnamefont {G.}~\bibnamefont {Trinchero}}, \ and\ \bibinfo {author}
  {\bibfnamefont {I.~I.}\ \bibnamefont {Yashin}},\ }\bibfield  {booktitle}
  {\emph {\bibinfo {booktitle} {{Proceedings, 15th International Symposium on
  Very High Energy Cosmic Ray Interactions (ISVHECRI 2008): Paris, France,
  September 1-6, 2008}}},\ }\href {\doibase 10.1016/j.nuclphysbps.2009.09.018}
  {\bibfield  {journal} {\bibinfo  {journal} {Nucl. Phys. Proc. Suppl.}\
  }\textbf {\bibinfo {volume} {196}},\ \bibinfo {pages} {106} (\bibinfo {year}
  {2009})}\BibitemShut {NoStop}%
\bibitem [{\citenamefont {Saavedra
  San~Martin}(2017)}]{SaavedraSanMartin:2017ito}%
  \BibitemOpen
  \bibfield  {author} {\bibinfo {author} {\bibfnamefont {O.}~\bibnamefont
  {Saavedra San~Martin}},\ }\bibfield  {booktitle} {\emph {\bibinfo {booktitle}
  {{Proceedings, 6th School on Cosmic Rays and Astrophysics: Tuxtla Gutiérrez,
  Chiapas, México, November 17-25, 2015}}},\ }\href {\doibase
  10.1088/1742-6596/866/1/012010} {\bibfield  {journal} {\bibinfo  {journal}
  {J. Phys. Conf. Ser.}\ }\textbf {\bibinfo {volume} {866}},\ \bibinfo {pages}
  {012010} (\bibinfo {year} {2017})}\BibitemShut {NoStop}%
\bibitem [{\citenamefont {Glushkov}\ \emph {et~al.}(2008)\citenamefont
  {Glushkov}, \citenamefont {Makarov}, \citenamefont {Pravdin}, \citenamefont
  {Sleptsov}, \citenamefont {Gorbunov}, \citenamefont {Rubtsov},\ and\
  \citenamefont {Troitsky}}]{Glushkov:2007gd}%
  \BibitemOpen
  \bibfield  {author} {\bibinfo {author} {\bibfnamefont {A.~V.}\ \bibnamefont
  {Glushkov}}, \bibinfo {author} {\bibfnamefont {I.~T.}\ \bibnamefont
  {Makarov}}, \bibinfo {author} {\bibfnamefont {M.~I.}\ \bibnamefont
  {Pravdin}}, \bibinfo {author} {\bibfnamefont {I.~E.}\ \bibnamefont
  {Sleptsov}}, \bibinfo {author} {\bibfnamefont {D.~S.}\ \bibnamefont
  {Gorbunov}}, \bibinfo {author} {\bibfnamefont {G.~I.}\ \bibnamefont
  {Rubtsov}}, \ and\ \bibinfo {author} {\bibfnamefont {S.~V.}\ \bibnamefont
  {Troitsky}},\ }\href {\doibase 10.1134/S0021364008040024} {\bibfield
  {journal} {\bibinfo  {journal} {JETP Lett.}\ }\textbf {\bibinfo {volume}
  {87}},\ \bibinfo {pages} {190} (\bibinfo {year} {2008})}\BibitemShut
  {NoStop}%
\bibitem [{\citenamefont {Fomin}\ \emph {et~al.}(2017)\citenamefont {Fomin},
  \citenamefont {Kalmykov}, \citenamefont {Karpikov}, \citenamefont {Kulikov},
  \citenamefont {Kuznetsov}, \citenamefont {Rubtsov}, \citenamefont {Sulakov},\
  and\ \citenamefont {Troitsky}}]{Fomin:2016kul}%
  \BibitemOpen
  \bibfield  {author} {\bibinfo {author} {\bibfnamefont {{\relax Yu}.~A.}\
  \bibnamefont {Fomin}}, \bibinfo {author} {\bibfnamefont {N.~N.}\ \bibnamefont
  {Kalmykov}}, \bibinfo {author} {\bibfnamefont {I.~S.}\ \bibnamefont
  {Karpikov}}, \bibinfo {author} {\bibfnamefont {G.~V.}\ \bibnamefont
  {Kulikov}}, \bibinfo {author} {\bibfnamefont {M.~{\relax Yu}.}\ \bibnamefont
  {Kuznetsov}}, \bibinfo {author} {\bibfnamefont {G.~I.}\ \bibnamefont
  {Rubtsov}}, \bibinfo {author} {\bibfnamefont {V.~P.}\ \bibnamefont
  {Sulakov}}, \ and\ \bibinfo {author} {\bibfnamefont {S.~V.}\ \bibnamefont
  {Troitsky}},\ }\href {\doibase 10.1016/j.astropartphys.2017.04.001}
  {\bibfield  {journal} {\bibinfo  {journal} {Astropart. Phys.}\ }\textbf
  {\bibinfo {volume} {92}},\ \bibinfo {pages} {1} (\bibinfo {year}
  {2017})}\BibitemShut {NoStop}%
\bibitem [{\citenamefont {Dedenko}\ \emph
  {et~al.}(2015{\natexlab{a}})\citenamefont {Dedenko}, \citenamefont
  {Lukyashin}, \citenamefont {Fedorova},\ and\ \citenamefont
  {Roganova}}]{Dedenko:2015qga}%
  \BibitemOpen
  \bibfield  {author} {\bibinfo {author} {\bibfnamefont {L.~G.}\ \bibnamefont
  {Dedenko}}, \bibinfo {author} {\bibfnamefont {A.~V.}\ \bibnamefont
  {Lukyashin}}, \bibinfo {author} {\bibfnamefont {G.~F.}\ \bibnamefont
  {Fedorova}}, \ and\ \bibinfo {author} {\bibfnamefont {T.~M.}\ \bibnamefont
  {Roganova}},\ }\bibfield  {booktitle} {\emph {\bibinfo {booktitle}
  {{Proceedings, 18th International Symposium on Very High Energy Cosmic Ray
  Interactions (ISVHECRI 2014): Geneva, Switzerland, August 18-22, 2014}}},\
  }\href {\doibase 10.1051/epjconf/20159910003} {\bibfield  {journal} {\bibinfo
   {journal} {EPJ Web Conf.}\ }\textbf {\bibinfo {volume} {99}},\ \bibinfo
  {pages} {10003} (\bibinfo {year} {2015}{\natexlab{a}})}\BibitemShut {NoStop}%
\bibitem [{\citenamefont {Dedenko}\ \emph
  {et~al.}(2015{\natexlab{b}})\citenamefont {Dedenko}, \citenamefont
  {Roganova},\ and\ \citenamefont {Fedorova}}]{Dedenko:2015hna}%
  \BibitemOpen
  \bibfield  {author} {\bibinfo {author} {\bibfnamefont {L.~G.}\ \bibnamefont
  {Dedenko}}, \bibinfo {author} {\bibfnamefont {T.~M.}\ \bibnamefont
  {Roganova}}, \ and\ \bibinfo {author} {\bibfnamefont {G.~F.}\ \bibnamefont
  {Fedorova}},\ }\href {\doibase 10.1134/S1063778815060083} {\bibfield
  {journal} {\bibinfo  {journal} {Phys. Atom. Nucl.}\ }\textbf {\bibinfo
  {volume} {78}},\ \bibinfo {pages} {840} (\bibinfo {year}
  {2015}{\natexlab{b}})}\BibitemShut {NoStop}%
\bibitem [{\citenamefont {Arteaga-Vel\'azquez}\ \emph
  {et~al.}(2015)\citenamefont {Arteaga-Vel\'azquez} \emph
  {et~al.}}]{Arteaga-Velazquez:2015jpa}%
  \BibitemOpen
  \bibfield  {author} {\bibinfo {author} {\bibfnamefont {J.~C.}\ \bibnamefont
  {Arteaga-Vel\'azquez}} \emph {et~al.},\ }\bibfield  {booktitle} {\emph
  {\bibinfo {booktitle} {{Proceedings, 18th International Symposium on Very
  High Energy Cosmic Ray Interactions (ISVHECRI 2014): Geneva, Switzerland,
  August 18-22, 2014}}},\ }\href {\doibase 10.1051/epjconf/20159912002}
  {\bibfield  {journal} {\bibinfo  {journal} {EPJ Web Conf.}\ }\textbf
  {\bibinfo {volume} {99}},\ \bibinfo {pages} {12002} (\bibinfo {year}
  {2015})}\BibitemShut {NoStop}%
\bibitem [{\citenamefont {Aartsen}\ \emph {et~al.}(2017)\citenamefont {Aartsen}
  \emph {et~al.}}]{Aartsen:2016nxy}%
  \BibitemOpen
  \bibfield  {author} {\bibinfo {author} {\bibfnamefont {M.~G.}\ \bibnamefont
  {Aartsen}} \emph {et~al.} (\bibinfo {collaboration} {IceCube
  Collaboration}),\ }\href {\doibase 10.1088/1748-0221/12/03/P03012} {\bibfield
   {journal} {\bibinfo  {journal} {JINST}\ }\textbf {\bibinfo {volume} {12}},\
  \bibinfo {pages} {P03012} (\bibinfo {year} {2017})}\BibitemShut {NoStop}%
\bibitem [{\citenamefont {Aartsen}\ \emph {et~al.}(2016)\citenamefont {Aartsen}
  \emph {et~al.}}]{Aartsen:2015nss}%
  \BibitemOpen
  \bibfield  {author} {\bibinfo {author} {\bibfnamefont {M.~G.}\ \bibnamefont
  {Aartsen}} \emph {et~al.} (\bibinfo {collaboration} {IceCube
  Collaboration}),\ }\href {\doibase 10.1016/j.astropartphys.2016.01.006}
  {\bibfield  {journal} {\bibinfo  {journal} {Astropart. Phys.}\ }\textbf
  {\bibinfo {volume} {78}},\ \bibinfo {pages} {1} (\bibinfo {year}
  {2016})}\BibitemShut {NoStop}%
\bibitem [{\citenamefont {Abbasi}\ \emph
  {et~al.}(2013{\natexlab{a}})\citenamefont {Abbasi} \emph
  {et~al.}}]{Abbasi:2012kza}%
  \BibitemOpen
  \bibfield  {author} {\bibinfo {author} {\bibfnamefont {R.}~\bibnamefont
  {Abbasi}} \emph {et~al.} (\bibinfo {collaboration} {IceCube Collaboration}),\
  }\href {\doibase 10.1103/PhysRevD.87.012005} {\bibfield  {journal} {\bibinfo
  {journal} {Phys. Rev.}\ }\textbf {\bibinfo {volume} {D87}},\ \bibinfo {pages}
  {012005} (\bibinfo {year} {2013}{\natexlab{a}})}\BibitemShut {NoStop}%
\bibitem [{\citenamefont {Soldin}(2019)}]{Soldin:2018vak}%
  \BibitemOpen
  \bibfield  {author} {\bibinfo {author} {\bibfnamefont {D.}~\bibnamefont
  {Soldin}} (\bibinfo {collaboration} {IceCube Collaboration}),\ }\bibfield
  {booktitle} {\emph {\bibinfo {booktitle} {{Proceedings, 20th International
  Symposium on Very High Energy Cosmic Ray Interactions (ISVHECRI 2018):
  Nagoya, Japan, May 21-25, 2018}}},\ }\href {\doibase
  10.1051/epjconf/201920808007} {\bibfield  {journal} {\bibinfo  {journal} {EPJ
  Web Conf.}\ }\textbf {\bibinfo {volume} {208}},\ \bibinfo {pages} {08007}
  (\bibinfo {year} {2019})}\BibitemShut {NoStop}%
\bibitem [{\citenamefont {Abbasi}\ \emph
  {et~al.}(2013{\natexlab{b}})\citenamefont {Abbasi} \emph
  {et~al.}}]{IceCube:2012nn}%
  \BibitemOpen
  \bibfield  {author} {\bibinfo {author} {\bibfnamefont {R.}~\bibnamefont
  {Abbasi}} \emph {et~al.} (\bibinfo {collaboration} {IceCube Collaboration}),\
  }\href {\doibase 10.1016/j.nima.2012.10.067} {\bibfield  {journal} {\bibinfo
  {journal} {Nucl. Instrum. Meth.}\ }\textbf {\bibinfo {volume} {A700}},\
  \bibinfo {pages} {188} (\bibinfo {year} {2013}{\natexlab{b}})}\BibitemShut
  {NoStop}%
\bibitem [{\citenamefont {Aartsen}\ \emph {et~al.}(2020)\citenamefont {Aartsen}
  \emph {et~al.}}]{IceCube:2020yct}%
  \BibitemOpen
  \bibfield  {author} {\bibinfo {author} {\bibfnamefont {M.~G.}\ \bibnamefont
  {Aartsen}} \emph {et~al.} (\bibinfo {collaboration} {IceCube
  Collaboration}),\ }\href {\doibase 10.1103/PhysRevD.102.122001} {\bibfield
  {journal} {\bibinfo  {journal} {Phys. Rev. D}\ }\textbf {\bibinfo {volume}
  {102}},\ \bibinfo {pages} {122001} (\bibinfo {year} {2020})}\BibitemShut
  {NoStop}%
\bibitem [{\citenamefont {Aartsen}\ \emph {et~al.}(2013)\citenamefont {Aartsen}
  \emph {et~al.}}]{Aartsen:2013wda}%
  \BibitemOpen
  \bibfield  {author} {\bibinfo {author} {\bibfnamefont {M.~G.}\ \bibnamefont
  {Aartsen}} \emph {et~al.} (\bibinfo {collaboration} {IceCube
  Collaboration}),\ }\href {\doibase 10.1103/PhysRevD.88.042004} {\bibfield
  {journal} {\bibinfo  {journal} {Phys. Rev.}\ }\textbf {\bibinfo {volume}
  {D88}},\ \bibinfo {pages} {042004} (\bibinfo {year} {2013})}\BibitemShut
  {NoStop}%
\bibitem [{\citenamefont {Aartsen}\ \emph {et~al.}(2019)\citenamefont {Aartsen}
  \emph {et~al.}}]{IceCube:2019}%
  \BibitemOpen
  \bibfield  {author} {\bibinfo {author} {\bibfnamefont {M.~G.}\ \bibnamefont
  {Aartsen}} \emph {et~al.} (\bibinfo {collaboration} {IceCube
  Collaboration}),\ }\href {\doibase 10.1103/PhysRevD.100.082002} {\bibfield
  {journal} {\bibinfo  {journal} {Phys. Rev.}\ }\textbf {\bibinfo {volume}
  {D100}},\ \bibinfo {pages} {082002} (\bibinfo {year} {2019})}\BibitemShut
  {NoStop}%
\bibitem [{\citenamefont {Abbasi}\ \emph
  {et~al.}(2013{\natexlab{c}})\citenamefont {Abbasi} \emph
  {et~al.}}]{IceCube:2012vv}%
  \BibitemOpen
  \bibfield  {author} {\bibinfo {author} {\bibfnamefont {R.}~\bibnamefont
  {Abbasi}} \emph {et~al.} (\bibinfo {collaboration} {IceCube Collaboration}),\
  }\href {\doibase 10.1016/j.astropartphys.2012.11.003} {\bibfield  {journal}
  {\bibinfo  {journal} {Astropart. Phys.}\ }\textbf {\bibinfo {volume} {42}},\
  \bibinfo {pages} {15} (\bibinfo {year} {2013}{\natexlab{c}})}\BibitemShut
  {NoStop}%
\bibitem [{\citenamefont {Abbasi}\ \emph {et~al.}(2009)\citenamefont {Abbasi}
  \emph {et~al.}}]{Abbasi:2008aa}%
  \BibitemOpen
  \bibfield  {author} {\bibinfo {author} {\bibfnamefont {R.}~\bibnamefont
  {Abbasi}} \emph {et~al.} (\bibinfo {collaboration} {IceCube Collaboration}),\
  }\href {\doibase 10.1016/j.nima.2009.01.001} {\bibfield  {journal} {\bibinfo
  {journal} {Nucl. Instrum. Meth.}\ }\textbf {\bibinfo {volume} {A601}},\
  \bibinfo {pages} {294} (\bibinfo {year} {2009})}\BibitemShut {NoStop}%
\bibitem [{\citenamefont {Abbasi}\ \emph {et~al.}(2010)\citenamefont {Abbasi}
  \emph {et~al.}}]{Abbasi:2010vc}%
  \BibitemOpen
  \bibfield  {author} {\bibinfo {author} {\bibfnamefont {R.}~\bibnamefont
  {Abbasi}} \emph {et~al.} (\bibinfo {collaboration} {IceCube Collaboration}),\
  }\href {\doibase 10.1016/j.nima.2010.03.102} {\bibfield  {journal} {\bibinfo
  {journal} {Nucl. Instrum. Meth.}\ }\textbf {\bibinfo {volume} {A618}},\
  \bibinfo {pages} {139} (\bibinfo {year} {2010})}\BibitemShut {NoStop}%
\bibitem [{\citenamefont {Abbasi}\ \emph
  {et~al.}(2013{\natexlab{d}})\citenamefont {Abbasi} \emph
  {et~al.}}]{Abbasi:2012wn}%
  \BibitemOpen
  \bibfield  {author} {\bibinfo {author} {\bibfnamefont {R.}~\bibnamefont
  {Abbasi}} \emph {et~al.} (\bibinfo {collaboration} {IceCube Collaboration}),\
  }\href {\doibase 10.1016/j.astropartphys.2013.01.016} {\bibfield  {journal}
  {\bibinfo  {journal} {Astropart. Phys.}\ }\textbf {\bibinfo {volume} {44}},\
  \bibinfo {pages} {40} (\bibinfo {year} {2013}{\natexlab{d}})}\BibitemShut
  {NoStop}%
\bibitem [{\citenamefont {Gaisser}(2012)}]{Gaisser:2011cc}%
  \BibitemOpen
  \bibfield  {author} {\bibinfo {author} {\bibfnamefont {T.~K.}\ \bibnamefont
  {Gaisser}},\ }\href {\doibase 10.1016/j.astropartphys.2012.02.010} {\bibfield
   {journal} {\bibinfo  {journal} {Astropart. Phys.}\ }\textbf {\bibinfo
  {volume} {35}},\ \bibinfo {pages} {801} (\bibinfo {year} {2012})}\BibitemShut
  {NoStop}%
\bibitem [{\citenamefont {Heck}\ \emph {et~al.}(1998)\citenamefont {Heck},
  \citenamefont {Schatz}, \citenamefont {Thouw}, \citenamefont {Knapp},\ and\
  \citenamefont {Capdevielle}}]{corsika}%
  \BibitemOpen
  \bibfield  {author} {\bibinfo {author} {\bibfnamefont {D.}~\bibnamefont
  {Heck}}, \bibinfo {author} {\bibfnamefont {G.}~\bibnamefont {Schatz}},
  \bibinfo {author} {\bibfnamefont {T.}~\bibnamefont {Thouw}}, \bibinfo
  {author} {\bibfnamefont {J.}~\bibnamefont {Knapp}}, \ and\ \bibinfo {author}
  {\bibfnamefont {J.~N.}\ \bibnamefont {Capdevielle}},\ }\href@noop {}
  {\bibfield  {journal} {\bibinfo  {journal} {FZKA-6019}\ } (\bibinfo {year}
  {1998})}\BibitemShut {NoStop}%
\bibitem [{\citenamefont {Agostinelli}\ \emph {et~al.}(2003)\citenamefont
  {Agostinelli} \emph {et~al.}}]{geant4}%
  \BibitemOpen
  \bibfield  {author} {\bibinfo {author} {\bibfnamefont {S.}~\bibnamefont
  {Agostinelli}} \emph {et~al.} (\bibinfo {collaboration} {GEANT4
  Collaboration}),\ }\href {\doibase 10.1016/S0168-9002(03)01368-8} {\bibfield
  {journal} {\bibinfo  {journal} {Nucl. Instrum. Meth.}\ }\textbf {\bibinfo
  {volume} {A506}},\ \bibinfo {pages} {250} (\bibinfo {year}
  {2003})}\BibitemShut {NoStop}%
\bibitem [{\citenamefont {De~Ridder}(2019)}]{sderidder_thesis}%
  \BibitemOpen
  \bibfield  {author} {\bibinfo {author} {\bibfnamefont {S.}~\bibnamefont
  {De~Ridder}},\ }\emph {\bibinfo {title} {Sensitivity of IceCube Cosmic Ray
  measurements to the hadronic interaction models}},\ \href@noop {} {Ph.D.
  thesis},\ \bibinfo  {school} {Ghent University} (\bibinfo {year}
  {2019})\BibitemShut {NoStop}%
\bibitem [{\citenamefont {Böhlen}\ \emph {et~al.}(2014)\citenamefont {Böhlen}
  \emph {et~al.}}]{fluka}%
  \BibitemOpen
  \bibfield  {author} {\bibinfo {author} {\bibfnamefont {T.~T.}\ \bibnamefont
  {Böhlen}} \emph {et~al.},\ }\href@noop {} {\bibfield  {journal} {\bibinfo
  {journal} {Nuclear Data Sheets}\ }\textbf {\bibinfo {volume} {120}},\
  \bibinfo {pages} {211} (\bibinfo {year} {2014})}\BibitemShut {NoStop}%
\bibitem [{\citenamefont {Ferrari}\ \emph {et~al.}(2005)\citenamefont
  {Ferrari}, \citenamefont {Sala}, \citenamefont {Fasso},\ and\ \citenamefont
  {Ranft}}]{Ferrari:2005zk}%
  \BibitemOpen
  \bibfield  {author} {\bibinfo {author} {\bibfnamefont {A.}~\bibnamefont
  {Ferrari}}, \bibinfo {author} {\bibfnamefont {P.~R.}\ \bibnamefont {Sala}},
  \bibinfo {author} {\bibfnamefont {A.}~\bibnamefont {Fasso}}, \ and\ \bibinfo
  {author} {\bibfnamefont {J.}~\bibnamefont {Ranft}},\ }\href@noop {}
  {\bibfield  {journal} {\bibinfo  {journal} {CERN-2005-010, SLAC-R-773,
  INFN-TC-05-11}\ } (\bibinfo {year} {2005})}\BibitemShut {NoStop}%
\bibitem [{\citenamefont {K\'egl}\ and\ \citenamefont
  {Veberi\u{c}}(2015)}]{single_mu_response}%
  \BibitemOpen
  \bibfield  {author} {\bibinfo {author} {\bibfnamefont {B.}~\bibnamefont
  {K\'egl}}\ and\ \bibinfo {author} {\bibfnamefont {D.}~\bibnamefont
  {Veberi\u{c}}},\ }\href@noop {} {\bibfield  {journal} {\bibinfo  {journal}
  {Pierre Auger Observatory internal notes}\ } (\bibinfo {year} {2015})},\
  \Eprint {http://arxiv.org/abs/1502.03347} {arXiv:1502.03347 [astro-ph.IM]}
  \BibitemShut {NoStop}%
\bibitem [{\citenamefont {Grushka}(1972)}]{exp_gaus}%
  \BibitemOpen
  \bibfield  {author} {\bibinfo {author} {\bibfnamefont {E.}~\bibnamefont
  {Grushka}},\ }\href {\doibase 10.1021/ac60319a011} {\bibfield  {journal}
  {\bibinfo  {journal} {Anal. Chem.}\ }\textbf {\bibinfo {volume} {44}},\
  \bibinfo {pages} {1733} (\bibinfo {year} {1972})}\BibitemShut {NoStop}%
\bibitem [{\citenamefont {Stanev}\ \emph {et~al.}(2014)\citenamefont {Stanev},
  \citenamefont {Gaisser},\ and\ \citenamefont {Tilav}}]{Stanev:2014mla}%
  \BibitemOpen
  \bibfield  {author} {\bibinfo {author} {\bibfnamefont {T.}~\bibnamefont
  {Stanev}}, \bibinfo {author} {\bibfnamefont {T.~K.}\ \bibnamefont {Gaisser}},
  \ and\ \bibinfo {author} {\bibfnamefont {S.}~\bibnamefont {Tilav}},\
  }\bibfield  {booktitle} {\emph {\bibinfo {booktitle} {{Proceedings, 4th Roma
  International Conference on Astro-Particle Physics (RICAP 13): Rome, Italy,
  May 22-24, 2013}}},\ }\href {\doibase 10.1016/j.nima.2013.11.094} {\bibfield
  {journal} {\bibinfo  {journal} {Nucl. Instrum. Meth.}\ }\textbf {\bibinfo
  {volume} {A742}},\ \bibinfo {pages} {42} (\bibinfo {year}
  {2014})}\BibitemShut {NoStop}%
\bibitem [{\citenamefont {Dembinski}\ \emph {et~al.}(2017)\citenamefont
  {Dembinski}, \citenamefont {Engel}, \citenamefont {Fedynitch}, \citenamefont
  {Gaisser}, \citenamefont {Riehn},\ and\ \citenamefont
  {Stanev}}]{Dembinski:2015xtn}%
  \BibitemOpen
  \bibfield  {author} {\bibinfo {author} {\bibfnamefont {H.~P.}\ \bibnamefont
  {Dembinski}}, \bibinfo {author} {\bibfnamefont {R.}~\bibnamefont {Engel}},
  \bibinfo {author} {\bibfnamefont {A.}~\bibnamefont {Fedynitch}}, \bibinfo
  {author} {\bibfnamefont {T.~K.}\ \bibnamefont {Gaisser}}, \bibinfo {author}
  {\bibfnamefont {F.}~\bibnamefont {Riehn}}, \ and\ \bibinfo {author}
  {\bibfnamefont {T.}~\bibnamefont {Stanev}},\ }\bibfield  {booktitle} {\emph
  {\bibinfo {booktitle} {{Proceedings, 35th International Cosmic Ray Conference
  (ICRC 2017): Busan, South Korea, July 12-20, 2017}}},\ }\href@noop {}
  {\bibfield  {journal} {\bibinfo  {journal} {PoS}\ }\textbf {\bibinfo {volume}
  {ICRC2017}},\ \bibinfo {pages} {533} (\bibinfo {year} {2017})}\BibitemShut
  {NoStop}%
\bibitem [{\citenamefont {Abbasi}\ \emph {et~al.}(2022)\citenamefont {Abbasi}
  \emph {et~al.}}]{DataRelease}%
  \BibitemOpen
  \bibfield  {author} {\bibinfo {author} {\bibfnamefont {R.}~\bibnamefont
  {Abbasi}} \emph {et~al.} (\bibinfo {collaboration} {IceCube Collaboration}),\
  }\href {\doibase 10.21234/sszj-qv50} {\bibfield  {journal} {\bibinfo
  {journal} {IceCube Public Data Release}\ } (\bibinfo {year} {2022}),\
  10.21234/sszj-qv50}\BibitemShut {NoStop}%
\bibitem [{\citenamefont {De~Ridder}\ \emph {et~al.}(2018)\citenamefont
  {De~Ridder}, \citenamefont {Dvorak},\ and\ \citenamefont
  {Gaisser}}]{DeRidder:2017alk}%
  \BibitemOpen
  \bibfield  {author} {\bibinfo {author} {\bibfnamefont {S.}~\bibnamefont
  {De~Ridder}}, \bibinfo {author} {\bibfnamefont {E.}~\bibnamefont {Dvorak}}, \
  and\ \bibinfo {author} {\bibfnamefont {T.~K.}\ \bibnamefont {Gaisser}}
  (\bibinfo {collaboration} {IceCube Collaboration}),\ }\bibfield  {booktitle}
  {\emph {\bibinfo {booktitle} {{Proceedings, 35th International Cosmic Ray
  Conference (ICRC 2017): Busan, South Korea, July 12-20, 2017}}},\ }\href
  {\doibase 10.22323/1.301.0319} {\bibfield  {journal} {\bibinfo  {journal}
  {PoS}\ }\textbf {\bibinfo {volume} {ICRC2017}},\ \bibinfo {pages} {319}
  (\bibinfo {year} {2018})}\BibitemShut {NoStop}%
\bibitem [{\citenamefont {Verpoest}\ \emph {et~al.}(2021)\citenamefont
  {Verpoest}, \citenamefont {Soldin},\ and\ \citenamefont
  {De~Ridder}}]{IceCube:2021ixw}%
  \BibitemOpen
  \bibfield  {author} {\bibinfo {author} {\bibfnamefont {S.}~\bibnamefont
  {Verpoest}}, \bibinfo {author} {\bibfnamefont {D.}~\bibnamefont {Soldin}}, \
  and\ \bibinfo {author} {\bibfnamefont {S.}~\bibnamefont {De~Ridder}}
  (\bibinfo {collaboration} {IceCube Collaboration}),\ }\bibfield  {booktitle}
  {\emph {\bibinfo {booktitle} {{Proceedings, 37th International Cosmic Ray
  Conference (ICRC 2021): Berlin, Germany, July 12-23, 2021}}},\ }\href
  {\doibase 10.22323/1.395.0357} {\bibfield  {journal} {\bibinfo  {journal}
  {PoS}\ }\textbf {\bibinfo {volume} {ICRC2021}},\ \bibinfo {pages} {357}
  (\bibinfo {year} {2021})}\BibitemShut {NoStop}%
\bibitem [{\citenamefont {Dembinski}\ \emph {et~al.}(2019)\citenamefont
  {Dembinski} \emph {et~al.}}]{Dembinski:2019uta}%
  \BibitemOpen
  \bibfield  {author} {\bibinfo {author} {\bibfnamefont {H.~P.}\ \bibnamefont
  {Dembinski}} \emph {et~al.} (\bibinfo {collaboration} {EAS-MSU, IceCube,
  KASCADE-Grande, NEVOD-DECOR, Pierre Auger, SUGAR, Telescope Array, Yakutsk
  EAS Array Collaborations}),\ }\bibfield  {booktitle} {\emph {\bibinfo
  {booktitle} {{Proceedings, Ultra High Energy Cosmic Rays (UHECR 2018): Paris,
  France, October 8-12, 2018}}},\ }\href {\doibase
  10.1051/epjconf/201921002004} {\bibfield  {journal} {\bibinfo  {journal} {EPJ
  Web Conf.}\ }\textbf {\bibinfo {volume} {210}},\ \bibinfo {pages} {02004}
  (\bibinfo {year} {2019})}\BibitemShut {NoStop}%
\bibitem [{\citenamefont {Soldin}(2021)}]{Soldin:2021wyv}%
  \BibitemOpen
  \bibfield  {author} {\bibinfo {author} {\bibfnamefont {D.}~\bibnamefont
  {Soldin}} (\bibinfo {collaboration} {EAS-MSU, IceCube, KASCADE-Grande,
  NEVOD-DECOR, Pierre Auger, SUGAR, Telescope Array, Yakutsk EAS Array
  Collaborations}),\ }\bibfield  {booktitle} {\emph {\bibinfo {booktitle}
  {{Proceedings, 37th International Cosmic Ray Conference (ICRC 2021): Berlin,
  Germany, July 12-23, 2021}}},\ }\href {\doibase 10.22323/1.395.0349}
  {\bibfield  {journal} {\bibinfo  {journal} {PoS}\ }\textbf {\bibinfo {volume}
  {ICRC2021}},\ \bibinfo {pages} {349} (\bibinfo {year} {2021})}\BibitemShut
  {NoStop}%
\bibitem [{\citenamefont {Cazon}(2020)}]{Cazon:2020zhx}%
  \BibitemOpen
  \bibfield  {author} {\bibinfo {author} {\bibfnamefont {L.}~\bibnamefont
  {Cazon}} (\bibinfo {collaboration} {EAS-MSU, IceCube, KASCADE Grande,
  NEVOD-DECOR, Pierre Auger, SUGAR, Telescope Array, Yakutsk EAS Array
  Collaborations}),\ }\bibfield  {booktitle} {\emph {\bibinfo {booktitle}
  {{Proceedings, 36th International Cosmic Ray Conference (ICRC 2019): Madison,
  USA, July 21-August 1, 2019}}},\ }\href {\doibase 10.22323/1.358.0214}
  {\bibfield  {journal} {\bibinfo  {journal} {PoS}\ }\textbf {\bibinfo {volume}
  {ICRC2019}},\ \bibinfo {pages} {214} (\bibinfo {year} {2020})}\BibitemShut
  {NoStop}%
\end{thebibliography}%

\end{document}